\documentclass[final,twocolumn,12pt]{elsarticle}
\usepackage[margin=1.5cm,includefoot]{geometry}
\usepackage[refcheck=false,todonotes=false]{autopaper}

\newcommand{\mytitle}{High-dimensional Bayesian Optimization of Hyperparameters for an Attention-based Network to Predict Materials Property: a Case Study on CrabNet using Ax and SAASBO}
\makeglossaries
\GlsXtrEnableEntryCounting{abbreviation}{3} 
\glssetcategoryattribute{abbreviation}{nohyper}{true}

\newabbreviation{discover}{DiSCoVeR}{Descending from Stochastic Clustering Variance Regression}
\newabbreviation{tgain}{T-GAIN}{Transformer-Generative Adaptive Inference Network}
\newabbreviation{hdbscan}{HDBSCAN*}{Hierarchical Density-based Spatial Clustering of Applications with Noise}
\newabbreviation{dbscan}{DBSCAN}{Density-based Spatial Clustering of Applications with Noise}
\newabbreviation{elmd}{ElMD}{Element Mover’s Distance}
\newabbreviation{umap}{UMAP}{Uniform Manifold Approximation and Projection}
\newabbreviation{densmap}{DensMAP}{Density-preserving Uniform Manifold Approximation and Projection}
\newabbreviation{kde}{KDE}{kernel density estimation}
\newabbreviation{tsne}{t-SNE}{t-distributed stochastic neighbor embeddings}
\newabbreviation{crabnet}{CrabNet}{Compositionally-Restricted Attention-Based Network}
\newabbreviation{dft}{DFT}{density functional theory}
\newabbreviation{fem}{FEM}{finite element modeling}
\newabbreviation{xrd}{XRD}{X-ray diffraction}
\newabbreviation{pfic}{PFIC}{predicted fraction of improved candidates}
\newabbreviation{cmli}{CMLI}{cumulative maximum likelihood of improvement}
\newabbreviation{loco-cv}{LOCO-CV}{leave-one-cluster-out cross-validation}
\newabbreviation{loo-cv}{LOO-CV}{leave-one-out cross-validation}
\newabbreviation{cdf}{CDF}{cumulative density function}
\newabbreviation{rmse}{RMSE}{root-mean-square error}
\newabbreviation{mae}{MAE}{mean absolute error}
\newabbreviation{mad}{MAD}{mean absolute deviation}
\newabbreviation{pdf}{PDF}{probability density function}
\newabbreviation{ml}{ML}{machine learning}
\newabbreviation{knn}{kNN}{k-nearest neighbor}
\newabbreviation{bowsr}{BOWSR}{Bayesian Optimization With Symmetry Relaxation}

\newabbreviation{ad}{AD}{adaptive design}
\newabbreviation{bo}{BO}{Bayesian optimization}
\newabbreviation{cv}{CV}{cross-validation}
\newabbreviation{ei}{EI}{expected improvement}
\newabbreviation{saasbo}{SAASBO}{sparse axis-aligned subspaces Bayesian optimization}
\newabbreviation{ard}{ARD}{automatic relevance determination}
\newabbreviation{moo}{MOO}{multi-objective optimization}

\newabbreviation{api}{API}{application programming interface}
\newabbreviation{gpu}{GPU}{graphical processing unit}
\newabbreviation{uq}{UQ}{uncertainty quantification}
\newabbreviation{uqq}{UQQ}{uncertainty quantification quality}
\newabbreviation{gpei}{GPEI}{Gaussian process expected improvement}

\zexternaldocument*{supp}

\begin{document}

\sloppy

\begin{frontmatter}

\title{\mytitle{}}

\author[myu]{Sterling G. Baird\corref{cor1}}
\ead{sterling.baird@utah.edu}
\author[myu,west]{Marianne Liu}
\author[myu]{Taylor D. Sparks}
\ead{sparks@eng.utah.edu}

\address[myu]{Department of Materials Science and Engineering, University of Utah, Salt Lake City, UT 84108, USA}
\address[west]{West High School, Salt Lake City, Utah 84112, USA}

\cortext[cor1]{Corresponding author.}

\date{October 2021}

\begin{abstract}
Expensive-to-train deep learning models can benefit from an optimization of the hyperparameters that determine the model architecture. We optimize 23 hyperparameters of a materials informatics model, \gls{crabnet}, over 100 adaptive design iterations using two models within the Adaptive Experimentation (Ax) Platform. This includes a recently developed \gls{bo} algorithm, \gls{saasbo}, which has shown exciting performance on high-dimensional optimization tasks. Using \gls{saasbo} to optimize \gls{crabnet} hyperparameters, we demonstrate a new state-of-the-art on the experimental band gap regression task within the materials informatics benchmarking platform, Matbench ($\sim$\SI{4.5}{\percent} decrease in \gls{mae} relative to incumbent). Characteristics of the adaptive design scheme as well as feature importances are described for each of the Ax models. \Gls{saasbo} has great potential to both improve existing surrogate models, as shown in this work, and in future work, to efficiently discover new, high-performing materials in high-dimensional materials science search spaces.
\end{abstract}


\end{frontmatter}

\glsresetall

\listoftodos[Notes]

\section{Introduction} \label{sec:intro}
\Gls{bo} is an iterative sequential learning\footnote{This is also referred to as adaptive design and active learning.} algorithm that simultaneously improves model accuracy through exploration of high-uncertainty regions and exploitation of high-performing parameter combinations. It is best-suited for expensive-to-evaluate models with a limited budget of design iterations, and has seen increasing usage in materials informatics \cite{dongInverseDesignComposite2021, espinosa3DOrthogonalVisionbased2022, juDesigningNanostructuresPhonon2017, karasuyamaComputationalDesignStable2020, sakuraiUltranarrowBandWavelengthSelectiveThermal2019, talapatraAutonomousEfficientExperiment2018, wakabayashiMachinelearningassistedThinfilmGrowth2019}.

In materials informatics, many structure-property regression and classification models have been developed to accelerate understanding and design of new materials. \Gls{crabnet} is one such model that leverages the transformer network architecture popularized in natural language processing (e.g. text prediction) \cite{wangCompositionallyRestrictedAttentionBasedNetwork2021}. The self-attention mechanism allows components (e.g. periodic elements, words) to be interpreted in the context of other components. For example, how does oxygen behave in the context of aluminum vs. nitrogen? How does the word bat behave in the context of cave vs. baseball? \Gls{crabnet} has achieved state-of-the-art performance on several materials informatics benchmark datasets while only relying composition alone, whereas competing algorithms benefit from having structural features as well.

Hyperparameter optimization can be used to improve model performance through optimization of model hyperparameters such as number of layers in a neural network, depth of a random forest, learning rates, and number of training epochs. Many hyperparameter optimization algorithms exist which can be used to tune materials informatics models: grid search, random sampling, Sobol sampling, genetic algorithms, and \gls{bo} (in order of decreasing iteration "budget" constraints) as well to search through computational \cite{juDesigningNanostructuresPhonon2017, talapatraAutonomousEfficientExperiment2018} and experimental \cite{balachandranExperimentalSearchHightemperature2018, caoHowOptimizeMaterials2018, chenMachineLearningAssisted2020, hommaOptimizationHeterogeneousTernary2020, houMachineLearningAssistedDevelopmentTheoretical2019, liEfficientOptimizationPerformance2018, raccugliaMachinelearningassistedMaterialsDiscovery2016, sakuraiUltranarrowBandWavelengthSelectiveThermal2019, wakabayashiMachinelearningassistedThinfilmGrowth2019} materials design spaces. For example, \citet{zhangFindingNextSuperhard2021} used a grid search to tune a random forest algorithm to predict Vickers hardness and \citet{dunnBenchmarkingMaterialsProperty2020} used a genetic algorithm to search through extensive combinations of compositional and/or structural features and model architectures to create a general purpose materials informatics algorithm. 

Because training deep learning architectures such as \gls{crabnet} is an expensive-to-evaluate process, here we focus on \gls{bo}.\todo[inline]{timing information for CrabNet on Matbench} To our knowledge\footnote{Manual hyperparameter tuning was employed during the initial development of CrabNet.}, sophisticated hyperparameter optimization has never been performed on \gls{crabnet} before and is rare for materials informatics deep learning models, again since they are usually already expensive to train. We note that \gls{crabnet} recently achieved state-of-the-art predictive performance on a materials science benchmark within the Matbench framework \cite{dunnBenchmarkingMaterialsProperty2020} on the experimental band gap (\verb|matbench_expt_gap|) task \cite{zhuoPredictingBandGaps2018}. \verb|matbench_expt_gap| is a composition-based (i.e. chemical formula as inputs) regression task consisting of 4604 literature-based datapoints and 5 \gls{cv} folds. For additional information, please refer to the Matbench publication \citet{dunnBenchmarkingMaterialsProperty2020} and the \href{Matbench website}{https://matbench.materialsproject.org/} \cite{MatBench2022}. \\

In this work, we pose the question:

\begin{quote}
For a model that already exhibits state-of-the-art performance, to what extent can the predictive performance benefit from hyperparameter optimization?
\end{quote}

We choose to focus on \gls{crabnet} and \verb|matbench_expt_gap| as the case study for hyperparameter optimization because it exhibits state-of-the-art performance without sophisticated hyperparameter optimization and because of our familiarity with the internals of the \gls{crabnet} architecture and codebase. Other sophisticated compositional models such as RooSt \cite{goodallPredictingMaterialsProperties2020} and ElemNet \cite{jhaElemNetDeepLearning2018, jhaEnhancingMaterialsProperty2019} as well as a variety of other advanced compositional and structure-based regression and classification models (e.g. \cite{chenGraphNetworksUniversal2019, debreuckMaterialsPropertyPrediction2021, dejongStatisticalLearningFramework2016, dunnBenchmarkingMaterialsProperty2020, goodallWyckoffSetRegression2020, klicperaFastUncertaintyAwareDirectional2020, louisGraphConvolutionalNeural2020, parkDevelopingImprovedCrystal2020, wangCompositionallyRestrictedAttentionBasedNetwork2021, xieCrystalGraphConvolutional2018}) could have been used with compatible datasets instead.


\section{Methods}
\begin{figure*}
    \centering
    \includegraphics[width=0.95\textwidth]{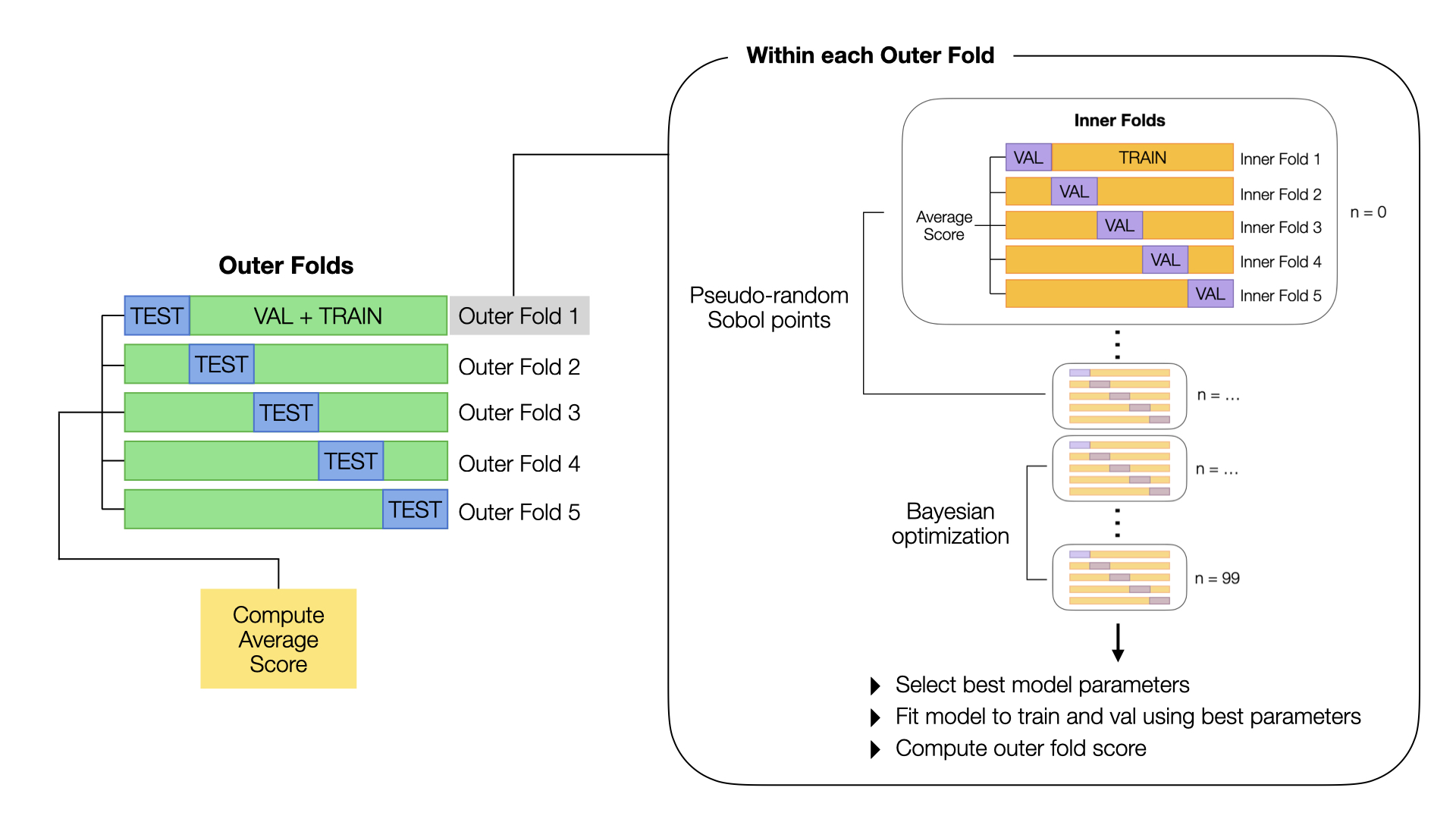}
    \caption{Schematic for \glsxtrfull{bo}, including the nested \glsxtrfull{cv} method. For each of the five outer folds, an adaptive design budget of 100 iterations is used to find optimal hyperparameters. The loss function for the hyperparameter optimization is the average score of the five inner folds. The scores are composed of the \gls{mae} for the validation data divided by the \gls{mae} of a dummy regressor (\gls{mad}) which allows for comparison across different materials science tasks. The optimization starts out with a pseudo-random Sobol point generation phase followed by a Bayesian optimization phase. The Sobol points exist to create a rough initial model prior to Bayesian optimization, consistent with convention and for performance reasons. After the Bayesian optimization is complete, the hyperparameter combination with the best averaged inner score is selected, and a model is refitted to both the training and validation data, with the outer fold score computed based on the outer fold test set. This process is repeated for the other four outer folds, finally resulting in an averaged outer fold score. This final score (along with other metadata and statistics) is reported on the Matbench website. Nested \gls{cv} is computationally expensive but helps ensure model generalizability and comparability with other models. This figure is inspired by \url{https://hackingmaterials.lbl.gov/automatminer/advanced.html}. }
    \label{fig:schematic}
\end{figure*}

We describe two models that we use from the Ax Bayesian Optimization framework (\gls{gpei} and \gls{saasbo}) (\cref{sec:methods:ax}), the \verb|matbench_expt_gap| dataset and Matbench nested \gls{cv} scheme (\cref{sec:methods:data}), and the \gls{crabnet} hyperparameters that we chose to optimize (\cref{sec:methods:hyper}). We summarize the methods in \cref{fig:schematic}.

\subsection{Ax Bayesian Optimization} \label{sec:methods:ax}
Of the many excellent packages for \gls{bo}, we choose the Adaptive Experimentation (Ax) platform due to its relative ease-of-use, modularity, developer support, and model sophistication. We refer to this as Ax. In particular, we utilize two models from Ax.

First is a single task Gaussian process regression model with \gls{ard} (which allows for feature importances), a Matérn 5/2 kernel, and an \gls{ei} acquisition function (\url{https://ax.dev/docs/models.html}). We refer to this model as \texttt{GPEI}.

Second is a recently introduced high-dimensional \gls{bo} scheme: \gls{saasbo} \cite{erikssonHighDimensionalBayesianOptimization2021} which "places strong priors on the inverse lengthscales to avoid overfitting in high-dimensional spaces" and performs well even on problems with hundreds of dimensions \cite{HighDimensionalBayesianOptimization2022}. \citet{erikssonHighDimensionalBayesianOptimization2021} demonstrated that \gls{saasbo} outperforms a slew of other high-dimensional optimization algorithms on a variety of general machine learning tasks. We are excited to bridge this model's use to the field of materials informatics which is rich in adaptive design problems. We refer to this model as \texttt{SAASBO}.\footnote{23 hyperparameters might not seem large at first; but for a design budget of only 100 iterations, finding optimal hyperparameter is a difficult problem.}

For both models, we allow the model to infer noise levels rather than imposing a constant noise as an \textit{a-priori} constraint or standard deviations supplied on a per-datapoint basis during the optimization\footnote{Repeat measurements may also be supplied directly rather than imposing the assumption of a particular noise distribution, e.g. Gaussian.}; however, Ax is capable of handling these and other tasks including \gls{moo}, risk-averse \gls{bo}, batch-wise optimization, and custom surrogate models as well as a variety of other non-algorithm related features\footnote{Examples of non-algorithm related features within Ax are field experiments (e.g. wetlab experimental adaptive design) and interfacing with external databases.}. Because the computational overhead of \gls{saasbo} scales cubically with the number of datapoints, it is typically limited to several hundred adaptive design iterations which is appropriate for our expensive hyperparameter optimization use-case.

One-hundred sequential design iterations were used for both models. For \texttt{GPEI} and \texttt{SAASBO}, $2\times n$ (where $n$ is the number of parameters) and $10$ initial (quasi-random) Sobol\footnote{Sobol sequences are a quasi-random sampling method that produces relatively equally spaced points across all dimensions compared with both random search and grid search. Random search tends to do well at avoiding systematic biases in the distance as with grid search; however, certain regions may still be sparsely populated. Sobol sequences tend to produce higher quality searches than both random and grid searches (this may be especially true for high-dimensional spaces). Sobol sequences are constrained to $2^m$ sampling points, where $m$ is a non-negative integer. For more information, see \href{https://docs.scipy.org/doc/scipy/reference/generated/scipy.stats.qmc.Sobol.html}{\texttt{scipy.stats.qmc.Sobol}} (\texttt{v1.8.0}) } \todo[inline]{include footnote about what Sobol is and why 2*n and 10} iterations were used to create a rough initial model, respectively. These choices are consistent with convention and defaults within Ax.

\subsection{Data and Validation} \label{sec:methods:data}

As mentioned earlier (\cref{sec:intro}), we choose the experimental band gap Matbench dataset \cite{dunnBenchmarkingMaterialsProperty2020, zhuoPredictingBandGaps2018} (\verb|matbench_expt_gap|), a composition-based regression task consisting of 4604 literature-based datapoints and 5 \gls{cv} outer folds. Additionally, we perform nested \gls{cv} such that hyperparameter optimization iterations are performed based on minimizing the average \gls{mae} across each of the five inner folds (this is the Ax objective). The best parameter set was then trained on all the inner fold data and used to predict on the test set (never before seen during training or hyperparameter optimization). In pseudo-code, this appears as follows:

\begin{python}
for outer_fold in outer_folds:
    for iteration in iterations:
        for inner_fold in inner_folds:
            compute score
        compute average score
    select best model parameters
    train model using best parameters
    compute score
compute average score
\end{python}

For the inner folds, we follow the Matbench recommendation (\url{https://hackingmaterials.lbl.gov/automatminer/datasets.html}) of using the following inner \gls{cv} folds\footnote{The code was formatted in Black code style via an online formatter: \url{https://black.vercel.app/}.}:

\begin{python}
from sklearn.model_selection import KFold
kf = KFold(
    n_splits=n_splits,
    shuffle=True,
    random_state=18012019,
)
for train_index, val_index in kf.split(
    train_val_df
):
    train_df, val_df = (
        train_val_df.loc[train_index],
        train_val_df.loc[val_index],
    )
\end{python}

For a single matbench task, \gls{crabnet} undergoes model instantiation and fitting $5 \times 5 \times 100=2500$ times. Use of nested \gls{cv} helps ensure that the model results are both generalizable and comparable with other models.\todo[inline]{REF} The nested \gls{cv} results bolster our confidence that the results in this work are both reproducible\footnote{Our remark about reproducibility assumes that the specified software versions for \gls{crabnet}, Matbench, and each of their dependencies are used.} and generalizable to in-domain experimental band gap predictions. While effective, nested \gls{cv} is also computationally expensive. With relatively sophisticated \glspl{gpu} (e.g. NVIDIA 2080-Ti), nested \gls{cv} for \verb|matbench_expt_gap| using \gls{crabnet} predictions takes over a day to complete and resides somewhere in the gray area between requiring consumer hardware vs. high-performance computing. See \href{https://hackingmaterials.lbl.gov/automatminer/advanced.html}{Automatminer: running a benchmark} and \citet{dunnBenchmarkingMaterialsProperty2020} for more information on nested \gls{cv}.


\subsection{CrabNet hyperparameters} \label{sec:methods:hyper}

We use two Ax Bayesian adaptive design models (\cref{sec:methods:ax}) to simultaneously optimize 23 hyperparameters of \gls{crabnet} (\cref{tab:param}). Parameters were surfaced in the top-level \gls{api} from several thousand lines of nested code within \gls{crabnet}, and search space constraints were imposed based on a combination of intuition and algorithm/data constraints. Some trial-and-error was involved to figure out invalid combinations of hyperparameters. For example, the model architecture dimensions (\verb|d_model|) needed to be divisible by the number of attention heads (\verb|heads|). As another example, the \verb|jarvis|, \verb|oliynyk|, and \verb|ptable| elemental featurizers (\verb|elem_prop| keyword argument) were eliminated either due to an incomplete representation of elements contained in \verb|matbench_expt_gap| chemical formulas or an incompatible datatype (in our case, \verb|string|).

An additional consideration is that certain CrabNet hyperparameters required reparameterization to be compatible with the Ax \gls{api} or remove degenerate dimensions. We give an example of each.

As an example of \gls{api} compatibility, the number of hidden dimensions in the \gls{crabnet} recurrent neural network (\verb|out_hidden|) is passed as a list of integers whereas the Ax \gls{api} would require each of these integers as individual parameters. In this case for both \gls{api} compatibility and simplicity, we chose the last integer in this list as \verb|out_hidden4| and chose the size of the preceding three layers as twice the size of the next layer (e.g. \texttt{[1024, 512, 256, 128]}). Likewise, \verb|betas| was split into \verb|betas1| and \verb|betas2| with the constraint that \verb|betas1 < betas2|.

As an example of removing degenerate dimensions, the embedding, fractional prevalence encoding, and log fractional prevalence encoding undergo a weighted average via \verb|emb_scaler|, \verb|pe_scaler|, and \verb|ple_scaler|, respectively. Despite having three parameters associated with them, this search subspace has only two non-degenerate dimensions (degrees-of-freedom) due to the normalization linear equality constraint of the weighted average (\verb|emb_scaler + pe_scaler + ple_scaler == 1|). Imposing equality constraints can lead to numerical instabilities for optimization schemes that rely on volume-based sampling, so in the reparameterization, this is represented without the degenerate dimension as an inequality constraint (\verb|emb_scaler + pe_scaler <= 1|). This is an important aspect of preventing unnecessarily large search spaces in an already high-dimensional optimization scheme.

We note that in addition to the rules mentioned in this section, \verb|epochs| was set to $4 \times \texttt{epochs\_step}$ during reparameterization and \verb|heads| was constrained to be even except in the case that \verb|heads==1|. For a full workflow of the reparameterization procedure, see \url{https://github.com/sparks-baird/crabnet-hyperparameter/blob/b720d5fd6cadb286079dc8c49c1a5733e263585e/utils/parameterization.py#L10-L48}.

A summary of the 23 CrabNet hyperparameters optimized in this work is presented in \cref{tab:param}.

\begin{table*}[]
    \centering
    \caption{Table of 23 \gls{crabnet} parameter names, ranges/possible values, short description, and default parameter values. FC stands for fully-connected. \todo[inline]{REF for mat2vec and magpie} }
    \begin{tabular}{@{}llllll@{}}
    \toprule
    Parameter   & \multicolumn{1}{l}{Min.} & Max. & Default         & Description                                                           \\ \midrule
    \texttt{N}                & \multicolumn{1}{l}{1}             & 10            & 3                     & Number of attention layers                                            \\
    \texttt{alpha}            & \multicolumn{1}{l}{0.0}           & 1.0           & 0.5                   & \texttt{Lookahead} "slow update" rate                                                                      \\
    \texttt{bias}             & \multicolumn{2}{l}{\small\texttt{[False, True]}}                                & \small\texttt{False}                 & Whether to bias residual network                                  \\
    \texttt{criterion}        & \multicolumn{2}{l}{\small{["RobustL1", "RobustL2"]}}                & \small{"RobustL1"}                  & Loss function                                                 \\
    \texttt{d\_model}         & \multicolumn{1}{l}{100}           & 1024          & 512                   & FC network output dimension                                              \\
    \texttt{dim\_feedforward} & \multicolumn{1}{l}{1024}          & 4096          & 2048                  & Feedforward network dimension         \\
    \texttt{dropout}          & \multicolumn{1}{l}{0.0}           & 1.0           & 0.1                   & Feedforward dropout fraction \\
    \texttt{elem\_prop}       & \multicolumn{2}{l}{\footnotesize{["mat2vec", "magpie", "onehot"]}} & \small{"mat2vec"}             & Elemental feature vector                               \\ 
    \texttt{emb\_scaler}      & \multicolumn{1}{l}{0.0}           & 1.0           & 1.0                   & Elemental embeddings weight                                   \\
    \texttt{epochs\_step}     & \multicolumn{1}{l}{5}             & 20            & 10                    & Step size of epochs                                                   \\
    \texttt{eps}              & \multicolumn{1}{l}{0.0000001}     & 0.0001        & 0.000001              & Prevents zero-division in \texttt{LAMB}                                                                      \\
    \texttt{fudge}            & \multicolumn{1}{l}{0.0}           & 0.1           & 0.02                  & Jitter to fractional encodings                                 \\
    \texttt{heads}            & \multicolumn{1}{l}{1}             & 10            & 4                     & Number of attention heads                                        \\
    \texttt{k}                & \multicolumn{1}{l}{2}             & 10            & 6                     & Number of \texttt{Lookahead} steps                                                                      \\
    \texttt{lr}               & \multicolumn{1}{l}{0.0001}        & 0.006         & 0.001                 & Learning rate                                                         \\
    \texttt{pe\_resolution}   & \multicolumn{1}{l}{2500}          & 10000         & 5000                  & Prevalence encoding resolution                  \\
    \texttt{ple\_resolution}  & \multicolumn{1}{l}{2500}          & 10000         & 5000                  & Prevalence log encoding resolution              \\
    \texttt{pos\_scaler}      & \multicolumn{1}{l}{0.0}           & 1.0           & 1.0                   & Fractional encodings weight                                   \\
    \texttt{weight\_decay}    & \multicolumn{1}{l}{0.0}           & 1.0           & 0                     & L2 penalty in \texttt{LAMB}                                                                       \\
    \texttt{batch\_size}      & \multicolumn{1}{l}{32}            & 256           & 32 & Training batch size                                     \\
    \texttt{out\_hidden4}   & \multicolumn{1}{l}{32}            & 512           & 128                & 4th layer size of residual network                 \\
    \texttt{betas1}         & \multicolumn{1}{l}{0.5}           & 0.9999        & 0.9                   & Gradient coefficent in \texttt{LAMB}                                                                      \\
    \texttt{betas2}         & \multicolumn{1}{l}{0.5}           & 0.9999        & 0.999                 & Squared gradient coefficient in \texttt{LAMB}                                                                      \\ \bottomrule
    \end{tabular}
    \label{tab:param}
\end{table*}

\section{Results and Discussion}

We discuss improvements in model performance offered by \texttt{GPEI} and \texttt{SAASBO} models relative to a baseline model (\cref{sec:results:err}) and analyze the characteristics of the hyperparameter optimization across iterations and for individual hyperparameters (\cref{sec:results:hyperopt}). Finally, we discuss future work (\cref{sec:results:future}).

\subsection{Improvement in Model Performance} \label{sec:results:err}

\begin{figure}
    \centering
    \includegraphics[width=0.48\textwidth]{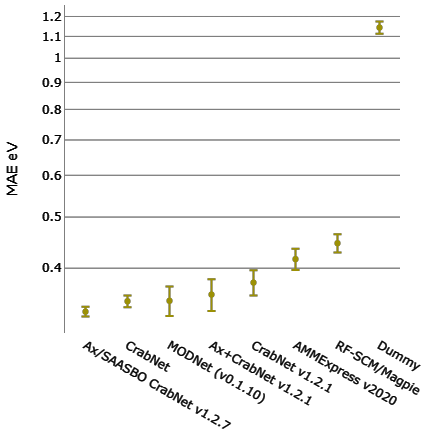}
    \caption{Comparison of mean \gls{mae} (eV) with standard deviation \gls{mae} (eV) across five folds for \texttt{CrabNet v1.2.1} with default hyperparameters, \texttt{Ax+CrabNet v1.2.1}, \texttt{Ax/SAASBO CrabNet v1.2.7}, and other \texttt{matbench} submissions on the \texttt{matbench\_expt\_gap v0.1} task. The \texttt{CrabNet} submission (\texttt{mae==0.3463}) was submitted by a separate party, likely with a large number of epochs (e.g. \texttt{epochs==300}) relative to the upper limit of \texttt{epochs} (80) in this work. As of 2022-03-16, the results shown in this figure are available on the \href{https://matbench.materialsproject.org/}{Matbench leaderboard website} \cite{MatBench2022} under "Leaderboards Per Task" for \texttt{matbench\_expt\_gap}. } 
    \label{fig:err}
\end{figure}

\texttt{GPEI} hyperparameter optimization resulted in a moderate performance improvement relative to the default \gls{crabnet} hyperparameters\footnote{Manual trial and error was used to tune \gls{crabnet} hyperparameters prior to publication of \gls{crabnet} \cite{wangCompositionallyRestrictedAttentionBasedNetwork2021}.}. Notably, \texttt{SAASBO} hyperparameter optimization exhibited nearly twice the performance improvement compared to the improvement of \texttt{GPEI} over default hyperparameters. Additionally, \texttt{SAASBO} set the new state-of-the-art benchmark for \verb|matbench_expt_gap|, which further strengthens its evidence as an effective high-dimensional Bayesian optimization scheme. We summarize the predictive performance of \texttt{GPEI}, \texttt{SAASBO}, and our baseline model against other Matbench submissions in \cref{fig:err} and \cref{tab:err}. As of 2022-03-16, the results of this work are available on the \href{https://matbench.materialsproject.org/}{Matbench leaderboard website} \cite{MatBench2022} under "Leaderboards Per Task" for \verb|matbench_expt_gap|. Reproducible scripts are made available in the Matbench GitHub repository under the \texttt{benchmarks} directory (\url{https://github.com/materialsproject/matbench}).

\begin{table*}
\centering
\caption{Comparison of mean \gls{mae}, standard deviation of \gls{mae}, mean \gls{rmse}, and max \texttt{max\_error} across five folds for \texttt{CrabNet v1.2.1} with default hyperparameters, \texttt{Ax+CrabNet v1.2.1}, \texttt{Ax/SAASBO CrabNet v1.2.7}, and other \texttt{matbench} submissions on the \texttt{matbench\_expt\_gap v0.1} task. \texttt{CrabNet} submission (\texttt{mae==0.3463}) was submitted by a separate party, likely with a large number of epochs (e.g. \texttt{epochs==300}) relative to the upper limit of \texttt{epochs} (80) in this work. As of 2022-03-16, the results shown in this figure are available on the \href{https://matbench.materialsproject.org/}{Matbench leaderboard website} \cite{MatBench2022} under "Leaderboards Per Task" for \texttt{matbench\_expt\_gap}. }
\label{tab:err}
\begin{tabular}{@{}lllll@{}}
\toprule
algorithm                & mean mae (eV) & std mae (eV) & mean rmse (eV) & max max\_error (eV) \\ \midrule
Ax/SAASBO CrabNet v1.2.7 & 0.3310   & 0.0071  & 0.8123    & 11.1001       \\
CrabNet                  & 0.3463   & 0.0088  & 0.8504    & 9.8002        \\
MODNet (v0.1.10)         & 0.3470   & 0.0222  & 0.7437    & 9.8567        \\
Ax+CrabNet v1.2.1        & 0.3566   & 0.0248  & 0.8673    & 11.0998       \\
CrabNet v1.2.1           & 0.3757   & 0.0207  & 0.8805    & 10.2572       \\
AMMExpress v2020         & 0.4161   & 0.0194  & 0.9918    & 12.7533       \\
RF-SCM/Magpie            & 0.4461   & 0.0177  & 0.8243    & 9.5428        \\
Dummy                    & 1.1435   & 0.0310  & 1.4438    & 10.7354       \\ \bottomrule
\end{tabular}
\end{table*}

\subsection{Hyperparameter Optimization} \label{sec:results:hyperopt}

We describe characteristics of the hyperparameter optimization results for both \texttt{GPEI} and \texttt{SAASBO} models in the context of best parameter combinations as a function of iteration (\cref{sec:results:hyperopt:iter}), interpretable model characteristics such as the selected hyperparameter combinations and feature importances (\cref{sec:results:hyperopt:model}), and \gls{cv} results for the predicted vs. actual CrabNet \glspl{mae} (\cref{sec:results:hyperopt:cv}).

\subsubsection{Best Objective vs. Iteration} \label{sec:results:hyperopt:iter}

\begin{figure*}
     \centering
     \begin{subfigure}[b]{0.3\textwidth}
         \includegraphics[width=\textwidth]{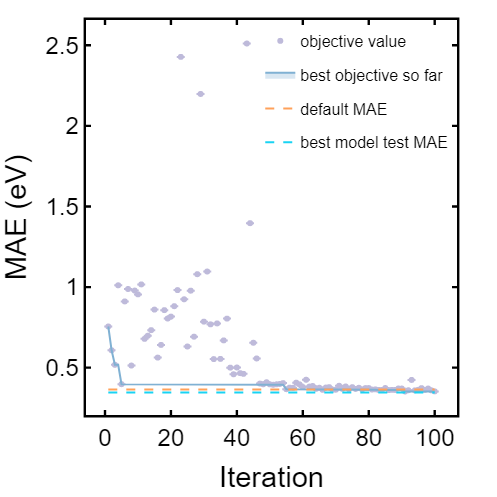}
         \caption{}
         \label{fig:ax-best-obj-0}
     \end{subfigure}
     \hfill
     \begin{subfigure}[b]{0.3\textwidth}
         \includegraphics[width=\textwidth]{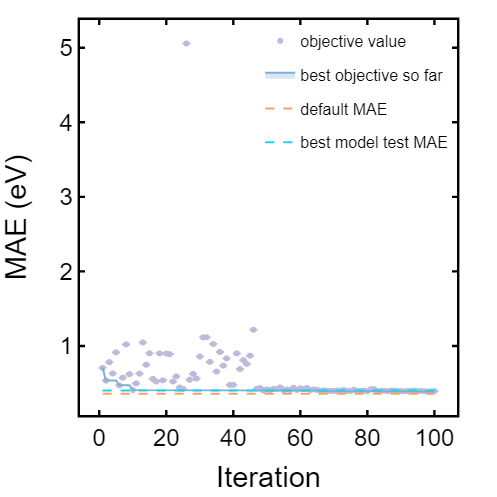}
         \caption{}
         \label{fig:ax-best-obj-1}
     \end{subfigure}
     \hfill
     \begin{subfigure}[b]{0.3\textwidth}
         \includegraphics[width=\textwidth]{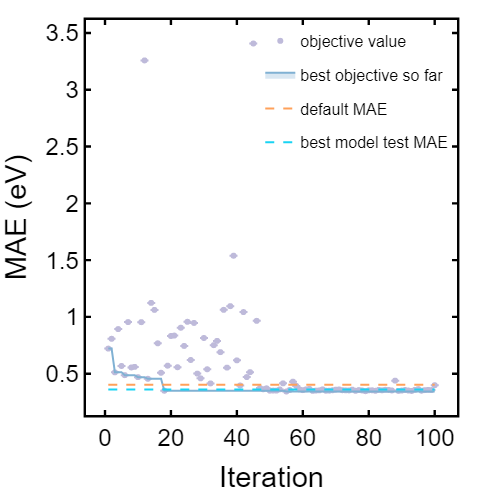}
         \caption{}
         \label{fig:ax-best-obj-2}
     \end{subfigure}
     
     \begin{subfigure}[b]{0.3\textwidth}
        \includegraphics[width=\textwidth]{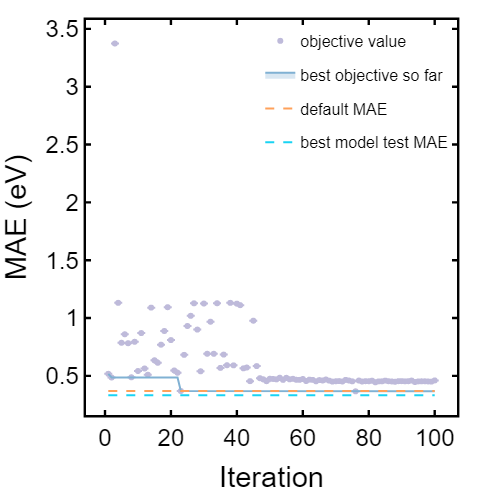}
         \caption{}
         \label{fig:ax-best-obj-3}
     \end{subfigure}
     \hspace{0.1cm}
     \begin{subfigure}[b]{0.3\textwidth}
        \includegraphics[width=\textwidth]{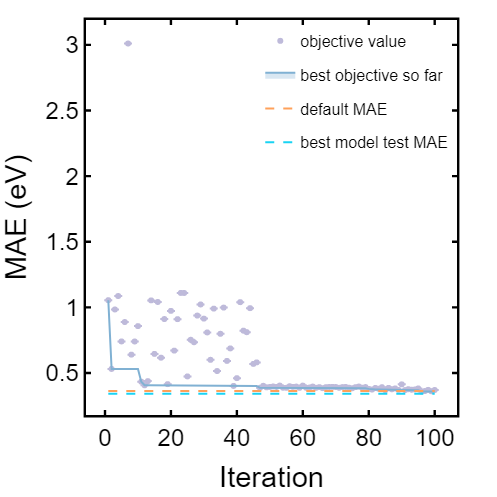}
         \caption{}
         \label{fig:ax-best-obj-4}
     \end{subfigure}
     
     \caption{Best objective vs. \texttt{GPEI} optimization iteration for each of the five \texttt{matbench\_expt\_gap} folds. First 46 iterations are Sobol points.}
     \label{fig:ax-best-obj}
\end{figure*}

\begin{figure*}
     \centering
     \begin{subfigure}[b]{0.3\textwidth}
         \includegraphics[width=\textwidth]{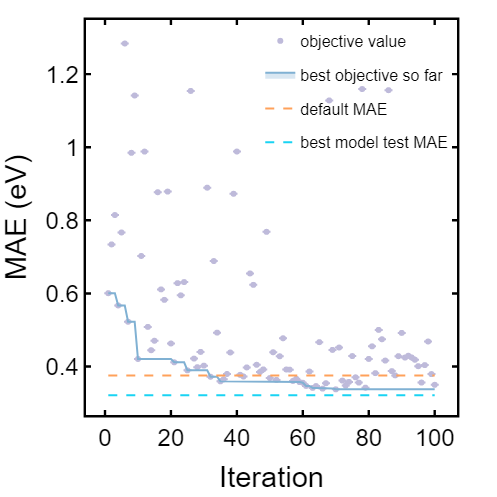}
         \caption{}
         \label{fig:saas-best-obj-0}
     \end{subfigure}
     \hfill
     \begin{subfigure}[b]{0.3\textwidth}
         \includegraphics[width=\textwidth]{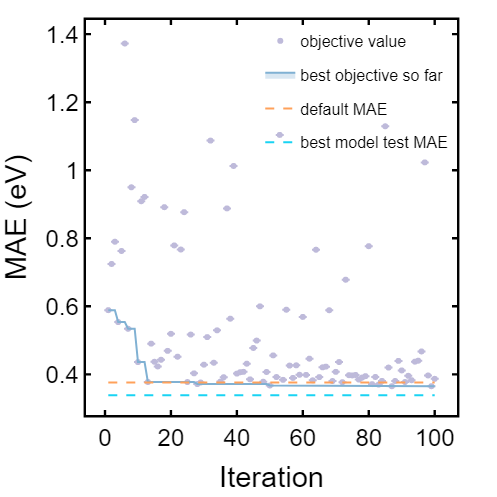}
         \caption{}
         \label{fig:saas-best-obj-1}
     \end{subfigure}
     \hfill
     \begin{subfigure}[b]{0.3\textwidth}
         \includegraphics[width=\textwidth]{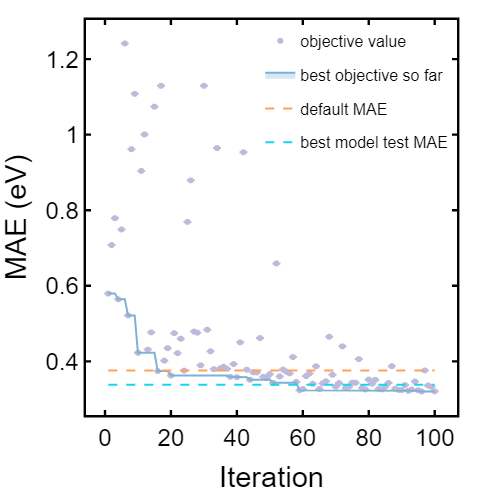}
         \caption{}
         \label{fig:saas-best-obj-2}
     \end{subfigure}
     
     \begin{subfigure}[b]{0.3\textwidth}
        \includegraphics[width=\textwidth]{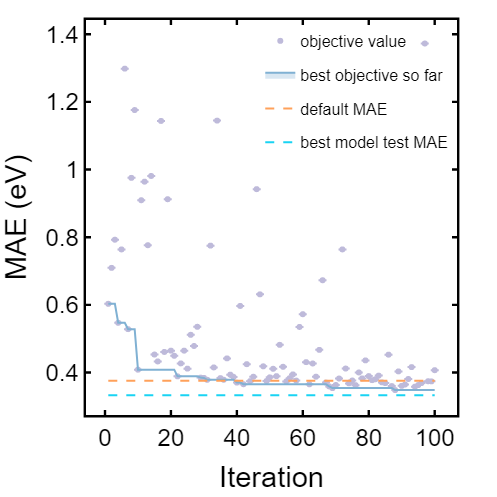}
         \caption{}
         \label{fig:saas-best-obj-3}
     \end{subfigure}
     \hspace{0.1cm}
     \begin{subfigure}[b]{0.3\textwidth}
        \includegraphics[width=\textwidth]{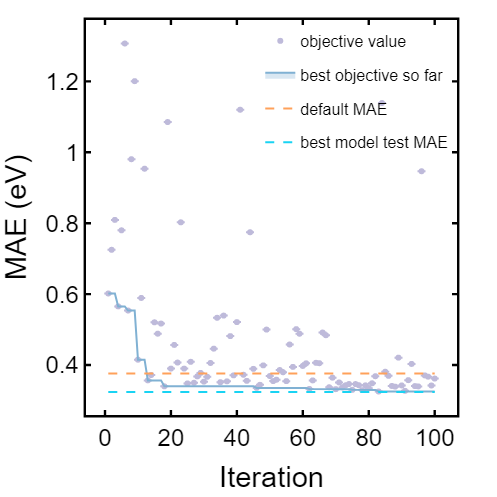}
         \caption{}
         \label{fig:saas-best-obj-4}
     \end{subfigure}
     
     \caption{Best objective vs. \texttt{SAASBO} optimization iteration for each of the five \texttt{matbench\_expt\_gap} folds. First 10 iterations are Sobol points.}
     \label{fig:saas-best-obj}
\end{figure*}

Best objective vs. iteration plots for \texttt{GPEI} and \texttt{SAASBO} are shown for each Matbench fold in \cref{fig:ax-best-obj} and \cref{fig:saas-best-obj}, respectively. The first 46 \texttt{GPEI} Sobol points have a large scatter, but improve the accuracy of the internal Gaussian process regression model. 

At iteration 47, the Bayesian optimization iterations begin, at which point the \glspl{mae} have a much tighter spread.

In the case of \texttt{SAASBO}, only 10 initial Sobol points are used, consistent with convention within Ax. We notice that during the \texttt{SAASBO} iterations, there is a much larger spread in \gls{mae} than with \texttt{GPEI} indicating perhaps that \texttt{SAASBO} is less likely to get trapped searching in a local minimum. In other words, \texttt{SAASBO} seems to favor exploration more than \texttt{GPEI} during later iterations.

We also note that in every \texttt{SAASBO} \verb|matbench_expt_gap| fold, the best model test \gls{mae} is better than the default \gls{mae}, whereas in one case with \texttt{GPEI} the best model test \gls{mae} is worse than the default \gls{mae}.

\subsubsection{Interpretable Model Characteristics} \label{sec:results:hyperopt:model}

\begin{table*}
\centering
\caption{Table of best parameterization for the default values and each of the 5 outer folds for \texttt{GPEI} and \texttt{SAASBO} with the mean absolute error as the last row (25 rows $\times$ 11 cols). \todo[inline]{report mode and standard deviations for ax and saas}}
\label{tab:best-par}
\resizebox{\textwidth}{!}{
\begin{tabular}{@{}llllllllllll@{}}
\toprule
parameter & default  & ax\_fold\_0 & ax\_fold\_1 & ax\_fold\_2 & ax\_fold\_3 & ax\_fold\_4 & saas\_fold\_0 & saas\_fold\_1 & saas\_fold\_2 & saas\_fold\_3 & saas\_fold\_4 \\ \midrule
\texttt{N}                & 3       & 4       & 5       & 3       & 3        & 6       & 2        & 3        & 4        & 5        & 2        \\
\texttt{alpha}            & 0.500   & 0.8791  & 0.7990  & 0.8042  & 1.000    & 0.7345  & 1.000    & 0.8019   & 0.666    & 0.6209   & 0.9403   \\
\texttt{bias}             & False   & True    & False   & False   & True     & False   & True     & False    & False    & False    & False    \\
\texttt{criterion} & RobustL1 & RobustL1    & RobustL1    & RobustL1    & RobustL1    & RobustL1    & RobustL1      & RobustL1      & RobustL1      & RobustL1      & RobustL1      \\
\texttt{d\_model}         & 512     & 860     & 516     & 660     & 940      & 288     & 890      & 690      & 1024     & 1024     & 1024     \\
\texttt{dim\_feedforward} & 2048    & 3498    & 2663    & 3469    & 1981     & 1393    & 4096     & 1179     & 1903     & 2322     & 2074     \\
\texttt{dropout}          & 0.1000  & 0.2105  & 0.2226  & 0.03475 & 0.003479 & 0.02590 & 0.02887  & 1.79e-15 & 0.09428  & 0.0954   & 0.00e+00 \\
\texttt{elem\_prop}       & mat2vec & mat2vec & onehot  & onehot  & mat2vec  & mat2vec & mat2vec  & onehot   & mat2vec  & onehot   & mat2vec  \\
\texttt{emb\_scaler}      & 1.000   & 0.2547  & 0.3182  & 0.5283  & 0.6800   & 0.4326  & 0.7786   & 0.2733   & 0.3257   & 0.2925   & 0.6672   \\
\texttt{epochs}           & 40      & 40      & 60      & 60      & 68       & 68      & 80       & 80       & 80       & 80       & 80       \\
\texttt{epochs\_step}     & 10      & 10      & 15      & 15      & 17       & 17      & 20       & 20       & 20       & 20       & 20       \\
\texttt{eps}       & 1.00e-06 & 0.00007087  & 4.48e-05    & 2.62e-05    & 0.00008816  & 3.95e-05    & 2.78e-05      & 2.39e-05      & 1.53e-05      & 1.00e-07      & 1.00e-07      \\
\texttt{fudge}            & 0.02000 & 0.07413 & 0.01511 & 0.08701 & 0.05812  & 0.06170 & 7.93e-08 & 2.83e-16 & 0.01194  & 0.03473  & 0.00e+00 \\
\texttt{heads}            & 4       & 4       & 6       & 10      & 10       & 8       & 10       & 10       & 2        & 8        & 8        \\
\texttt{k}                & 6       & 6       & 2       & 3       & 10       & 5       & 2        & 2        & 6        & 2        & 2        \\
\texttt{lr}        & 0.001000 & 0.002053    & 0.002904    & 0.002141    & 0.004779    & 0.004934    & 0.0003422     & 0.0001000     & 0.006         & 0.001696      & 0.005231      \\
\texttt{pe\_resolution}   & 5000    & 7185    & 8609    & 7277    & 6354     & 5426    & 2500     & 5403     & 2652     & 8735     & 2500     \\
\texttt{ple\_resolution}  & 5000    & 4322    & 6758    & 7289    & 4584     & 8203    & 4952     & 2500     & 10000    & 4814     & 10000    \\
\texttt{pos\_scaler}      & 1.000   & 0.2043  & 0.2064  & 0.1782  & 0.3012   & 0.5221  & 2.47e-07 & 0.005612 & 3.01e-12 & 0.3951   & 0.3328   \\
\texttt{pos\_scaler\_log} & 1.000   & 0.5411  & 0.4754  & 0.2936  & 0.01877  & 0.04523 & 0.2214   & 0.7211   & 0.6743   & 0.3124   & 4.78e-10 \\
\texttt{weight\_decay}    & 0       & 0.1264  & 0.5729  & 0.07740 & 0.7779   & 0.4129  & 1.44e-09 & 2.37e-15 & 1.10e-13 & 3.81e-17 & 0.00e+00 \\
\texttt{batch\_size}      & 256     & 69      & 165     & 63      & 241      & 125     & 32       & 32       & 32       & 104      & 32       \\
\texttt{out\_hidden0}   & 1024    & 1584    & 3048    & 3720    & 3392     & 984     & 1264     & 256      & 256      & 1904     & 2904     \\
\texttt{betas0}         & 0.9000  & 0.5216  & 0.6461  & 0.7111  & 0.5592   & 0.5574  & 0.5166   & 0.5000   & 0.5000   & 0.7642   & 0.5213   \\
\texttt{betas1}         & 0.999   & 0.7118  & 0.7283  & 0.9476  & 0.5830   & 0.9347  & 0.5265   & 0.5000   & 0.5000   & 0.7978   & 0.9999   \\
test \gls{mae}              & 0.3757  & 0.3465  & 0.4029  & 0.3599  & 0.3324   & 0.3412  & 0.3214   & 0.3385   & 0.3383   & 0.3327   & 0.3239   \\ \bottomrule
\end{tabular}
} 
\end{table*}

A summary of the best hyperparameter combinations out of 100 iterations for the \texttt{GPEI} and \texttt{SAASBO} models is shown in \cref{tab:best-par}. It is interesting to note that in every case, \texttt{SAASBO} pushed the number of epochs to the upper limit of $80$, whereas \texttt{GPEI} ranged from $40$ to $68$. In other words, with a limited budget of 100 iterations, \texttt{SAASBO} seems to have determined that maximizing the number of epochs tends to reduce the test error, especially since \gls{crabnet} implements early stopping. While this may be obvious to \gls{ml} practitioners, it is noteworthy that \texttt{SAASBO} identified this trend without prior knowledge other than the user-chosen constraints on the search space.


\begin{figure*}
    \centering
    \begin{subfigure}[b]{0.475\textwidth}
        \includegraphics[width=\textwidth,trim={0.4cm 0 0.85cm 0.5cm},clip]{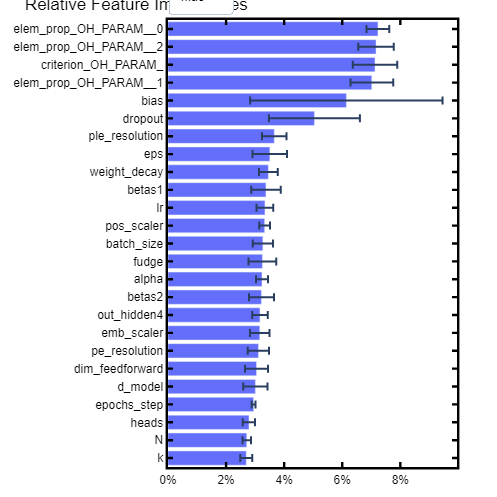}
        \caption{\texttt{GPEI}}
        \label{fig:ax-avg-feat}
    \end{subfigure}
    \hfill
    \begin{subfigure}[b]{0.475\textwidth}
        \includegraphics[width=\textwidth,trim={0.4cm 0 0.85cm 0.5cm},clip]{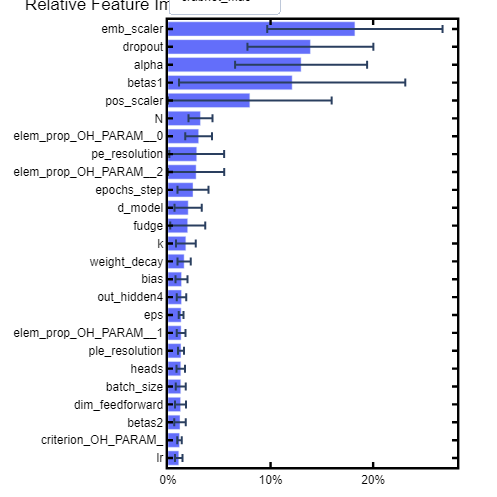}
        \caption{\texttt{SAASBO}}
        \label{fig:saas-avg-feat}
    \end{subfigure}
    \caption{Relative \texttt{GPEI} (\subref{fig:ax-avg-feat}) and \texttt{SAASBO} (\subref{fig:saas-avg-feat}) feature importances of \gls{crabnet} hyperparameters averaged over the five \texttt{matbench\_expt\_gap} folds based on data from 100 iterations per fold. For \texttt{GPEI}, the first 46 iterations were Sobol points, and the remaining 54 iterations were Bayesian optimization iterations. For \texttt{SAASBO}, the first 10 iterations were Sobol points, and the remaining 90 iterations were Bayesian optimization iterations. Standard deviations are given as error bars with lower error bars truncated. \texttt{\_OH\_PARAM\_\_\#} refers to choices for categorical variables. }
    \label{fig:avg-feat}
\end{figure*}

A summary of the \texttt{GPEI} and \texttt{SAASBO} feature importances with standard deviations across the 5 Matbench folds is given in \cref{fig:avg-feat}. Relative feature importances for each of the five folds for the two models are given in \cref{fig:ax-feat} and \cref{fig:saas-feat}, respectively. The \texttt{GPEI} feature importances have a more gradual decay compared with the \texttt{SAASBO} features. The minimum and maximum relative feature importances for \texttt{GPEI} are typically $\sim$\SI{2}{\percent} and $\sim$\SI{10}{\percent}, respectively, whereas the minimum and maximum relative feature importances for \texttt{SAASBO} range from $\sim$\SI{1}{\percent} to $\sim$\SI{25}{\percent}, respectively. This is indicative of the strong priors on the inverse lengthscales within the \gls{saasbo} framework and previous results indicating a relatively small number of features being recognized as important to the model as the adaptive design process progresses (see Figure 5. of \citet{erikssonHighDimensionalBayesianOptimization2021}).

Additionally, we notice that \texttt{GPEI} tends to place much higher weights on the categorical parameters than \texttt{SAASBO} and that \texttt{SAASBO} places greater weight on the relative contributions of the embedding, fractional encoding, and log-fractional encoding than \texttt{GPEI}. A commonality between the two methods is that \verb|dropout| tends to be ranked highly in terms of feature importance in either method.

\begin{figure*}
    \centering
    \verb|epochs_step|

     \begin{subfigure}[b]{0.1925\textwidth}
         \caption{\texttt{GPEI} fold 0}
         \includegraphics[width=\textwidth,trim={1.09cm 1.04cm 0.86cm 0.35cm},clip]{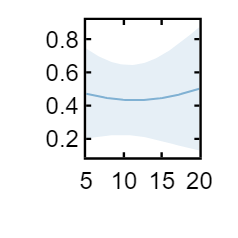}
         \label{fig:ax-0-epochs_step_main}
     \end{subfigure}
     \hfill
     \begin{subfigure}[b]{0.1925\textwidth}
         \caption{\texttt{GPEI} fold 1}
         \includegraphics[width=\textwidth,trim={1.09cm 1.04cm 0.86cm 0.35cm},clip]{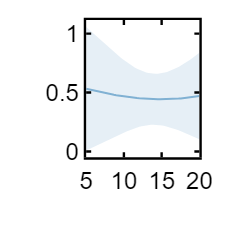}
         \label{fig:ax-1-epochs_step_main}
     \end{subfigure}
     \hfill
     \begin{subfigure}[b]{0.1925\textwidth}
         \caption{\texttt{GPEI} fold 2}
         \includegraphics[width=\textwidth,trim={1.09cm 1.04cm 0.86cm 0.35cm},clip]{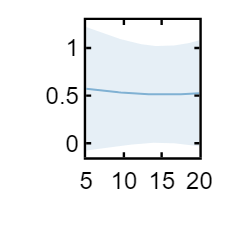}
         \label{fig:ax-2-epochs_step_main}
     \end{subfigure}
     \hfill
     \begin{subfigure}[b]{0.1925\textwidth}
         \caption{\texttt{GPEI} fold 3}
         \includegraphics[width=\textwidth,trim={1.09cm 1.04cm 0.86cm 0.35cm},clip]{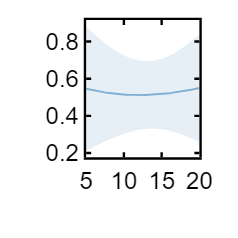}
         \label{fig:ax-3-epochs_step_main}
     \end{subfigure}
     \hfill
     \begin{subfigure}[b]{0.1925\textwidth}
         \caption{\texttt{GPEI} fold 4}
         \includegraphics[width=\textwidth,trim={1.09cm 1.04cm 0.86cm 0.35cm},clip]{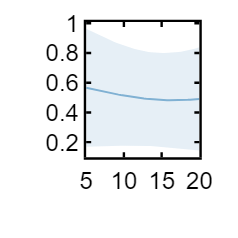}
         \label{fig:ax-4-epochs_step_main}
     \end{subfigure}
     
     \begin{subfigure}[b]{0.1925\textwidth}
         \caption{\texttt{SAASBO} fold 0}
         \includegraphics[width=\textwidth,trim={1.09cm 1.04cm 0.86cm 0.35cm},clip]{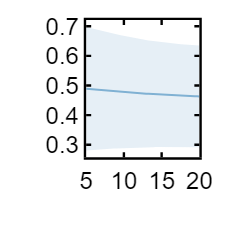}
         \label{fig:saas-0-epochs_step_main}
     \end{subfigure}
     \hfill
     \begin{subfigure}[b]{0.1925\textwidth}
         \caption{\texttt{SAASBO} fold 1}
         \includegraphics[width=\textwidth,trim={1.09cm 1.04cm 0.86cm 0.35cm},clip]{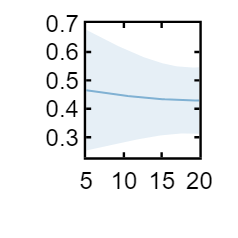}
         \label{fig:saas-1-epochs_step_main}
     \end{subfigure}
     \hfill
     \begin{subfigure}[b]{0.1925\textwidth}
         \caption{\texttt{SAASBO} fold 2}
         \includegraphics[width=\textwidth,trim={1.09cm 1.04cm 0.86cm 0.35cm},clip]{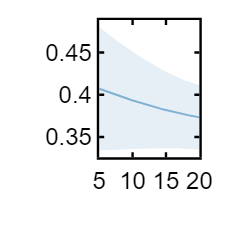}
         \label{fig:saas-2-epochs_step_main}
     \end{subfigure}
     \hfill
     \begin{subfigure}[b]{0.1925\textwidth}
         \caption{\texttt{SAASBO} fold 3}
         \includegraphics[width=\textwidth,trim={1.09cm 1.04cm 0.86cm 0.35cm},clip]{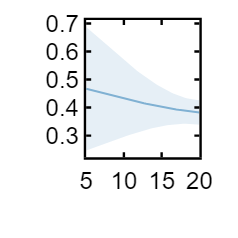}
         \label{fig:saas-3-epochs_step_main}
     \end{subfigure}
     \hfill
     \begin{subfigure}[b]{0.1925\textwidth}
         \caption{\texttt{SAASBO} fold 4}
         \includegraphics[width=\textwidth,trim={1.09cm 1.04cm 0.86cm 0.35cm},clip]{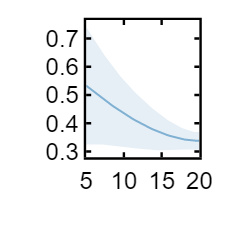}
         \label{fig:saas-4-epochs_step_main}
     \end{subfigure}

     \caption{One-dimensional (1D) slices of \gls{mae} (eV) vs. \texttt{epochs\_step} through the \texttt{GPEI} and \texttt{SAASBO} parameter spaces with the rest of the parameters fixed to the mean and mode of the 100 sampled numeric and categorical parameters, respectively, for each of the models. } 
     \label{fig:1d-epochs_step_main}
\end{figure*}

Another interpretable aspect is the plotting of 1D slices through the model parameter space. In this work, we fix all other parameters than the one being varied to the mean and mode of the numerical and categorical parameters across the 100 iterations for each of the models. For \verb|epochs_step|, we notice that in general, \texttt{SAASBO} captures the intuitive trend that as number of training epochs increases, model error decreases until it plateaus due to early stopping better than \texttt{GPEI}. With \texttt{GPEI}, local minima and flatter trends dominate, indicating that the effect of this parameter was not properly learned within the 100 iterations across 23 hyperparameters. The \texttt{SAASBO} results are significant; no \textit{a-priori} information was made available to either of the models excepting the allowed ranges that constrain the search space. In other words, \texttt{SAASBO} identified an important trend with limited observations in a high-dimensional design space. This is further supported by the observation that \texttt{SAASBO} uncertainties are generally tighter for a larger number of epochs than \texttt{GPEI}, which indicates that \texttt{SAASBO} devoted more search power to the regions of the model parameter space with a large number of epochs (which we expect to have favorable performance).

The superiority of \texttt{SAASBO} relative to \texttt{GPEI} in recognizing the true influence of \verb|epochs_step| is well-supported; however, this does not preclude \texttt{SAASBO} from missing important trends and fine details depending on the complexity and dimensionality of the search space as well as the design budget. For interested readers, 1D slices for all 23 parameters for both models across each of the 5 Matbench folds are available in supporting information (\cref{supp:slice}).

\subsubsection{Predicted Model Error Cross-validation Results} \label{sec:results:hyperopt:cv}

\begin{figure*}
     \centering
     \begin{subfigure}[b]{0.32\textwidth}
         \includegraphics[width=\textwidth,trim={0 0 4.25cm 1.75cm},clip]{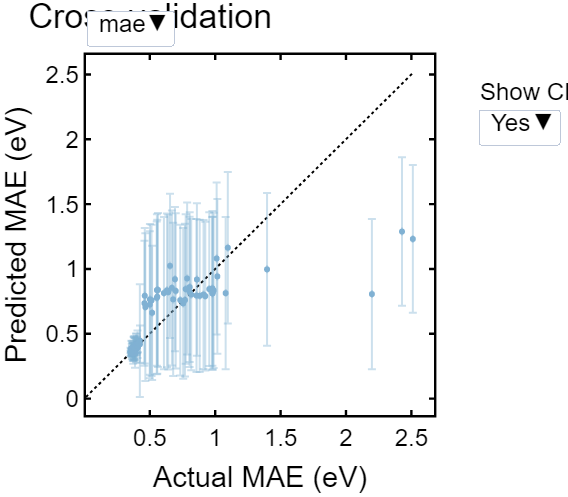}
         \caption{}
         \label{fig:ax-cv-0}
     \end{subfigure}
     \hfill
     \begin{subfigure}[b]{0.32\textwidth}
         \includegraphics[width=\textwidth,trim={0 0 4.25cm 1.75cm},clip]{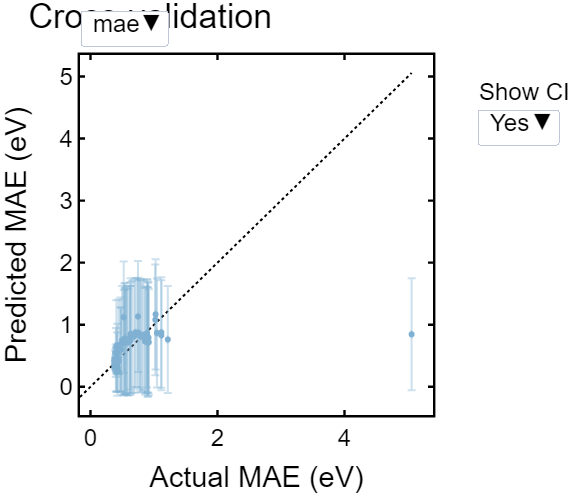}
         \caption{}
         \label{fig:ax-cv-1}
     \end{subfigure}
     \hfill
     \begin{subfigure}[b]{0.32\textwidth}
         \includegraphics[width=\textwidth,trim={0 0 4.25cm 1.75cm},clip]{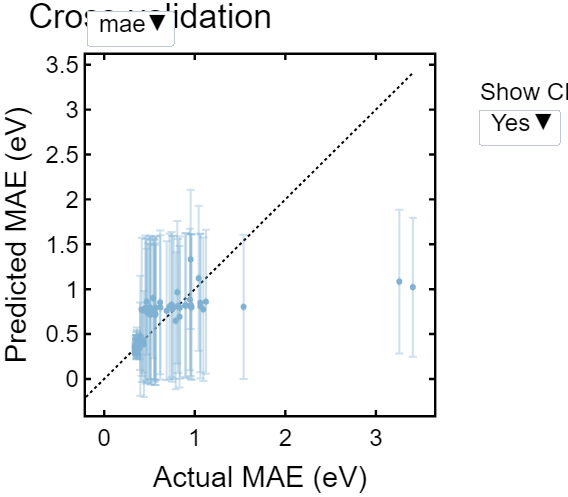}
         \caption{}
         \label{fig:ax-cv-2}
     \end{subfigure}
     
     \begin{subfigure}[b]{0.32\textwidth}
        \includegraphics[width=\textwidth,trim={0 0 4.25cm 1.75cm},clip]{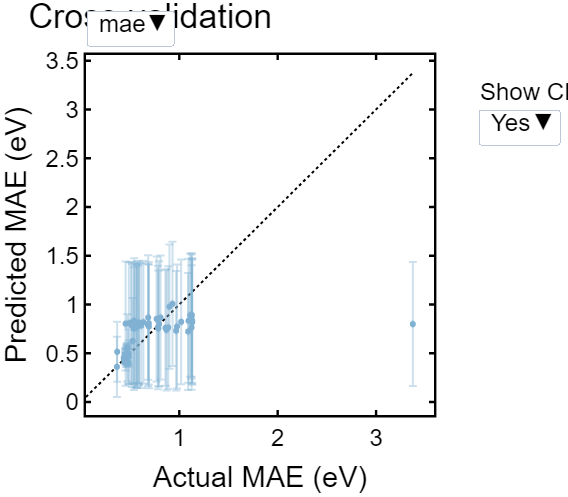}
         \caption{}
         \label{fig:ax-cv-3}
     \end{subfigure}
     \begin{subfigure}[b]{0.32\textwidth}
        \includegraphics[width=\textwidth,trim={0 0 4.25cm 1.75cm},clip]{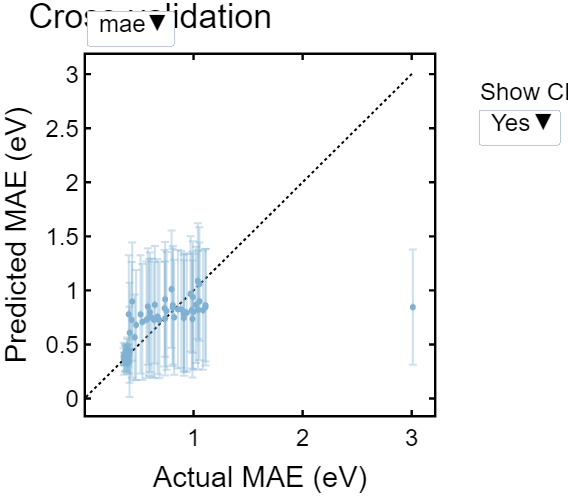}
         \caption{}
         \label{fig:ax-cv-4}
     \end{subfigure}
     
     \caption{\Gls{loo-cv} results for \texttt{GPEI} \gls{crabnet} hyperparameter combinations for each of the five \texttt{matbench\_expt\_gap}. The first 46 iterations were Sobol points, and the remaining 54 iterations were Bayesian optimization iterations.}
     \label{fig:ax-cv}
\end{figure*}

\begin{figure*}
     \centering
     \begin{subfigure}[b]{0.32\textwidth}
         \includegraphics[width=\textwidth,trim={0 0 4.25cm 1.75cm},clip]{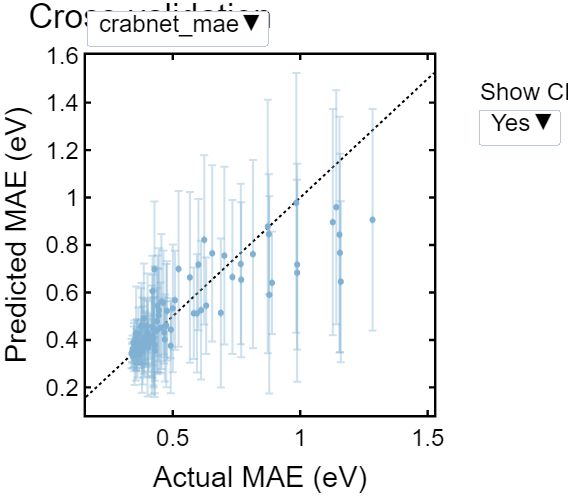}
         \caption{}
         \label{fig:saas-cv-0}
     \end{subfigure}
     \hfill
     \begin{subfigure}[b]{0.32\textwidth}
         \includegraphics[width=\textwidth,trim={0 0 4.25cm 1.75cm},clip]{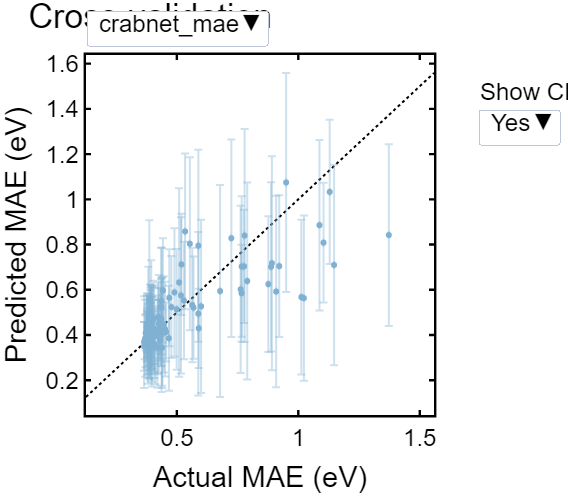}
         \caption{}
         \label{fig:saas-cv-1}
     \end{subfigure}
     \hfill
     \begin{subfigure}[b]{0.32\textwidth}
         \includegraphics[width=\textwidth,trim={0 0 4.25cm 1.75cm},clip]{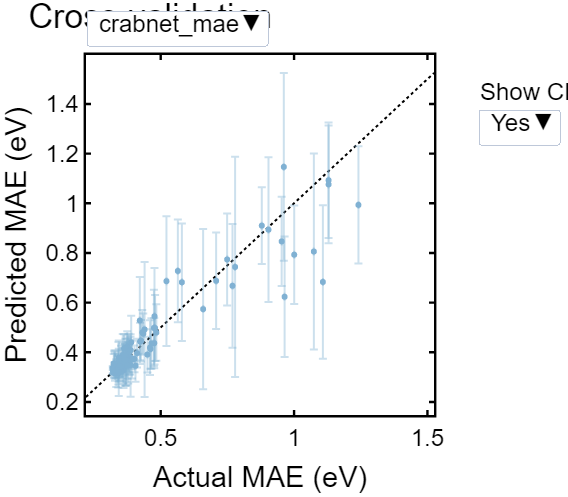}
         \caption{}
         \label{fig:saas-cv-2}
     \end{subfigure}
     
     \begin{subfigure}[b]{0.32\textwidth}
        \includegraphics[width=\textwidth,trim={0 0 4.25cm 1.75cm},clip]{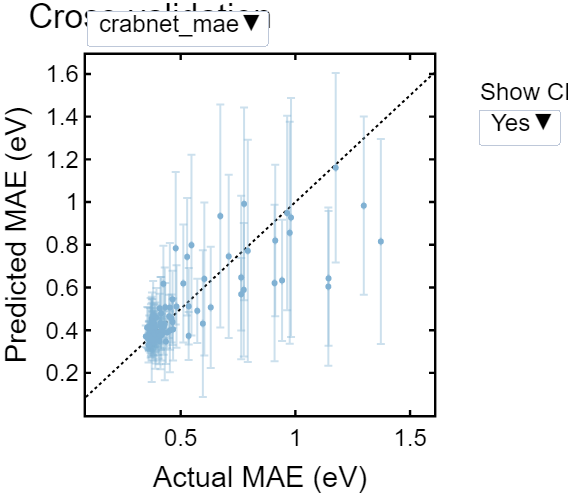}
         \caption{}
         \label{fig:saas-cv-3}
     \end{subfigure}
     \begin{subfigure}[b]{0.32\textwidth}
        \includegraphics[width=\textwidth,trim={0 0 4.25cm 1.75cm},clip]{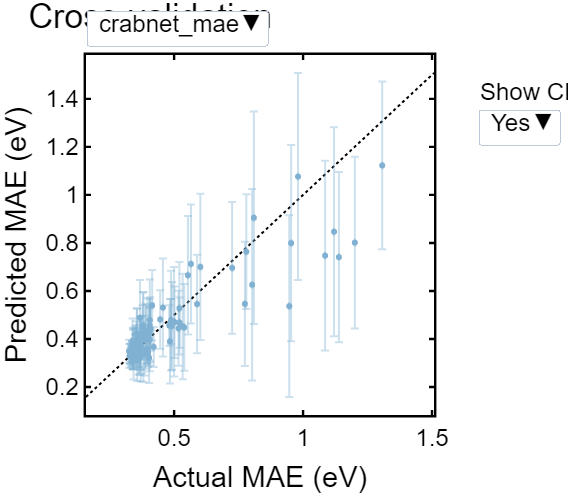}
         \caption{}
         \label{fig:saas-cv-4}
     \end{subfigure}
     
     \caption{\Gls{loo-cv} results for \texttt{SAASBO} \gls{crabnet} hyperparameter combinations for each of the five \texttt{matbench\_expt\_gap}. The first 10 iterations were Sobol points, and the remaining 90 were \gls{saasbo} iterations.}
     \label{fig:saas-cv}
\end{figure*}

The \gls{loo-cv} results for \texttt{GPEI} (\cref{fig:ax-cv}) and \texttt{SAASBO} (\cref{fig:saas-cv}) exhibit qualitative differences. For example, the \texttt{GPEI} results tend to exhibit a plateau from \SIrange{0.4}{1.2}{\eV} and actual \gls{mae} values as large as $\sim$\SI{5}{\eV}. For \texttt{SAASBO}, there are no distinct plateaus, and the typical maximum actual \gls{mae} is $\sim$\SI{1.5}{\eV}. In general, the \texttt{GPEI} results seem to have a higher frequency of overconfident predictions in the high error regions compared with \texttt{SAASBO}.



\subsection{Future Work} \label{sec:results:future}

A number of questions may be interesting to explore in future work:

\begin{itemize}
    \item Is the improvement in predictive performance worth the computational and implementation cost?
    \item How does \gls{moo} \gls{saasbo} perform when also considering \gls{rmse}, \gls{uq} quality (e.g. interval score \cite{chungUncertaintyToolboxOpenSource2021}), model size, and/or computational runtime as additional objectives?
    \item Does the best combination of hyperparameters for one property (e.g. experimental band gap) retain its superior predictive performance when applied to separate properties (e.g. bulk modulus, hardness, Coulombic efficiency) or different data modalities (computational vs. experimental), or are the best parameterizations specific to the particular optimization task?
    \item Can CrabNet hyperparameter optimization be successfully framed as an artificial ( cheaper-than-\gls{dft}) materials discovery benchmark task? 

\end{itemize}

While we focus on hyperparameter optimization, \gls{saasbo} is also highly applicable to finding extraordinary candidates during materials discovery campaigns assuming a relatively small number of initial datapoints and adaptive design iterations characteristic of many materials discovery tasks. In the near-term, we plan to apply \gls{saasbo} to a real materials discovery campaign in comparison with other materials discovery algorithms (Open Citrine Platform \cite{CitrineInformaticsAIPowered2022} and DiSCoVeR \cite{bairdDiSCoVeRMaterialsDiscovery2022}).


\section{Conclusion} 
A recently introduced high-dimensional \gls{bo} scheme (\gls{saasbo}) as well as a classical \gls{bo} scheme are successfully applied to the task of optimizing hyperparameters for a materials informatics model (\gls{crabnet}). To illuminate our initially posed question (\cref{sec:intro}), use of \gls{saasbo} to optimize \gls{crabnet} sets a new state-of-the-art for an experimental band gap regression task (\verb|matbench_expt_gap|) by a margin of $\sim$\SI{0.015}{\eV} ($\sim$\SI{4.5}{\percent} decrease). In terms of model interpretability, \gls{saasbo} successfully identifies number of epochs\footnote{In \gls{crabnet}, early stopping is implemented, and so we expect the performance to eventually plateau as number of epochs is increased.} as a parameter to be maximized; in each of the 5 \gls{cv} folds, the \texttt{SAASBO} optimized number of epochs reached the limit of the max allowed epochs (80). Additionally, we find that \gls{saasbo} seems to favor exploration over exploitation during later design iterations, indicative of the model's ability to avoid overfitting for high-dimensional problems. We believe that \gls{saasbo} and other high-dimensional \gls{bo} algorithms will accelerate deployment of better materials informatics models and efficient searches in high-dimensional materials discovery design spaces.

\printglossaries 

\section*{Conflicts of Interest}
There are no conflicts of interest to declare.

\section*{Acknowledgement}

We thank Anthony Wang for useful discussion on CrabNet hyperparameters. Plots were produced via Plotly \cite{plotly} and Ax's plotting wrapper functions. Several tables were formatted via an \href{https://www.tablesgenerator.com/}{online formatter} \cite{CreateLaTeXTables2021} and the \href{https://github.com/sparks-baird/auto-paper}{auto-paper methodology} \cite{Autopaper2021} was used. This work was supported by the National Science Foundation under Grant No. DMR-1651668.

\section*{CRediT Statement}
\textbf{Sterling G. Baird}: Supervision, Project administration, Conceptualization, Methodology, Software, Validation, Formal analysis, Investigation, Writing - Original Draft, Writing - Review \& Editing, Visualization. \textbf{Taylor D. Sparks}: Supervision, Project administration, Funding acquisition, Resources, Writing - Review \& Editing. \textbf{Marianne Liu}: Methodology, Software, Writing - Review \& Editing, Visualization

\section*{Data Availability}
The raw data required to reproduce these findings is available to download via the python \texttt{matbench v0.5} package (\url{https://pypi.org/project/matbench/0.5/}). The data is also available at \url{https://hackingmaterials.lbl.gov/automatminer/datasets.html}.

The processed data required to reproduce these findings is available to download from \url{https://github.com/sparks-baird/crabnet-hyperparameter} as \href{https://doi.org/10.5281/zenodo.6355045}{v0.1.0} \cite{sterling_baird_2022_6355045}.

The code required to reproduce these findings is hosted at \url{https://github.com/sparks-baird/crabnet-hyperparameter} as \href{https://doi.org/10.5281/zenodo.6355045}{v0.1.0} \cite{sterling_baird_2022_6355045}.

Additionally, the results are published on the Matbench webpage (\url{https://matbench.materialsproject.org/Leaderboards\%20Per-Task/matbench_v0.1_matbench_expt_gap/}), and reproducer Jupyter notebooks and full metadata are available on the Matbench GitHub repository (\url{https://github.com/materialsproject/matbench}) under the \texttt{benchmarks} folder.

\bibliographystyle{elsarticle-num-names}
\bibliography{crabnet-hyperparameter}

\begin{thebibliography}{39}
\expandafter\ifx\csname natexlab\endcsname\relax\def\natexlab#1{#1}\fi
\providecommand{\url}[1]{\texttt{#1}}
\providecommand{\href}[2]{#2}
\providecommand{\path}[1]{#1}
\providecommand{\DOIprefix}{doi:}
\providecommand{\ArXivprefix}{arXiv:}
\providecommand{\URLprefix}{URL: }
\providecommand{\Pubmedprefix}{pmid:}
\providecommand{\doi}[1]{\href{http://dx.doi.org/#1}{\path{#1}}}
\providecommand{\Pubmed}[1]{\href{pmid:#1}{\path{#1}}}
\providecommand{\bibinfo}[2]{#2}
\ifx\xfnm\relax \def\xfnm[#1]{\unskip,\space#1}\fi
\bibitem[{Dong et~al.(2021)Dong, Dan, Li, and
  Hu}]{dongInverseDesignComposite2021}
\bibinfo{author}{R.~Dong}, \bibinfo{author}{Y.~Dan}, \bibinfo{author}{X.~Li},
  \bibinfo{author}{J.~Hu},
\newblock \bibinfo{title}{Inverse {{Design}} of {{Composite Metal Oxide Optical
  Materials}} based on {{Deep Transfer Learning}}},
\newblock \bibinfo{journal}{Computational Materials Science}
  \bibinfo{volume}{188} (\bibinfo{year}{2021}) \bibinfo{pages}{110166}.
  \DOIprefix\doi{10.1016/j.commatsci.2020.110166}.
  \href{http://arxiv.org/abs/2008.10618}{{\tt arXiv:2008.10618}}.
\bibitem[{Espinosa et~al.(2022)Espinosa, Ponce, and
  {Ortiz-Medina}}]{espinosa3DOrthogonalVisionbased2022}
\bibinfo{author}{R.~Espinosa}, \bibinfo{author}{H.~Ponce},
  \bibinfo{author}{J.~{Ortiz-Medina}},
\newblock \bibinfo{title}{A {{3D}} orthogonal vision-based band-gap prediction
  using deep learning: {{A}} proof of concept},
\newblock \bibinfo{journal}{Computational Materials Science}
  \bibinfo{volume}{202} (\bibinfo{year}{2022}) \bibinfo{pages}{110967}.
  \DOIprefix\doi{10.1016/j.commatsci.2021.110967}.
\bibitem[{Ju et~al.(2017)Ju, Shiga, Feng, Hou, Tsuda, and
  Shiomi}]{juDesigningNanostructuresPhonon2017}
\bibinfo{author}{S.~Ju}, \bibinfo{author}{T.~Shiga}, \bibinfo{author}{L.~Feng},
  \bibinfo{author}{Z.~Hou}, \bibinfo{author}{K.~Tsuda},
  \bibinfo{author}{J.~Shiomi},
\newblock \bibinfo{title}{Designing {{Nanostructures}} for {{Phonon Transport}}
  via {{Bayesian Optimization}}},
\newblock \bibinfo{journal}{Phys. Rev. X} \bibinfo{volume}{7}
  (\bibinfo{year}{2017}) \bibinfo{pages}{021024}.
  \DOIprefix\doi{10.1103/PhysRevX.7.021024}.
\bibitem[{Karasuyama et~al.(2020)Karasuyama, Kasugai, Tamura, and
  Shitara}]{karasuyamaComputationalDesignStable2020}
\bibinfo{author}{M.~Karasuyama}, \bibinfo{author}{H.~Kasugai},
  \bibinfo{author}{T.~Tamura}, \bibinfo{author}{K.~Shitara},
\newblock \bibinfo{title}{Computational design of stable and highly
  ion-conductive materials using multi-objective bayesian optimization:
  {{Case}} studies on diffusion of oxygen and lithium},
\newblock \bibinfo{journal}{Computational Materials Science}
  \bibinfo{volume}{184} (\bibinfo{year}{2020}) \bibinfo{pages}{109927}.
  \DOIprefix\doi{10.1016/j.commatsci.2020.109927}.
\bibitem[{Sakurai et~al.(2019)Sakurai, Yada, Simomura, Ju, Kashiwagi, Okada,
  Nagao, Tsuda, and
  Shiomi}]{sakuraiUltranarrowBandWavelengthSelectiveThermal2019}
\bibinfo{author}{A.~Sakurai}, \bibinfo{author}{K.~Yada},
  \bibinfo{author}{T.~Simomura}, \bibinfo{author}{S.~Ju},
  \bibinfo{author}{M.~Kashiwagi}, \bibinfo{author}{H.~Okada},
  \bibinfo{author}{T.~Nagao}, \bibinfo{author}{K.~Tsuda},
  \bibinfo{author}{J.~Shiomi},
\newblock \bibinfo{title}{Ultranarrow-{{Band Wavelength-Selective Thermal
  Emission}} with {{Aperiodic Multilayered Metamaterials Designed}} by
  {{Bayesian Optimization}}},
\newblock \bibinfo{journal}{ACS Cent. Sci.} \bibinfo{volume}{5}
  (\bibinfo{year}{2019}) \bibinfo{pages}{319--326}.
  \DOIprefix\doi{10.1021/acscentsci.8b00802}.
\bibitem[{Talapatra et~al.(2018)Talapatra, Boluki, Duong, Qian, Dougherty, and
  Arr{\'o}yave}]{talapatraAutonomousEfficientExperiment2018}
\bibinfo{author}{A.~Talapatra}, \bibinfo{author}{S.~Boluki},
  \bibinfo{author}{T.~Duong}, \bibinfo{author}{X.~Qian},
  \bibinfo{author}{E.~Dougherty}, \bibinfo{author}{R.~Arr{\'o}yave},
\newblock \bibinfo{title}{Autonomous efficient experiment design for materials
  discovery with {{Bayesian}} model averaging},
\newblock \bibinfo{journal}{Physical Review Materials} \bibinfo{volume}{2}
  (\bibinfo{year}{2018}) \bibinfo{pages}{113803}.
  \DOIprefix\doi{10.1103/PhysRevMaterials.2.113803}.
  \href{http://arxiv.org/abs/1803.05460}{{\tt arXiv:1803.05460}}.
\bibitem[{Wakabayashi et~al.(2019)Wakabayashi, Otsuka, Krockenberger, Sawada,
  Taniyasu, and
  Yamamoto}]{wakabayashiMachinelearningassistedThinfilmGrowth2019}
\bibinfo{author}{Y.~K. Wakabayashi}, \bibinfo{author}{T.~Otsuka},
  \bibinfo{author}{Y.~Krockenberger}, \bibinfo{author}{H.~Sawada},
  \bibinfo{author}{Y.~Taniyasu}, \bibinfo{author}{H.~Yamamoto},
\newblock \bibinfo{title}{Machine-learning-assisted thin-film growth:
  {{Bayesian}} optimization in molecular beam epitaxy of {{SrRuO3}} thin
  films},
\newblock \bibinfo{journal}{APL Materials} \bibinfo{volume}{7}
  (\bibinfo{year}{2019}) \bibinfo{pages}{101114}.
  \DOIprefix\doi{10.1063/1.5123019}.
  \href{http://arxiv.org/abs/1908.00739}{{\tt arXiv:1908.00739}}.
\bibitem[{Wang et~al.(2021)Wang, Kauwe, Murdock, and
  Sparks}]{wangCompositionallyRestrictedAttentionBasedNetwork2021}
\bibinfo{author}{A.~Y.-T. Wang}, \bibinfo{author}{S.~K. Kauwe},
  \bibinfo{author}{R.~J. Murdock}, \bibinfo{author}{D.~Sparks},
\newblock \bibinfo{title}{Compositionally-{{Restricted Attention-Based
  Network}} for {{Materials Property Predictions}}},
\newblock \bibinfo{journal}{npj Computational Materials}
  (\bibinfo{year}{2021}) \bibinfo{pages}{33}.
  \DOIprefix\doi{10.1038/s41524-021-00545-1}.
\bibitem[{Balachandran et~al.(2018)Balachandran, Kowalski, Sehirlioglu, and
  Lookman}]{balachandranExperimentalSearchHightemperature2018}
\bibinfo{author}{P.~V. Balachandran}, \bibinfo{author}{B.~Kowalski},
  \bibinfo{author}{A.~Sehirlioglu}, \bibinfo{author}{T.~Lookman},
\newblock \bibinfo{title}{Experimental search for high-temperature
  ferroelectric perovskites guided by two-step machine learning},
\newblock \bibinfo{journal}{Nature Communications} \bibinfo{volume}{9}
  (\bibinfo{year}{2018}) \bibinfo{pages}{1668}.
  \DOIprefix\doi{10.1038/s41467-018-03821-9}.
\bibitem[{Cao et~al.(2018)Cao, Adutwum, Oliynyk, Luber, Olsen, Mar, and
  Buriak}]{caoHowOptimizeMaterials2018}
\bibinfo{author}{B.~Cao}, \bibinfo{author}{L.~A. Adutwum},
  \bibinfo{author}{A.~O. Oliynyk}, \bibinfo{author}{E.~J. Luber},
  \bibinfo{author}{B.~C. Olsen}, \bibinfo{author}{A.~Mar},
  \bibinfo{author}{J.~M. Buriak},
\newblock \bibinfo{title}{How to optimize materials and devices via design of
  experiments and machine learning: {{Demonstration}} using organic
  photovoltaics},
\newblock \bibinfo{journal}{ACS Nano} \bibinfo{volume}{12}
  (\bibinfo{year}{2018}) \bibinfo{pages}{7434--7444}.
  \DOIprefix\doi{10.1021/acsnano.8b04726}.
\bibitem[{Chen et~al.(2020)Chen, Tian, Zhou, Fang, Ding, Sun, and
  Xue}]{chenMachineLearningAssisted2020}
\bibinfo{author}{Y.~Chen}, \bibinfo{author}{Y.~Tian},
  \bibinfo{author}{Y.~Zhou}, \bibinfo{author}{D.~Fang},
  \bibinfo{author}{X.~Ding}, \bibinfo{author}{J.~Sun},
  \bibinfo{author}{D.~Xue},
\newblock \bibinfo{title}{Machine learning assisted multi-objective
  optimization for materials processing parameters: {{A}} case study in {{Mg}}
  alloy},
\newblock \bibinfo{journal}{Journal of Alloys and Compounds}
  \bibinfo{volume}{844} (\bibinfo{year}{2020}) \bibinfo{pages}{156159}.
  \DOIprefix\doi{10.1016/j.jallcom.2020.156159}.
\bibitem[{Homma et~al.(2020)Homma, Liu, Sumita, Tamura, Fushimi, Iwata, Tsuda,
  and Kaneta}]{hommaOptimizationHeterogeneousTernary2020}
\bibinfo{author}{K.~Homma}, \bibinfo{author}{Y.~Liu},
  \bibinfo{author}{M.~Sumita}, \bibinfo{author}{R.~Tamura},
  \bibinfo{author}{N.~Fushimi}, \bibinfo{author}{J.~Iwata},
  \bibinfo{author}{K.~Tsuda}, \bibinfo{author}{C.~Kaneta},
\newblock \bibinfo{title}{Optimization of a {{Heterogeneous Ternary
  Li3PO4-Li3BO3-Li2SO4Mixture}} for {{Li-Ion Conductivity}} by {{Machine
  Learning}}},
\newblock \bibinfo{journal}{Journal of Physical Chemistry C}
  \bibinfo{volume}{124} (\bibinfo{year}{2020}) \bibinfo{pages}{12865--12870}.
  \DOIprefix\doi{10.1021/acs.jpcc.9b11654}.
  \href{http://arxiv.org/abs/1911.12576}{{\tt arXiv:1911.12576}}.
\bibitem[{Hou et~al.(2019)Hou, Takagiwa, Shinohara, Xu, and
  Tsuda}]{houMachineLearningAssistedDevelopmentTheoretical2019}
\bibinfo{author}{Z.~Hou}, \bibinfo{author}{Y.~Takagiwa},
  \bibinfo{author}{Y.~Shinohara}, \bibinfo{author}{Y.~Xu},
  \bibinfo{author}{K.~Tsuda},
\newblock \bibinfo{title}{Machine-{{Learning-Assisted Development}} and
  {{Theoretical Consideration}} for the {{Al}} 2 {{Fe}} 3 {{Si}} 3
  {{Thermoelectric Material}}},
\newblock \bibinfo{journal}{ACS Applied Materials and Interfaces}
  \bibinfo{volume}{11} (\bibinfo{year}{2019}) \bibinfo{pages}{11545--11554}.
  \DOIprefix\doi{10.1021/acsami.9b02381}.
\bibitem[{Li et~al.(2018)Li, Hou, Gao, Zeng, Ao, Zhou, Da, Liu, Sun, and
  Zhang}]{liEfficientOptimizationPerformance2018}
\bibinfo{author}{X.~Li}, \bibinfo{author}{Z.~Hou}, \bibinfo{author}{S.~Gao},
  \bibinfo{author}{Y.~Zeng}, \bibinfo{author}{J.~Ao},
  \bibinfo{author}{Z.~Zhou}, \bibinfo{author}{B.~Da}, \bibinfo{author}{W.~Liu},
  \bibinfo{author}{Y.~Sun}, \bibinfo{author}{Y.~Zhang},
\newblock \bibinfo{title}{Efficient {{Optimization}} of the {{Performance}} of
  {{Mn2}}+-{{Doped Kesterite Solar Cell}}: {{Machine Learning Aided Synthesis}}
  of {{High Efficient Cu2}}({{Mn}},{{Zn}}){{Sn}}({{S}},{{Se}})4 {{Solar
  Cells}}},
\newblock \bibinfo{journal}{Solar RRL} \bibinfo{volume}{2}
  (\bibinfo{year}{2018}) \bibinfo{pages}{1800198}.
  \DOIprefix\doi{10.1002/solr.201800198}.
\bibitem[{Raccuglia et~al.(2016)Raccuglia, Elbert, Adler, Falk, Wenny, Mollo,
  Zeller, Friedler, Schrier, and
  Norquist}]{raccugliaMachinelearningassistedMaterialsDiscovery2016}
\bibinfo{author}{P.~Raccuglia}, \bibinfo{author}{K.~C. Elbert},
  \bibinfo{author}{P.~D. Adler}, \bibinfo{author}{C.~Falk},
  \bibinfo{author}{M.~B. Wenny}, \bibinfo{author}{A.~Mollo},
  \bibinfo{author}{M.~Zeller}, \bibinfo{author}{S.~A. Friedler},
  \bibinfo{author}{J.~Schrier}, \bibinfo{author}{A.~J. Norquist},
\newblock \bibinfo{title}{Machine-learning-assisted materials discovery using
  failed experiments},
\newblock \bibinfo{journal}{Nature} \bibinfo{volume}{533}
  (\bibinfo{year}{2016}) \bibinfo{pages}{73--76}.
  \DOIprefix\doi{10.1038/nature17439}.
\bibitem[{Zhang et~al.(2021)Zhang, Mansouri~Tehrani, Oliynyk, Day, and
  Brgoch}]{zhangFindingNextSuperhard2021}
\bibinfo{author}{Z.~Zhang}, \bibinfo{author}{A.~Mansouri~Tehrani},
  \bibinfo{author}{A.~O. Oliynyk}, \bibinfo{author}{B.~Day},
  \bibinfo{author}{J.~Brgoch},
\newblock \bibinfo{title}{Finding the {{Next Superhard Material}} through
  {{Ensemble Learning}}},
\newblock \bibinfo{journal}{Adv. Mater.} \bibinfo{volume}{33}
  (\bibinfo{year}{2021}) \bibinfo{pages}{2005112}.
  \DOIprefix\doi{10.1002/adma.202005112}.
\bibitem[{Dunn et~al.(2020)Dunn, Wang, Ganose, Dopp, and
  Jain}]{dunnBenchmarkingMaterialsProperty2020}
\bibinfo{author}{A.~Dunn}, \bibinfo{author}{Q.~Wang},
  \bibinfo{author}{A.~Ganose}, \bibinfo{author}{D.~Dopp},
  \bibinfo{author}{A.~Jain},
\newblock \bibinfo{title}{Benchmarking materials property prediction methods:
  The {{Matbench}} test set and {{Automatminer}} reference algorithm},
\newblock \bibinfo{journal}{npj Comput Mater} \bibinfo{volume}{6}
  (\bibinfo{year}{2020}) \bibinfo{pages}{138}.
  \DOIprefix\doi{10.1038/s41524-020-00406-3}.
\bibitem[{Zhuo et~al.(2018)Zhuo, Mansouri~Tehrani, and
  Brgoch}]{zhuoPredictingBandGaps2018}
\bibinfo{author}{Y.~Zhuo}, \bibinfo{author}{A.~Mansouri~Tehrani},
  \bibinfo{author}{J.~Brgoch},
\newblock \bibinfo{title}{Predicting the {{Band Gaps}} of {{Inorganic Solids}}
  by {{Machine Learning}}},
\newblock \bibinfo{journal}{J. Phys. Chem. Lett.} \bibinfo{volume}{9}
  (\bibinfo{year}{2018}) \bibinfo{pages}{1668--1673}.
  \DOIprefix\doi{10.1021/acs.jpclett.8b00124}.
\bibitem[{Mat(2022)}]{MatBench2022}
\bibinfo{title}{{{MatBench}}},
  \bibinfo{howpublished}{https://matbench.materialsproject.org/},
  \bibinfo{year}{2022}.
\bibitem[{Goodall and Lee(2020)}]{goodallPredictingMaterialsProperties2020}
\bibinfo{author}{R.~E.~A. Goodall}, \bibinfo{author}{A.~A. Lee},
\newblock \bibinfo{title}{Predicting materials properties without crystal
  structure: Deep representation learning from stoichiometry},
\newblock \bibinfo{journal}{Nat Commun} \bibinfo{volume}{11}
  (\bibinfo{year}{2020}) \bibinfo{pages}{6280}.
  \DOIprefix\doi{10.1038/s41467-020-19964-7}.
\bibitem[{Jha et~al.(2018)Jha, Ward, Paul, Liao, Choudhary, Wolverton, and
  Agrawal}]{jhaElemNetDeepLearning2018}
\bibinfo{author}{D.~Jha}, \bibinfo{author}{L.~Ward}, \bibinfo{author}{A.~Paul},
  \bibinfo{author}{W.-k. Liao}, \bibinfo{author}{A.~Choudhary},
  \bibinfo{author}{C.~Wolverton}, \bibinfo{author}{A.~Agrawal},
\newblock \bibinfo{title}{{{ElemNet}}: {{Deep Learning}} the {{Chemistry}} of
  {{Materials From Only Elemental Composition}}},
\newblock \bibinfo{journal}{Sci Rep} \bibinfo{volume}{8} (\bibinfo{year}{2018})
  \bibinfo{pages}{17593}. \DOIprefix\doi{10.1038/s41598-018-35934-y}.
\bibitem[{Jha et~al.(2019)Jha, Choudhary, Tavazza, Liao, Choudhary, Campbell,
  and Agrawal}]{jhaEnhancingMaterialsProperty2019}
\bibinfo{author}{D.~Jha}, \bibinfo{author}{K.~Choudhary},
  \bibinfo{author}{F.~Tavazza}, \bibinfo{author}{W.-k. Liao},
  \bibinfo{author}{A.~Choudhary}, \bibinfo{author}{C.~Campbell},
  \bibinfo{author}{A.~Agrawal},
\newblock \bibinfo{title}{Enhancing materials property prediction by leveraging
  computational and experimental data using deep transfer learning},
\newblock \bibinfo{journal}{Nat Commun} \bibinfo{volume}{10}
  (\bibinfo{year}{2019}) \bibinfo{pages}{5316}.
  \DOIprefix\doi{10.1038/s41467-019-13297-w}.
\bibitem[{Chen et~al.(2019)Chen, Ye, Zuo, Zheng, and
  Ong}]{chenGraphNetworksUniversal2019}
\bibinfo{author}{C.~Chen}, \bibinfo{author}{W.~Ye}, \bibinfo{author}{Y.~Zuo},
  \bibinfo{author}{C.~Zheng}, \bibinfo{author}{S.~P. Ong},
\newblock \bibinfo{title}{Graph {{Networks}} as a {{Universal Machine Learning
  Framework}} for {{Molecules}} and {{Crystals}}},
\newblock \bibinfo{journal}{Chem. Mater.} \bibinfo{volume}{31}
  (\bibinfo{year}{2019}) \bibinfo{pages}{3564--3572}.
  \DOIprefix\doi{10.1021/acs.chemmater.9b01294}.
\bibitem[{De~Breuck et~al.(2021)De~Breuck, Hautier, and
  Rignanese}]{debreuckMaterialsPropertyPrediction2021}
\bibinfo{author}{P.-P. De~Breuck}, \bibinfo{author}{G.~Hautier},
  \bibinfo{author}{G.-M. Rignanese},
\newblock \bibinfo{title}{Materials property prediction for limited datasets
  enabled by feature selection and joint learning with {{MODNet}}},
\newblock \bibinfo{journal}{npj Comput Mater} \bibinfo{volume}{7}
  (\bibinfo{year}{2021}) \bibinfo{pages}{83}.
  \DOIprefix\doi{10.1038/s41524-021-00552-2}.
\bibitem[{{de Jong} et~al.(2016){de Jong}, Chen, Notestine, Persson, Ceder,
  Jain, Asta, and Gamst}]{dejongStatisticalLearningFramework2016}
\bibinfo{author}{M.~{de Jong}}, \bibinfo{author}{W.~Chen},
  \bibinfo{author}{R.~Notestine}, \bibinfo{author}{K.~Persson},
  \bibinfo{author}{G.~Ceder}, \bibinfo{author}{A.~Jain},
  \bibinfo{author}{M.~Asta}, \bibinfo{author}{A.~Gamst},
\newblock \bibinfo{title}{A {{Statistical Learning Framework}} for {{Materials
  Science}}: {{Application}} to {{Elastic Moduli}} of k-nary {{Inorganic
  Polycrystalline Compounds}}},
\newblock \bibinfo{journal}{Sci Rep} \bibinfo{volume}{6} (\bibinfo{year}{2016})
  \bibinfo{pages}{34256}. \DOIprefix\doi{10.1038/srep34256}.
\bibitem[{Goodall et~al.(2020)Goodall, Parackal, Faber, and
  Armiento}]{goodallWyckoffSetRegression2020}
\bibinfo{author}{R.~E.~A. Goodall}, \bibinfo{author}{A.~S. Parackal},
  \bibinfo{author}{F.~A. Faber}, \bibinfo{author}{R.~Armiento},
\newblock \bibinfo{title}{Wyckoff {{Set Regression}} for {{Materials
  Discovery}}},
\newblock in: \bibinfo{booktitle}{Neural {{Information Processing Systems}}},
  \bibinfo{year}{2020}, p.~\bibinfo{pages}{7}.
\bibitem[{Klicpera et~al.(2020)Klicpera, Giri, Margraf, and
  G{\"u}nnemann}]{klicperaFastUncertaintyAwareDirectional2020}
\bibinfo{author}{J.~Klicpera}, \bibinfo{author}{S.~Giri},
  \bibinfo{author}{J.~T. Margraf}, \bibinfo{author}{S.~G{\"u}nnemann},
\newblock \bibinfo{title}{Fast and {{Uncertainty-Aware Directional Message
  Passing}} for {{Non-Equilibrium Molecules}}},
\newblock \bibinfo{journal}{arXiv:2011.14115 [physics]}
  (\bibinfo{year}{2020}). \href{http://arxiv.org/abs/2011.14115}{{\tt
  arXiv:2011.14115}}.
\bibitem[{Louis et~al.(2020)Louis, Zhao, Nasiri, Wang, Song, Liu, and
  Hu}]{louisGraphConvolutionalNeural2020}
\bibinfo{author}{S.-Y. Louis}, \bibinfo{author}{Y.~Zhao},
  \bibinfo{author}{A.~Nasiri}, \bibinfo{author}{X.~Wang},
  \bibinfo{author}{Y.~Song}, \bibinfo{author}{F.~Liu}, \bibinfo{author}{J.~Hu},
\newblock \bibinfo{title}{Graph convolutional neural networks with global
  attention for improved materials property prediction},
\newblock \bibinfo{journal}{Phys. Chem. Chem. Phys.} \bibinfo{volume}{22}
  (\bibinfo{year}{2020}) \bibinfo{pages}{18141--18148}.
  \DOIprefix\doi{10.1039/D0CP01474E}.
\bibitem[{Park and Wolverton(2020)}]{parkDevelopingImprovedCrystal2020}
\bibinfo{author}{C.~W. Park}, \bibinfo{author}{C.~Wolverton},
\newblock \bibinfo{title}{Developing an improved crystal graph convolutional
  neural network framework for accelerated materials discovery},
\newblock \bibinfo{journal}{Phys. Rev. Materials} \bibinfo{volume}{4}
  (\bibinfo{year}{2020}) \bibinfo{pages}{063801}.
  \DOIprefix\doi{10.1103/PhysRevMaterials.4.063801}.
\bibitem[{Xie and Grossman(2018)}]{xieCrystalGraphConvolutional2018}
\bibinfo{author}{T.~Xie}, \bibinfo{author}{J.~C. Grossman},
\newblock \bibinfo{title}{Crystal {{Graph Convolutional Neural Networks}} for
  an {{Accurate}} and {{Interpretable Prediction}} of {{Material Properties}}},
\newblock \bibinfo{journal}{Phys. Rev. Lett.} \bibinfo{volume}{120}
  (\bibinfo{year}{2018}) \bibinfo{pages}{145301}.
  \DOIprefix\doi{10.1103/PhysRevLett.120.145301}.
\bibitem[{Eriksson and
  Jankowiak(2021)}]{erikssonHighDimensionalBayesianOptimization2021}
\bibinfo{author}{D.~Eriksson}, \bibinfo{author}{M.~Jankowiak},
\newblock \bibinfo{title}{High-{{Dimensional Bayesian Optimization}} with
  {{Sparse Axis-Aligned Subspaces}}},
\newblock \bibinfo{journal}{arXiv:2103.00349 [cs, stat]}
  (\bibinfo{year}{2021}). \href{http://arxiv.org/abs/2103.00349}{{\tt
  arXiv:2103.00349}}.
\bibitem[{Hig(2022)}]{HighDimensionalBayesianOptimization2022}
\bibinfo{title}{High-{{Dimensional Bayesian Optimization}} with {{SAASBO}}},
  \bibinfo{howpublished}{https://ax.dev/tutorials/saasbo.html},
  \bibinfo{year}{2022}.
\bibitem[{Chung et~al.(2021)Chung, Char, Guo, Schneider, and
  Neiswanger}]{chungUncertaintyToolboxOpenSource2021}
\bibinfo{author}{Y.~Chung}, \bibinfo{author}{I.~Char},
  \bibinfo{author}{H.~Guo}, \bibinfo{author}{J.~Schneider},
  \bibinfo{author}{W.~Neiswanger},
\newblock \bibinfo{title}{Uncertainty {{Toolbox}}: An {{Open-Source Library}}
  for {{Assessing}}, {{Visualizing}}, and {{Improving Uncertainty
  Quantification}}},
\newblock \bibinfo{journal}{arXiv:2109.10254 [cs, stat]}
  (\bibinfo{year}{2021}). \href{http://arxiv.org/abs/2109.10254}{{\tt
  arXiv:2109.10254}}.
\bibitem[{Cit(2022)}]{CitrineInformaticsAIPowered2022}
\bibinfo{title}{Citrine {{Informatics}}: {{AI-Powered Materials Data
  Platform}}}, \bibinfo{howpublished}{https://citrination.com/},
  \bibinfo{year}{2022}.
\bibitem[{Baird et~al.(2022)Baird, Diep, and
  Sparks}]{bairdDiSCoVeRMaterialsDiscovery2022}
\bibinfo{author}{S.~G. Baird}, \bibinfo{author}{T.~Q. Diep},
  \bibinfo{author}{T.~D. Sparks},
\newblock \bibinfo{title}{{{DiSCoVeR}}: A {{Materials Discovery Screening
  Tool}} for {{High Performance}}, {{Unique Chemical Compositions}}},
\newblock \bibinfo{journal}{Digital Discovery}  (\bibinfo{year}{2022}).
  \DOIprefix\doi{10.1039/D1DD00028D}.
\bibitem[{Inc.(2015)}]{plotly}
\bibinfo{author}{P.~T. Inc.}, \bibinfo{title}{Collaborative data science},
  \bibinfo{howpublished}{https://plot.ly}, \bibinfo{year}{2015}.
\bibitem[{Cre(2021)}]{CreateLaTeXTables2021}
\bibinfo{title}{Create {{LaTeX}} tables online \textendash{}
  {{TablesGenerator}}.com},
  \bibinfo{howpublished}{https://www.tablesgenerator.com/},
  \bibinfo{year}{2021}.
\bibitem[{Aut(2021)}]{Autopaper2021}
\bibinfo{title}{Auto-paper}, \bibinfo{howpublished}{sparks-baird},
  \bibinfo{year}{2021}.
\bibitem[{Baird et~al.(2022)Baird, {sgbaird-alt}, and
  {mliu7051}}]{sterling_baird_2022_6355045}
\bibinfo{author}{S.~Baird}, \bibinfo{author}{{sgbaird-alt}},
  \bibinfo{author}{{mliu7051}},
  \bibinfo{title}{Sparks-baird/crabnet-hyperparameter:},
  \bibinfo{howpublished}{Zenodo}, \bibinfo{year}{2022}.
  \DOIprefix\doi{10.5281/zenodo.6355045}.

\end{thebibliography}

\end{document}



\begin{frontmatter}

\title{\mytitle{}: Supporting Information}

\author[myu]{Sterling G. Baird\corref{cor1}}
\ead{sterling.baird@utah.edu}
\author[myu,west]{Marianne Liu}
\author[myu]{Taylor D. Sparks}
\ead{sparks@eng.utah.edu}

\address[myu]{Department of Materials Science and Engineering, University of Utah, Salt Lake City, UT 84108, USA}
\address[west]{West High School, Salt Lake City, Utah 84112, USA}

\cortext[cor1]{Corresponding author.}

\date{October 2021}

\end{frontmatter}

\tableofcontents

\section{Fold-wise Feature Importances}
\label{supp:feat}

\begin{figure*}
     \centering
     \begin{subfigure}[b]{0.32\textwidth}
         \includegraphics[width=\textwidth,trim={0 0 0.85cm 0.5cm},clip]{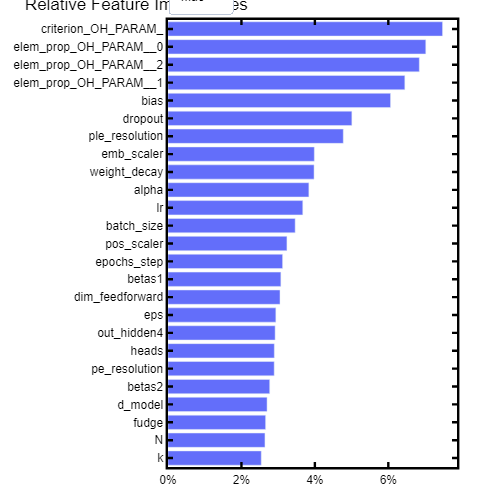}
         \caption{}
         \label{fig:ax-feat-0}
     \end{subfigure}
     \hfill
     \begin{subfigure}[b]{0.32\textwidth}
         \includegraphics[width=\textwidth,trim={0.4cm 0 0.85cm 0.5cm},clip]{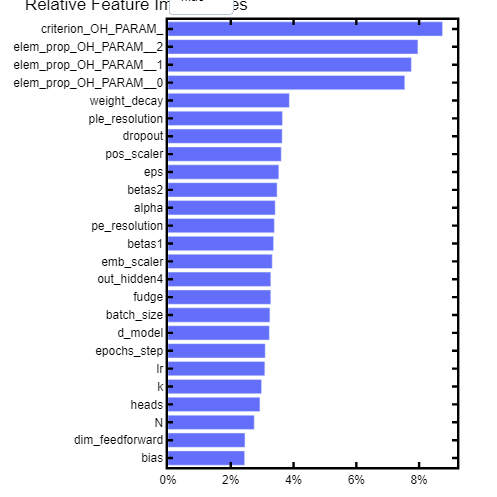}
         \caption{}
         \label{fig:ax-feat-1}
     \end{subfigure}
     \hfill
     \begin{subfigure}[b]{0.32\textwidth}
         \includegraphics[width=\textwidth,trim={0.4cm 0 0.85cm 0.5cm},clip]{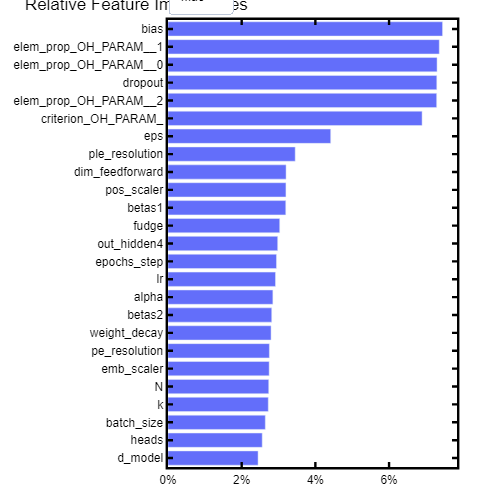}
         \caption{}
         \label{fig:ax-feat-2}
     \end{subfigure}
     
     \begin{subfigure}[b]{0.32\textwidth}
        \includegraphics[width=\textwidth,trim={0.4cm 0 0.85cm 0.5cm},clip]{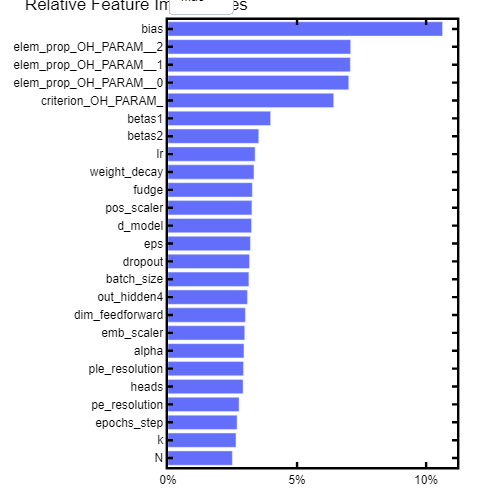}
         \caption{}
         \label{fig:ax-feat-3}
     \end{subfigure}
     \begin{subfigure}[b]{0.32\textwidth}
        \includegraphics[width=\textwidth,trim={0.4cm 0 0.85cm 0.5cm},clip]{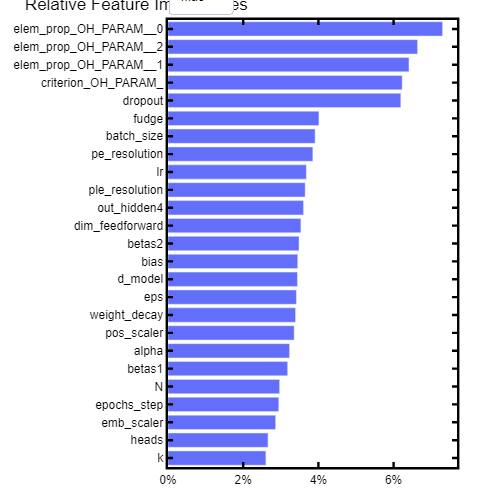}
         \caption{}
         \label{fig:ax-feat-4}
     \end{subfigure}
     
     \caption{Relative \texttt{GPEI} feature importances of \gls{crabnet} hyperparameters for each of the five \texttt{matbench\_expt\_gap} folds based on data from 100 iterations. \texttt{\_OH\_PARAM\_\_\#} refers to choices for categorical variables. The first 46 iterations were Sobol points, and the remaining 54 iterations were Bayesian optimization iterations. }
     \label{fig:ax-feat}
\end{figure*}

\begin{figure*}
     \centering
     \begin{subfigure}[b]{0.32\textwidth}
         \includegraphics[width=\textwidth,trim={0.4cm 0 0.85cm 0.5cm},clip]{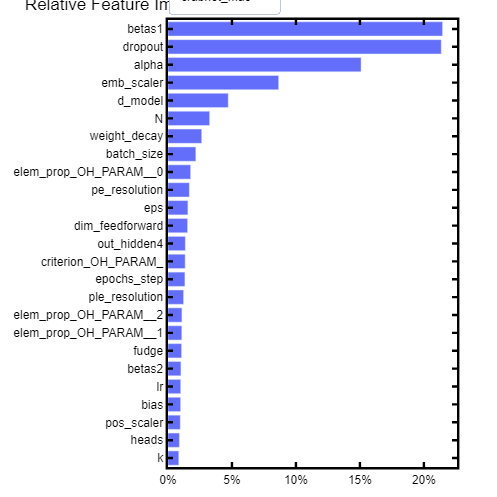}
         \caption{}
         \label{fig:saas-feat-0}
     \end{subfigure}
     \hfill
     \begin{subfigure}[b]{0.32\textwidth}
         \includegraphics[width=\textwidth,trim={0.4cm 0 0.85cm 0.5cm},clip]{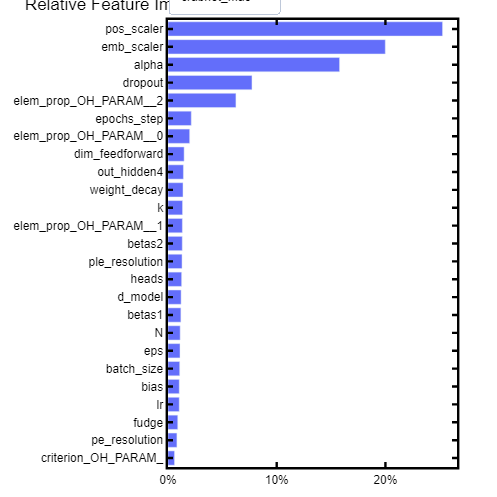}
         \caption{}
         \label{fig:saas-feat-1}
     \end{subfigure}
     \hfill
     \begin{subfigure}[b]{0.32\textwidth}
         \includegraphics[width=\textwidth,trim={0.4cm 0 0.85cm 0.5cm},clip]{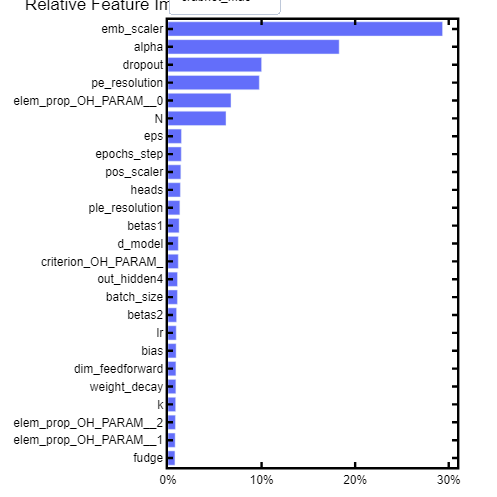}
         \caption{}
         \label{fig:saas-feat-2}
     \end{subfigure}
     
     \begin{subfigure}[b]{0.32\textwidth}
        \includegraphics[width=\textwidth,trim={0.4cm 0 0.85cm 0.5cm},clip]{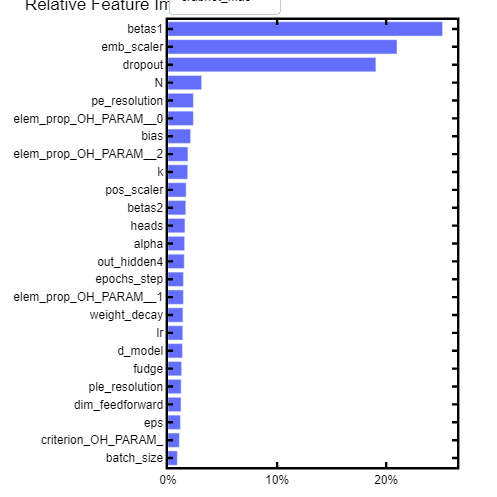}
         \caption{}
         \label{fig:saas-feat-3}
     \end{subfigure}
     \begin{subfigure}[b]{0.32\textwidth}
        \includegraphics[width=\textwidth,trim={0.4cm 0 0.85cm 0.5cm},clip]{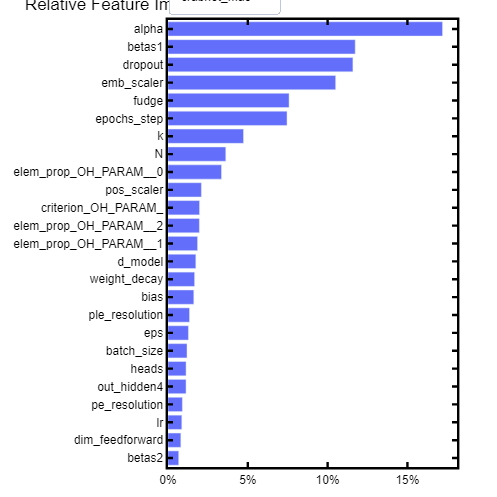}
         \caption{}
         \label{fig:saas-feat-4}
     \end{subfigure}
     
     \caption{Relative \texttt{SAASBO} feature importances of \gls{crabnet} hyperparameters for each of the five \texttt{matbench\_expt\_gap} folds based on data from 100 iterations. \texttt{\_OH\_PARAM\_\_\#} refers to choices for categorical variables. The first 10 iterations were Sobol points, and the remaining 90 were \gls{saasbo} iterations. Note that feature importances are tied with the constraints on the search space imposed by the user. A differently constrained search space (and by extension characteristics of the sampled points) may result in different feature importances. }
     \label{fig:saas-feat}
\end{figure*}

Feature importances for each of the 5 Matbench folds across the 23 hyperparameters for the \gls{gpei} (\texttt{GPEI}) and \gls{saasbo} models (\texttt{SAASBO}) are given in \cref{fig:ax-feat} and \cref{fig:saas-feat}, respectively.

\clearpage
\newpage

\section{1D Slices through Model Parameter Space}
\label{supp:slice}

One-dimensional slices through \texttt{GPEI} (top row) and \texttt{SAASBO} (bottom row) models for each of the 23 CrabNet hyperparameters (\cref{tab:param}) and for each of the 5 Matbench folds are shown following roughly the order in \cref{tab:param}. There is one figure per hyperparameter, labeled above each figure as well as mentioned within the caption.

\newpage

\begin{figure*}
    \centering
    \verb|N|

    \centering
     \begin{subfigure}[b]{0.1925\textwidth}
         \includegraphics[width=\textwidth,trim={1.09cm 1.04cm 0.86cm 0.35cm},clip]{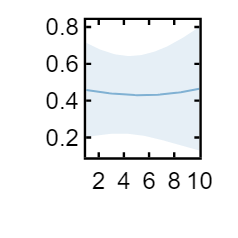}
         \label{fig:ax-0-N}
         \caption{\texttt{GPEI} fold 0}
     \end{subfigure}
     \hfill
     \begin{subfigure}[b]{0.1925\textwidth}
         \includegraphics[width=\textwidth,trim={1.09cm 1.04cm 0.86cm 0.35cm},clip]{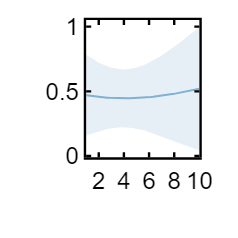}
         \label{fig:ax-1-N}
         \caption{\texttt{GPEI} fold 1}
     \end{subfigure}
     \hfill
     \begin{subfigure}[b]{0.1925\textwidth}
         \includegraphics[width=\textwidth,trim={1.09cm 1.04cm 0.86cm 0.35cm},clip]{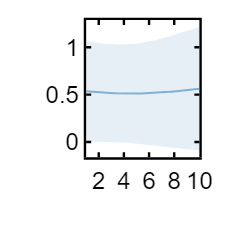}
         \label{fig:ax-2-N}
         \caption{\texttt{GPEI} fold 2}
     \end{subfigure}
     \hfill
     \begin{subfigure}[b]{0.1925\textwidth}
         \includegraphics[width=\textwidth,trim={1.09cm 1.04cm 0.86cm 0.35cm},clip]{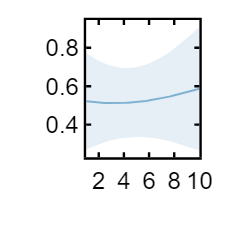}
         \label{fig:ax-3-N}
         \caption{\texttt{GPEI} fold 3}
     \end{subfigure}
     \hfill
     \begin{subfigure}[b]{0.1925\textwidth}
         \includegraphics[width=\textwidth,trim={1.09cm 1.04cm 0.86cm 0.35cm},clip]{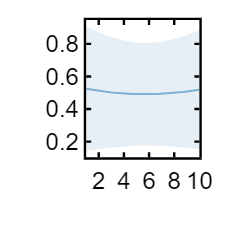}
         \label{fig:ax-4-N}
         \caption{\texttt{GPEI} fold 4}
     \end{subfigure}
     
     \begin{subfigure}[b]{0.1925\textwidth}
         \includegraphics[width=\textwidth,trim={1.09cm 1.04cm 0.86cm 0.35cm},clip]{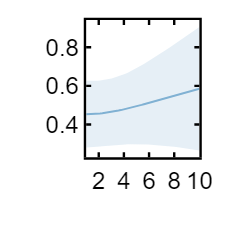}
         \label{fig:saas-0-N}
         \caption{\texttt{SAASBO} fold 0}
     \end{subfigure}
     \hfill
     \begin{subfigure}[b]{0.1925\textwidth}
         \includegraphics[width=\textwidth,trim={1.09cm 1.04cm 0.86cm 0.35cm},clip]{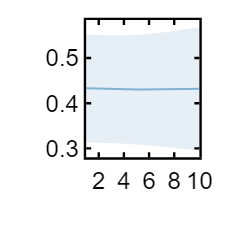}
         \label{fig:saas-1-N}
         \caption{\texttt{SAASBO} fold 1}
     \end{subfigure}
     \hfill
     \begin{subfigure}[b]{0.1925\textwidth}
         \includegraphics[width=\textwidth,trim={1.09cm 1.04cm 0.86cm 0.35cm},clip]{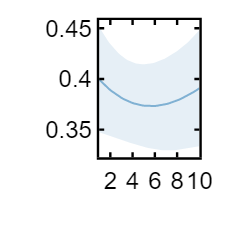}
         \label{fig:saas-2-N}
         \caption{\texttt{SAASBO} fold 2}
     \end{subfigure}
     \hfill
     \begin{subfigure}[b]{0.1925\textwidth}
         \includegraphics[width=\textwidth,trim={1.09cm 1.04cm 0.86cm 0.35cm},clip]{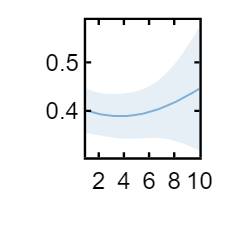}
         \label{fig:saas-3-N}
         \caption{\texttt{SAASBO} fold 3}
     \end{subfigure}
     \hfill
     \begin{subfigure}[b]{0.1925\textwidth}
         \includegraphics[width=\textwidth,trim={1.09cm 1.04cm 0.86cm 0.35cm},clip]{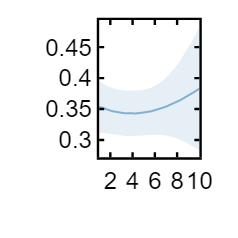}
         \label{fig:saas-4-N}
         \caption{\texttt{SAASBO} fold 4}
     \end{subfigure}

     \caption{One-dimensional (1D) slices of \gls{mae} (eV) vs. \texttt{N} through the \texttt{GPEI} and \texttt{SAASBO} parameter spaces with the rest of the parameters fixed to the mean and mode of the numeric and categorical parameters for each of the models, respectively. Shaded blue error bands give the standard deviation uncertainty predicted by the model. . }
     \label{fig:1d-N}
\end{figure*}

\begin{figure*}
    \centering
    \verb|alpha|

    \centering
     \begin{subfigure}[b]{0.1925\textwidth}
         \includegraphics[width=\textwidth,trim={1.09cm 1.04cm 0.86cm 0.35cm},clip]{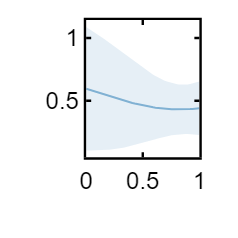}
         \label{fig:ax-0-alpha}
         \caption{\texttt{GPEI} fold 0}
     \end{subfigure}
     \hfill
     \begin{subfigure}[b]{0.1925\textwidth}
         \includegraphics[width=\textwidth,trim={1.09cm 1.04cm 0.86cm 0.35cm},clip]{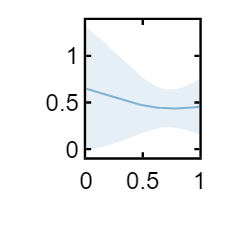}
         \label{fig:ax-1-alpha}
         \caption{\texttt{GPEI} fold 1}
     \end{subfigure}
     \hfill
     \begin{subfigure}[b]{0.1925\textwidth}
         \includegraphics[width=\textwidth,trim={1.09cm 1.04cm 0.86cm 0.35cm},clip]{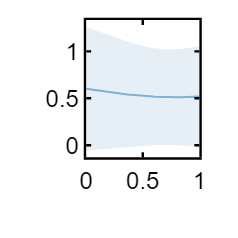}
         \label{fig:ax-2-alpha}
         \caption{\texttt{GPEI} fold 2}
     \end{subfigure}
     \hfill
     \begin{subfigure}[b]{0.1925\textwidth}
         \includegraphics[width=\textwidth,trim={1.09cm 1.04cm 0.86cm 0.35cm},clip]{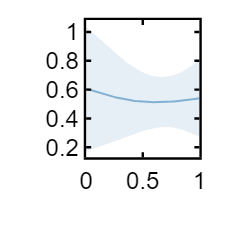}
         \label{fig:ax-3-alpha}
         \caption{\texttt{GPEI} fold 3}
     \end{subfigure}
     \hfill
     \begin{subfigure}[b]{0.1925\textwidth}
         \includegraphics[width=\textwidth,trim={1.09cm 1.04cm 0.86cm 0.35cm},clip]{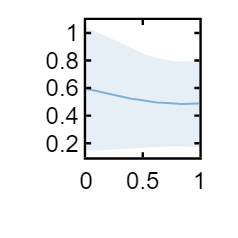}
         \label{fig:ax-4-alpha}
         \caption{\texttt{GPEI} fold 4}
     \end{subfigure}
     
     \begin{subfigure}[b]{0.1925\textwidth}
         \includegraphics[width=\textwidth,trim={1.09cm 1.04cm 0.86cm 0.35cm},clip]{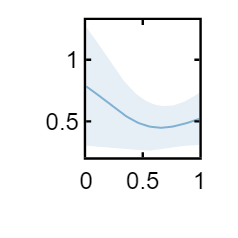}
         \label{fig:saas-0-alpha}
         \caption{\texttt{SAASBO} fold 0}
     \end{subfigure}
     \hfill
     \begin{subfigure}[b]{0.1925\textwidth}
         \includegraphics[width=\textwidth,trim={1.09cm 1.04cm 0.86cm 0.35cm},clip]{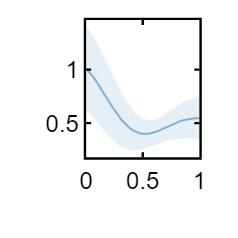}
         \label{fig:saas-1-alpha}
         \caption{\texttt{SAASBO} fold 1}
     \end{subfigure}
     \hfill
     \begin{subfigure}[b]{0.1925\textwidth}
         \includegraphics[width=\textwidth,trim={1.09cm 1.04cm 0.86cm 0.35cm},clip]{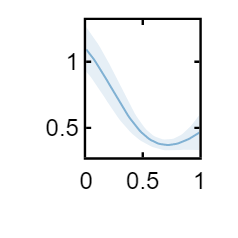}
         \label{fig:saas-2-alpha}
         \caption{\texttt{SAASBO} fold 2}
     \end{subfigure}
     \hfill
     \begin{subfigure}[b]{0.1925\textwidth}
         \includegraphics[width=\textwidth,trim={1.09cm 1.04cm 0.86cm 0.35cm},clip]{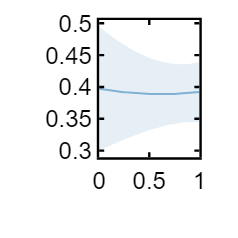}
         \label{fig:saas-3-alpha}
         \caption{\texttt{SAASBO} fold 3}
     \end{subfigure}
     \hfill
     \begin{subfigure}[b]{0.1925\textwidth}
         \includegraphics[width=\textwidth,trim={1.09cm 1.04cm 0.86cm 0.35cm},clip]{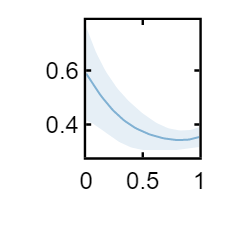}
         \label{fig:saas-4-alpha}
         \caption{\texttt{SAASBO} fold 4}
     \end{subfigure}

     \caption{One-dimensional (1D) slices of \gls{mae} (eV) vs. \texttt{alpha} through the \texttt{GPEI} and \texttt{SAASBO} parameter spaces with the rest of the parameters fixed to the mean and mode of the numeric and categorical parameters for each of the models, respectively. Shaded blue error bands give the standard deviation uncertainty predicted by the model. . }
     \label{fig:1d-alpha}
\end{figure*}

\begin{figure*}
    \centering
    \verb|d_model|

     \begin{subfigure}[b]{0.1925\textwidth}
         \includegraphics[width=\textwidth,trim={1.09cm 1.04cm 0.86cm 0.35cm},clip]{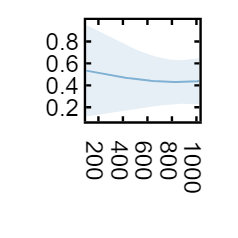}
         \label{fig:ax-0-d_model}
         \caption{\texttt{GPEI} fold 0}
     \end{subfigure}
     \hfill
     \begin{subfigure}[b]{0.1925\textwidth}
         \includegraphics[width=\textwidth,trim={1.09cm 1.04cm 0.86cm 0.35cm},clip]{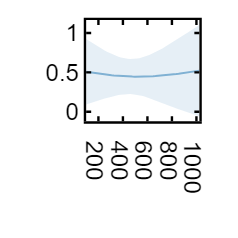}
         \label{fig:ax-1-d_model}
         \caption{\texttt{GPEI} fold 1}
     \end{subfigure}
     \hfill
     \begin{subfigure}[b]{0.1925\textwidth}
         \includegraphics[width=\textwidth,trim={1.09cm 1.04cm 0.86cm 0.35cm},clip]{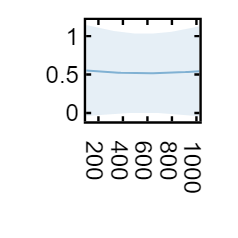}
         \label{fig:ax-2-d_model}
         \caption{\texttt{GPEI} fold 2}
     \end{subfigure}
     \hfill
     \begin{subfigure}[b]{0.1925\textwidth}
         \includegraphics[width=\textwidth,trim={1.09cm 1.04cm 0.86cm 0.35cm},clip]{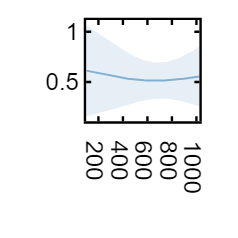}
         \label{fig:ax-3-d_model}
         \caption{\texttt{GPEI} fold 3}
     \end{subfigure}
     \hfill
     \begin{subfigure}[b]{0.1925\textwidth}
         \includegraphics[width=\textwidth,trim={1.09cm 1.04cm 0.86cm 0.35cm},clip]{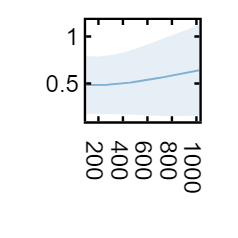}
         \label{fig:ax-4-d_model}
         \caption{\texttt{GPEI} fold 4}
     \end{subfigure}
     
     \begin{subfigure}[b]{0.1925\textwidth}
         \includegraphics[width=\textwidth,trim={1.09cm 1.04cm 0.86cm 0.35cm},clip]{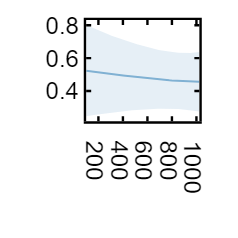}
         \label{fig:saas-0-d_model}
         \caption{\texttt{SAASBO} fold 0}
     \end{subfigure}
     \hfill
     \begin{subfigure}[b]{0.1925\textwidth}
         \includegraphics[width=\textwidth,trim={1.09cm 1.04cm 0.86cm 0.35cm},clip]{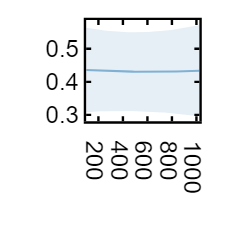}
         \label{fig:saas-1-d_model}
         \caption{\texttt{SAASBO} fold 1}
     \end{subfigure}
     \hfill
     \begin{subfigure}[b]{0.1925\textwidth}
         \includegraphics[width=\textwidth,trim={1.09cm 1.04cm 0.86cm 0.35cm},clip]{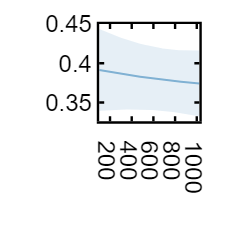}
         \label{fig:saas-2-d_model}
         \caption{\texttt{SAASBO} fold 2}
     \end{subfigure}
     \hfill
     \begin{subfigure}[b]{0.1925\textwidth}
         \includegraphics[width=\textwidth,trim={1.09cm 1.04cm 0.86cm 0.35cm},clip]{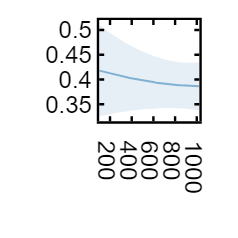}
         \label{fig:saas-3-d_model}
         \caption{\texttt{SAASBO} fold 3}
     \end{subfigure}
     \hfill
     \begin{subfigure}[b]{0.1925\textwidth}
         \includegraphics[width=\textwidth,trim={1.09cm 1.04cm 0.86cm 0.35cm},clip]{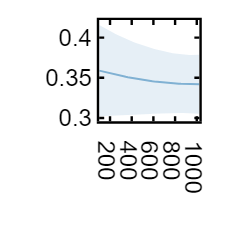}
         \label{fig:saas-4-d_model}
         \caption{\texttt{SAASBO} fold 4}
     \end{subfigure}

     \caption{One-dimensional (1D) slices of \gls{mae} (eV) vs. \texttt{d\_model} through the \texttt{GPEI} and \texttt{SAASBO} parameter spaces with the rest of the parameters fixed to the mean and mode of the numeric and categorical parameters for each of the models, respectively. Shaded blue error bands give the standard deviation uncertainty predicted by the model. . }
     \label{fig:1d-d_model}
\end{figure*}

\begin{figure*}
    \centering
    \verb|dim_feedforward|

     \begin{subfigure}[b]{0.1925\textwidth}
         \includegraphics[width=\textwidth,trim={1.09cm 1.04cm 0.86cm 0.35cm},clip]{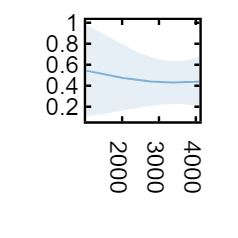}
         \label{fig:ax-0-dim_feedforward}
         \caption{\texttt{GPEI} fold 0}
     \end{subfigure}
     \hfill
     \begin{subfigure}[b]{0.1925\textwidth}
         \includegraphics[width=\textwidth,trim={1.09cm 1.04cm 0.86cm 0.35cm},clip]{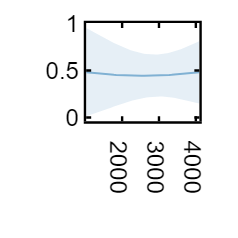}
         \label{fig:ax-1-dim_feedforward}
         \caption{\texttt{GPEI} fold 1}
     \end{subfigure}
     \hfill
     \begin{subfigure}[b]{0.1925\textwidth}
         \includegraphics[width=\textwidth,trim={1.09cm 1.04cm 0.86cm 0.35cm},clip]{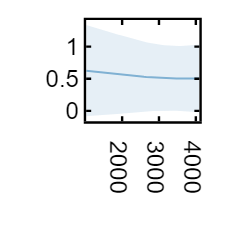}
         \label{fig:ax-2-dim_feedforward}
         \caption{\texttt{GPEI} fold 2}
     \end{subfigure}
     \hfill
     \begin{subfigure}[b]{0.1925\textwidth}
         \includegraphics[width=\textwidth,trim={1.09cm 1.04cm 0.86cm 0.35cm},clip]{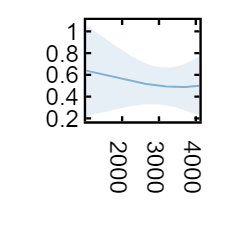}
         \label{fig:ax-3-dim_feedforward}
         \caption{\texttt{GPEI} fold 3}
     \end{subfigure}
     \hfill
     \begin{subfigure}[b]{0.1925\textwidth}
         \includegraphics[width=\textwidth,trim={1.09cm 1.04cm 0.86cm 0.35cm},clip]{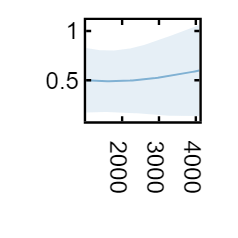}
         \label{fig:ax-4-dim_feedforward}
         \caption{\texttt{GPEI} fold 4}
     \end{subfigure}
     
     \begin{subfigure}[b]{0.1925\textwidth}
         \includegraphics[width=\textwidth,trim={1.09cm 1.04cm 0.86cm 0.35cm},clip]{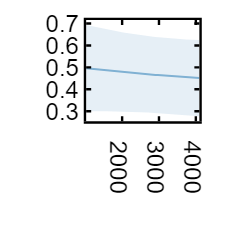}
         \label{fig:saas-0-dim_feedforward}
         \caption{\texttt{SAASBO} fold 0}
     \end{subfigure}
     \hfill
     \begin{subfigure}[b]{0.1925\textwidth}
         \includegraphics[width=\textwidth,trim={1.09cm 1.04cm 0.86cm 0.35cm},clip]{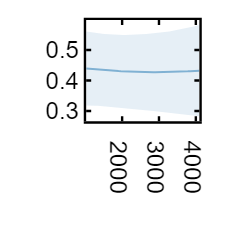}
         \label{fig:saas-1-dim_feedforward}
         \caption{\texttt{SAASBO} fold 1}
     \end{subfigure}
     \hfill
     \begin{subfigure}[b]{0.1925\textwidth}
         \includegraphics[width=\textwidth,trim={1.09cm 1.04cm 0.86cm 0.35cm},clip]{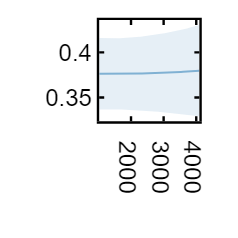}
         \label{fig:saas-2-dim_feedforward}
         \caption{\texttt{SAASBO} fold 2}
     \end{subfigure}
     \hfill
     \begin{subfigure}[b]{0.1925\textwidth}
         \includegraphics[width=\textwidth,trim={1.09cm 1.04cm 0.86cm 0.35cm},clip]{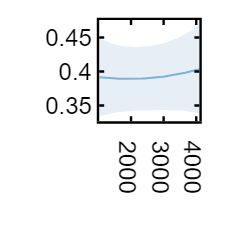}
         \label{fig:saas-3-dim_feedforward}
         \caption{\texttt{SAASBO} fold 3}
     \end{subfigure}
     \hfill
     \begin{subfigure}[b]{0.1925\textwidth}
         \includegraphics[width=\textwidth,trim={1.09cm 1.04cm 0.86cm 0.35cm},clip]{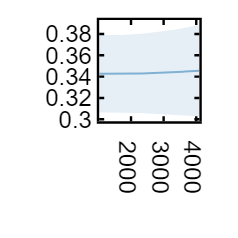}
         \label{fig:saas-4-dim_feedforward}
         \caption{\texttt{SAASBO} fold 4}
     \end{subfigure}

     \caption{One-dimensional (1D) slices of \gls{mae} (eV) vs. \texttt{dim\_feedforward} through the \texttt{GPEI} and \texttt{SAASBO} parameter spaces with the rest of the parameters fixed to the mean and mode of the numeric and categorical parameters for each of the models, respectively. Shaded blue error bands give the standard deviation uncertainty predicted by the model. . }
     \label{fig:1d-dim_feedforward}
\end{figure*}

\begin{figure*}
    \centering
    \verb|dropout|

     \begin{subfigure}[b]{0.1925\textwidth}
         \includegraphics[width=\textwidth,trim={1.09cm 1.04cm 0.86cm 0.35cm},clip]{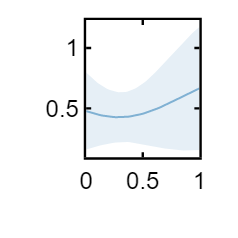}
         \label{fig:ax-0-dropout}
         \caption{\texttt{GPEI} fold 0}
     \end{subfigure}
     \hfill
     \begin{subfigure}[b]{0.1925\textwidth}
         \includegraphics[width=\textwidth,trim={1.09cm 1.04cm 0.86cm 0.35cm},clip]{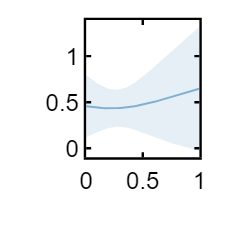}
         \label{fig:ax-1-dropout}
         \caption{\texttt{GPEI} fold 1}
     \end{subfigure}
     \hfill
     \begin{subfigure}[b]{0.1925\textwidth}
         \includegraphics[width=\textwidth,trim={1.09cm 1.04cm 0.86cm 0.35cm},clip]{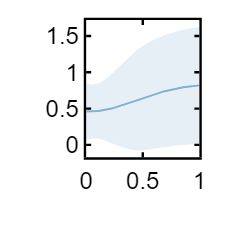}
         \label{fig:ax-2-dropout}
         \caption{\texttt{GPEI} fold 2}
     \end{subfigure}
     \hfill
     \begin{subfigure}[b]{0.1925\textwidth}
         \includegraphics[width=\textwidth,trim={1.09cm 1.04cm 0.86cm 0.35cm},clip]{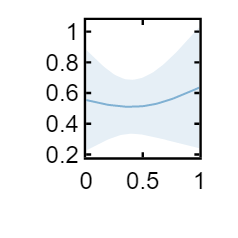}
         \label{fig:ax-3-dropout}
         \caption{\texttt{GPEI} fold 3}
     \end{subfigure}
     \hfill
     \begin{subfigure}[b]{0.1925\textwidth}
         \includegraphics[width=\textwidth,trim={1.09cm 1.04cm 0.86cm 0.35cm},clip]{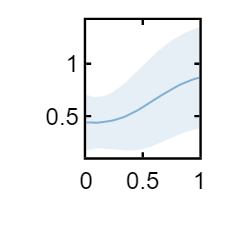}
         \label{fig:ax-4-dropout}
         \caption{\texttt{GPEI} fold 4}
     \end{subfigure}
     
     \begin{subfigure}[b]{0.1925\textwidth}
         \includegraphics[width=\textwidth,trim={1.09cm 1.04cm 0.86cm 0.35cm},clip]{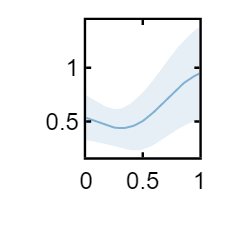}
         \label{fig:saas-0-dropout}
         \caption{\texttt{SAASBO} fold 0}
     \end{subfigure}
     \hfill
     \begin{subfigure}[b]{0.1925\textwidth}
         \includegraphics[width=\textwidth,trim={1.09cm 1.04cm 0.86cm 0.35cm},clip]{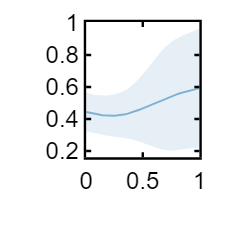}
         \label{fig:saas-1-dropout}
         \caption{\texttt{SAASBO} fold 1}
     \end{subfigure}
     \hfill
     \begin{subfigure}[b]{0.1925\textwidth}
         \includegraphics[width=\textwidth,trim={1.09cm 1.04cm 0.86cm 0.35cm},clip]{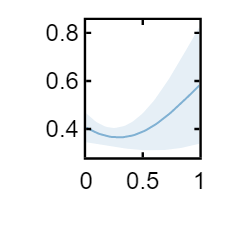}
         \label{fig:saas-2-dropout}
         \caption{\texttt{SAASBO} fold 2}
     \end{subfigure}
     \hfill
     \begin{subfigure}[b]{0.1925\textwidth}
         \includegraphics[width=\textwidth,trim={1.09cm 1.04cm 0.86cm 0.35cm},clip]{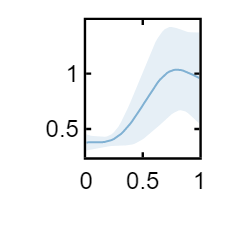}
         \label{fig:saas-3-dropout}
         \caption{\texttt{SAASBO} fold 3}
     \end{subfigure}
     \hfill
     \begin{subfigure}[b]{0.1925\textwidth}
         \includegraphics[width=\textwidth,trim={1.09cm 1.04cm 0.86cm 0.35cm},clip]{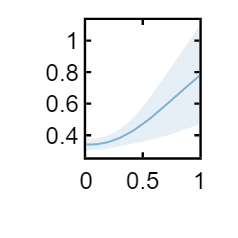}
         \label{fig:saas-4-dropout}
         \caption{\texttt{SAASBO} fold 4}
     \end{subfigure}

     \caption{One-dimensional (1D) slices of \gls{mae} (eV) vs. \texttt{dropout} through the \texttt{GPEI} and \texttt{SAASBO} parameter spaces with the rest of the parameters fixed to the mean and mode of the numeric and categorical parameters for each of the models, respectively. Shaded blue error bands give the standard deviation uncertainty predicted by the model. . }
     \label{fig:1d-dropout}
\end{figure*}

\begin{figure*}
    \centering
    \verb|emb_scaler|

     \begin{subfigure}[b]{0.1925\textwidth}
         \includegraphics[width=\textwidth,trim={1.09cm 1.04cm 0.86cm 0.35cm},clip]{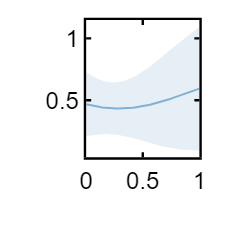}
         \label{fig:ax-0-emb_scaler}
         \caption{\texttt{GPEI} fold 0}
     \end{subfigure}
     \hfill
     \begin{subfigure}[b]{0.1925\textwidth}
         \includegraphics[width=\textwidth,trim={1.09cm 1.04cm 0.86cm 0.35cm},clip]{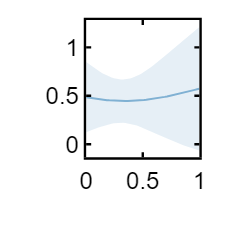}
         \label{fig:ax-1-emb_scaler}
         \caption{\texttt{GPEI} fold 1}
     \end{subfigure}
     \hfill
     \begin{subfigure}[b]{0.1925\textwidth}
         \includegraphics[width=\textwidth,trim={1.09cm 1.04cm 0.86cm 0.35cm},clip]{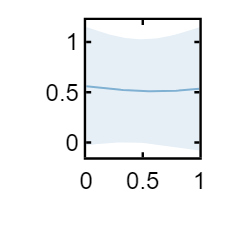}
         \label{fig:ax-2-emb_scaler}
         \caption{\texttt{GPEI} fold 2}
     \end{subfigure}
     \hfill
     \begin{subfigure}[b]{0.1925\textwidth}
         \includegraphics[width=\textwidth,trim={1.09cm 1.04cm 0.86cm 0.35cm},clip]{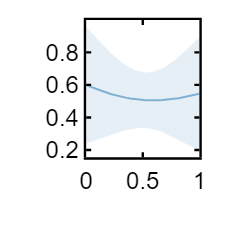}
         \label{fig:ax-3-emb_scaler}
         \caption{\texttt{GPEI} fold 3}
     \end{subfigure}
     \hfill
     \begin{subfigure}[b]{0.1925\textwidth}
         \includegraphics[width=\textwidth,trim={1.09cm 1.04cm 0.86cm 0.35cm},clip]{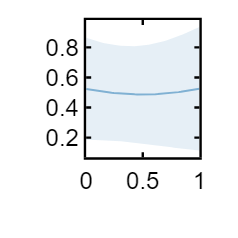}
         \label{fig:ax-4-emb_scaler}
         \caption{\texttt{GPEI} fold 4}
     \end{subfigure}
     
     \begin{subfigure}[b]{0.1925\textwidth}
         \includegraphics[width=\textwidth,trim={1.09cm 1.04cm 0.86cm 0.35cm},clip]{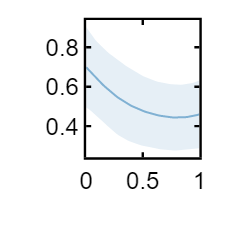}
         \label{fig:saas-0-emb_scaler}
         \caption{\texttt{SAASBO} fold 0}
     \end{subfigure}
     \hfill
     \begin{subfigure}[b]{0.1925\textwidth}
         \includegraphics[width=\textwidth,trim={1.09cm 1.04cm 0.86cm 0.35cm},clip]{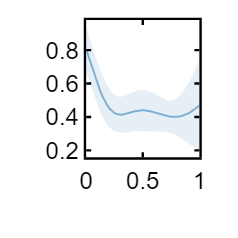}
         \label{fig:saas-1-emb_scaler}
         \caption{\texttt{SAASBO} fold 1}
     \end{subfigure}
     \hfill
     \begin{subfigure}[b]{0.1925\textwidth}
         \includegraphics[width=\textwidth,trim={1.09cm 1.04cm 0.86cm 0.35cm},clip]{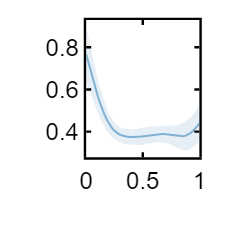}
         \label{fig:saas-2-emb_scaler}
         \caption{\texttt{SAASBO} fold 2}
     \end{subfigure}
     \hfill
     \begin{subfigure}[b]{0.1925\textwidth}
         \includegraphics[width=\textwidth,trim={1.09cm 1.04cm 0.86cm 0.35cm},clip]{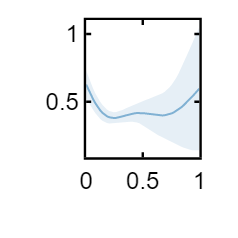}
         \label{fig:saas-3-emb_scaler}
         \caption{\texttt{SAASBO} fold 3}
     \end{subfigure}
     \hfill
     \begin{subfigure}[b]{0.1925\textwidth}
         \includegraphics[width=\textwidth,trim={1.09cm 1.04cm 0.86cm 0.35cm},clip]{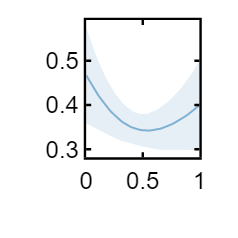}
         \label{fig:saas-4-emb_scaler}
         \caption{\texttt{SAASBO} fold 4}
     \end{subfigure}

     \caption{One-dimensional (1D) slices of \gls{mae} (eV) vs. \texttt{emb\_scaler} through the \texttt{GPEI} and \texttt{SAASBO} parameter spaces with the rest of the parameters fixed to the mean and mode of the numeric and categorical parameters for each of the models, respectively. Shaded blue error bands give the standard deviation uncertainty predicted by the model. . }
     \label{fig:1d-emb_scaler}
\end{figure*}

\begin{figure*}
    \centering
    \verb|epochs_step|

     \begin{subfigure}[b]{0.1925\textwidth}
         \includegraphics[width=\textwidth,trim={1.09cm 1.04cm 0.86cm 0.35cm},clip]{figures/ax/slice/slice_0_epochs_step.png}
         \label{fig:ax-0-epochs_step}
         \caption{\texttt{GPEI} fold 0}
     \end{subfigure}
     \hfill
     \begin{subfigure}[b]{0.1925\textwidth}
         \includegraphics[width=\textwidth,trim={1.09cm 1.04cm 0.86cm 0.35cm},clip]{figures/ax/slice/slice_1_epochs_step.png}
         \label{fig:ax-1-epochs_step}
         \caption{\texttt{GPEI} fold 1}
     \end{subfigure}
     \hfill
     \begin{subfigure}[b]{0.1925\textwidth}
         \includegraphics[width=\textwidth,trim={1.09cm 1.04cm 0.86cm 0.35cm},clip]{figures/ax/slice/slice_2_epochs_step.png}
         \label{fig:ax-2-epochs_step}
         \caption{\texttt{GPEI} fold 2}
     \end{subfigure}
     \hfill
     \begin{subfigure}[b]{0.1925\textwidth}
         \includegraphics[width=\textwidth,trim={1.09cm 1.04cm 0.86cm 0.35cm},clip]{figures/ax/slice/slice_3_epochs_step.png}
         \label{fig:ax-3-epochs_step}
         \caption{\texttt{GPEI} fold 3}
     \end{subfigure}
     \hfill
     \begin{subfigure}[b]{0.1925\textwidth}
         \includegraphics[width=\textwidth,trim={1.09cm 1.04cm 0.86cm 0.35cm},clip]{figures/ax/slice/slice_4_epochs_step.png}
         \label{fig:ax-4-epochs_step}
         \caption{\texttt{GPEI} fold 4}
     \end{subfigure}
     
     \begin{subfigure}[b]{0.1925\textwidth}
         \includegraphics[width=\textwidth,trim={1.09cm 1.04cm 0.86cm 0.35cm},clip]{figures/saas/slice/slice_0_epochs_step.png}
         \label{fig:saas-0-epochs_step}
         \caption{\texttt{SAASBO} fold 0}
     \end{subfigure}
     \hfill
     \begin{subfigure}[b]{0.1925\textwidth}
         \includegraphics[width=\textwidth,trim={1.09cm 1.04cm 0.86cm 0.35cm},clip]{figures/saas/slice/slice_1_epochs_step.png}
         \label{fig:saas-1-epochs_step}
         \caption{\texttt{SAASBO} fold 1}
     \end{subfigure}
     \hfill
     \begin{subfigure}[b]{0.1925\textwidth}
         \includegraphics[width=\textwidth,trim={1.09cm 1.04cm 0.86cm 0.35cm},clip]{figures/saas/slice/slice_2_epochs_step.png}
         \label{fig:saas-2-epochs_step}
         \caption{\texttt{SAASBO} fold 2}
     \end{subfigure}
     \hfill
     \begin{subfigure}[b]{0.1925\textwidth}
         \includegraphics[width=\textwidth,trim={1.09cm 1.04cm 0.86cm 0.35cm},clip]{figures/saas/slice/slice_3_epochs_step.png}
         \label{fig:saas-3-epochs_step}
         \caption{\texttt{SAASBO} fold 3}
     \end{subfigure}
     \hfill
     \begin{subfigure}[b]{0.1925\textwidth}
         \includegraphics[width=\textwidth,trim={1.09cm 1.04cm 0.86cm 0.35cm},clip]{figures/saas/slice/slice_4_epochs_step.png}
         \label{fig:saas-4-epochs_step}
         \caption{\texttt{SAASBO} fold 4}
     \end{subfigure}

     \caption{One-dimensional (1D) slices of \gls{mae} (eV) vs. \texttt{epochs\_step} through the \texttt{GPEI} and \texttt{SAASBO} parameter spaces with the rest of the parameters fixed to the mean and mode of the numeric and categorical parameters for each of the models, respectively. Shaded blue error bands give the standard deviation uncertainty predicted by the model. . }
     \label{fig:1d-epochs_step}
\end{figure*}

\begin{figure*}
    \centering
    \verb|eps|

     \begin{subfigure}[b]{0.1925\textwidth}
         \includegraphics[width=\textwidth,trim={1.09cm 1.04cm 0.86cm 0.35cm},clip]{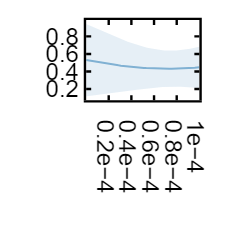}
         \label{fig:ax-0-eps}
         \caption{\texttt{GPEI} fold 0}
     \end{subfigure}
     \hfill
     \begin{subfigure}[b]{0.1925\textwidth}
         \includegraphics[width=\textwidth,trim={1.09cm 1.04cm 0.86cm 0.35cm},clip]{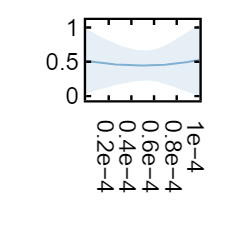}
         \label{fig:ax-1-eps}
         \caption{\texttt{GPEI} fold 1}
     \end{subfigure}
     \hfill
     \begin{subfigure}[b]{0.1925\textwidth}
         \includegraphics[width=\textwidth,trim={1.09cm 1.04cm 0.86cm 0.35cm},clip]{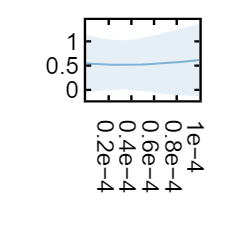}
         \label{fig:ax-2-eps}
         \caption{\texttt{GPEI} fold 2}
     \end{subfigure}
     \hfill
     \begin{subfigure}[b]{0.1925\textwidth}
         \includegraphics[width=\textwidth,trim={1.09cm 1.04cm 0.86cm 0.35cm},clip]{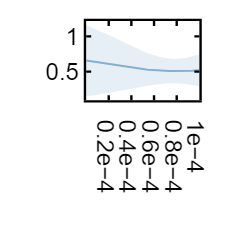}
         \label{fig:ax-3-eps}
         \caption{\texttt{GPEI} fold 3}
     \end{subfigure}
     \hfill
     \begin{subfigure}[b]{0.1925\textwidth}
         \includegraphics[width=\textwidth,trim={1.09cm 1.04cm 0.86cm 0.35cm},clip]{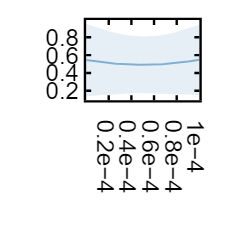}
         \label{fig:ax-4-eps}
         \caption{\texttt{GPEI} fold 4}
     \end{subfigure}
     
     \begin{subfigure}[b]{0.1925\textwidth}
         \includegraphics[width=\textwidth,trim={1.09cm 1.04cm 0.86cm 0.35cm},clip]{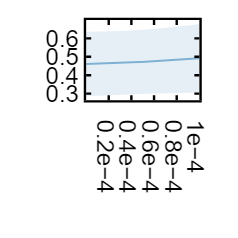}
         \label{fig:saas-0-eps}
         \caption{\texttt{SAASBO} fold 0}
     \end{subfigure}
     \hfill
     \begin{subfigure}[b]{0.1925\textwidth}
         \includegraphics[width=\textwidth,trim={1.09cm 1.04cm 0.86cm 0.35cm},clip]{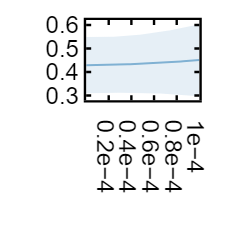}
         \label{fig:saas-1-eps}
         \caption{\texttt{SAASBO} fold 1}
     \end{subfigure}
     \hfill
     \begin{subfigure}[b]{0.1925\textwidth}
         \includegraphics[width=\textwidth,trim={1.09cm 1.04cm 0.86cm 0.35cm},clip]{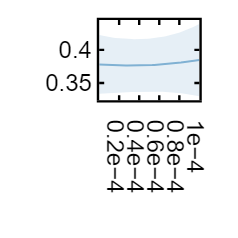}
         \label{fig:saas-2-eps}
         \caption{\texttt{SAASBO} fold 2}
     \end{subfigure}
     \hfill
     \begin{subfigure}[b]{0.1925\textwidth}
         \includegraphics[width=\textwidth,trim={1.09cm 1.04cm 0.86cm 0.35cm},clip]{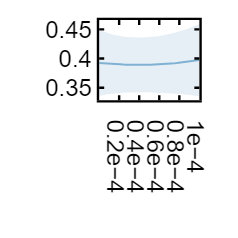}
         \label{fig:saas-3-eps}
         \caption{\texttt{SAASBO} fold 3}
     \end{subfigure}
     \hfill
     \begin{subfigure}[b]{0.1925\textwidth}
         \includegraphics[width=\textwidth,trim={1.09cm 1.04cm 0.86cm 0.35cm},clip]{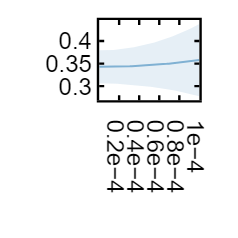}
         \label{fig:saas-4-eps}
         \caption{\texttt{SAASBO} fold 4}
     \end{subfigure}

     \caption{One-dimensional (1D) slices of \gls{mae} (eV) vs. \texttt{eps} through the \texttt{GPEI} and \texttt{SAASBO} parameter spaces with the rest of the parameters fixed to the mean and mode of the numeric and categorical parameters for each of the models, respectively. Shaded blue error bands give the standard deviation uncertainty predicted by the model. . }
     \label{fig:1d-eps}
\end{figure*}

\begin{figure*}
    \centering
    \verb|fudge|

     \begin{subfigure}[b]{0.1925\textwidth}
         \includegraphics[width=\textwidth,trim={1.09cm 1.04cm 0.86cm 0.35cm},clip]{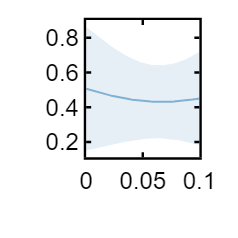}
         \label{fig:ax-0-fudge}
         \caption{\texttt{GPEI} fold 0}
     \end{subfigure}
     \hfill
     \begin{subfigure}[b]{0.1925\textwidth}
         \includegraphics[width=\textwidth,trim={1.09cm 1.04cm 0.86cm 0.35cm},clip]{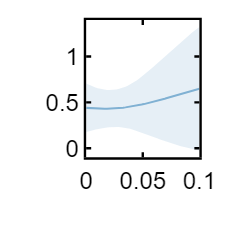}
         \label{fig:ax-1-fudge}
         \caption{\texttt{GPEI} fold 1}
     \end{subfigure}
     \hfill
     \begin{subfigure}[b]{0.1925\textwidth}
         \includegraphics[width=\textwidth,trim={1.09cm 1.04cm 0.86cm 0.35cm},clip]{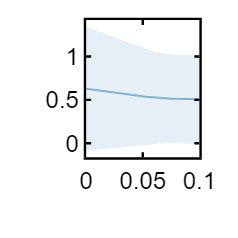}
         \label{fig:ax-2-fudge}
         \caption{\texttt{GPEI} fold 2}
     \end{subfigure}
     \hfill
     \begin{subfigure}[b]{0.1925\textwidth}
         \includegraphics[width=\textwidth,trim={1.09cm 1.04cm 0.86cm 0.35cm},clip]{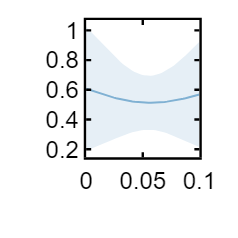}
         \label{fig:ax-3-fudge}
         \caption{\texttt{GPEI} fold 3}
     \end{subfigure}
     \hfill
     \begin{subfigure}[b]{0.1925\textwidth}
         \includegraphics[width=\textwidth,trim={1.09cm 1.04cm 0.86cm 0.35cm},clip]{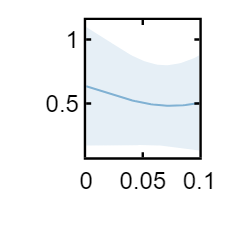}
         \label{fig:ax-4-fudge}
         \caption{\texttt{GPEI} fold 4}
     \end{subfigure}
     
     \begin{subfigure}[b]{0.1925\textwidth}
         \includegraphics[width=\textwidth,trim={1.09cm 1.04cm 0.86cm 0.35cm},clip]{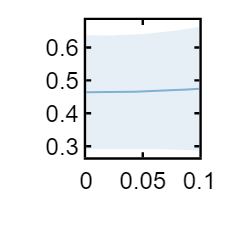}
         \label{fig:saas-0-fudge}
         \caption{\texttt{SAASBO} fold 0}
     \end{subfigure}
     \hfill
     \begin{subfigure}[b]{0.1925\textwidth}
         \includegraphics[width=\textwidth,trim={1.09cm 1.04cm 0.86cm 0.35cm},clip]{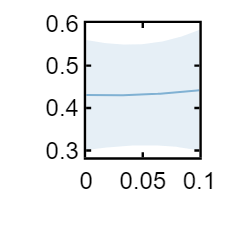}
         \label{fig:saas-1-fudge}
         \caption{\texttt{SAASBO} fold 1}
     \end{subfigure}
     \hfill
     \begin{subfigure}[b]{0.1925\textwidth}
         \includegraphics[width=\textwidth,trim={1.09cm 1.04cm 0.86cm 0.35cm},clip]{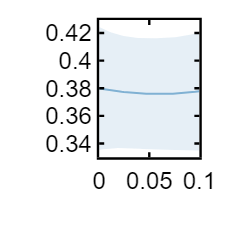}
         \label{fig:saas-2-fudge}
         \caption{\texttt{SAASBO} fold 2}
     \end{subfigure}
     \hfill
     \begin{subfigure}[b]{0.1925\textwidth}
         \includegraphics[width=\textwidth,trim={1.09cm 1.04cm 0.86cm 0.35cm},clip]{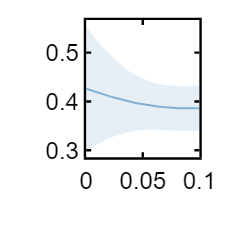}
         \label{fig:saas-3-fudge}
         \caption{\texttt{SAASBO} fold 3}
     \end{subfigure}
     \hfill
     \begin{subfigure}[b]{0.1925\textwidth}
         \includegraphics[width=\textwidth,trim={1.09cm 1.04cm 0.86cm 0.35cm},clip]{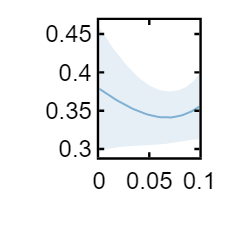}
         \label{fig:saas-4-fudge}
         \caption{\texttt{SAASBO} fold 4}
     \end{subfigure}

     \caption{One-dimensional (1D) slices of \gls{mae} (eV) vs. \texttt{fudge} through the \texttt{GPEI} and \texttt{SAASBO} parameter spaces with the rest of the parameters fixed to the mean and mode of the numeric and categorical parameters for each of the models, respectively. Shaded blue error bands give the standard deviation uncertainty predicted by the model. . }
     \label{fig:1d-fudge}
\end{figure*}

\begin{figure*}
    \centering
    \verb|heads|

     \begin{subfigure}[b]{0.1925\textwidth}
         \includegraphics[width=\textwidth,trim={1.09cm 1.04cm 0.86cm 0.35cm},clip]{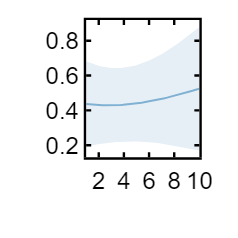}
         \label{fig:ax-0-heads}
         \caption{\texttt{GPEI} fold 0}
     \end{subfigure}
     \hfill
     \begin{subfigure}[b]{0.1925\textwidth}
         \includegraphics[width=\textwidth,trim={1.09cm 1.04cm 0.86cm 0.35cm},clip]{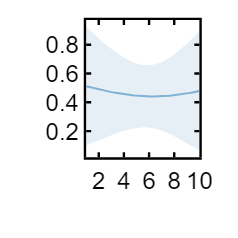}
         \label{fig:ax-1-heads}
         \caption{\texttt{GPEI} fold 1}
     \end{subfigure}
     \hfill
     \begin{subfigure}[b]{0.1925\textwidth}
         \includegraphics[width=\textwidth,trim={1.09cm 1.04cm 0.86cm 0.35cm},clip]{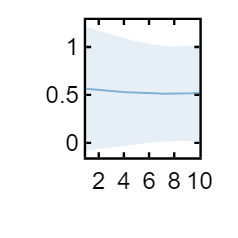}
         \label{fig:ax-2-heads}
         \caption{\texttt{GPEI} fold 2}
     \end{subfigure}
     \hfill
     \begin{subfigure}[b]{0.1925\textwidth}
         \includegraphics[width=\textwidth,trim={1.09cm 1.04cm 0.86cm 0.35cm},clip]{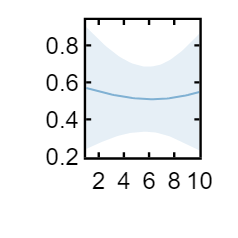}
         \label{fig:ax-3-heads}
         \caption{\texttt{GPEI} fold 3}
     \end{subfigure}
     \hfill
     \begin{subfigure}[b]{0.1925\textwidth}
         \includegraphics[width=\textwidth,trim={1.09cm 1.04cm 0.86cm 0.35cm},clip]{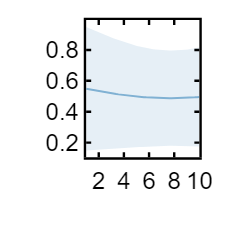}
         \label{fig:ax-4-heads}
         \caption{\texttt{GPEI} fold 4}
     \end{subfigure}
     
     \begin{subfigure}[b]{0.1925\textwidth}
         \includegraphics[width=\textwidth,trim={1.09cm 1.04cm 0.86cm 0.35cm},clip]{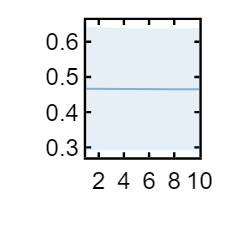}
         \label{fig:saas-0-heads}
         \caption{\texttt{SAASBO} fold 0}
     \end{subfigure}
     \hfill
     \begin{subfigure}[b]{0.1925\textwidth}
         \includegraphics[width=\textwidth,trim={1.09cm 1.04cm 0.86cm 0.35cm},clip]{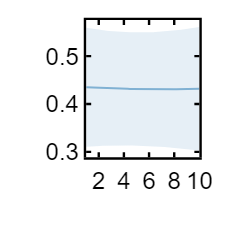}
         \label{fig:saas-1-heads}
         \caption{\texttt{SAASBO} fold 1}
     \end{subfigure}
     \hfill
     \begin{subfigure}[b]{0.1925\textwidth}
         \includegraphics[width=\textwidth,trim={1.09cm 1.04cm 0.86cm 0.35cm},clip]{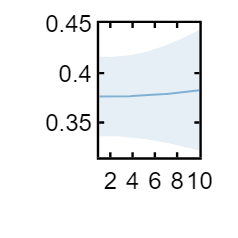}
         \label{fig:saas-2-heads}
         \caption{\texttt{SAASBO} fold 2}
     \end{subfigure}
     \hfill
     \begin{subfigure}[b]{0.1925\textwidth}
         \includegraphics[width=\textwidth,trim={1.09cm 1.04cm 0.86cm 0.35cm},clip]{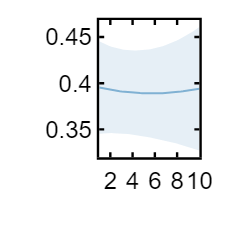}
         \label{fig:saas-3-heads}
         \caption{\texttt{SAASBO} fold 3}
     \end{subfigure}
     \hfill
     \begin{subfigure}[b]{0.1925\textwidth}
         \includegraphics[width=\textwidth,trim={1.09cm 1.04cm 0.86cm 0.35cm},clip]{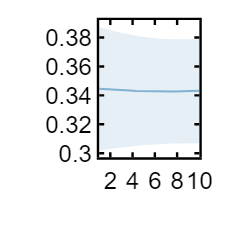}
         \label{fig:saas-4-heads}
         \caption{\texttt{SAASBO} fold 4}
     \end{subfigure}

     \caption{One-dimensional (1D) slices of \gls{mae} (eV) vs. \texttt{heads} through the \texttt{GPEI} and \texttt{SAASBO} parameter spaces with the rest of the parameters fixed to the mean and mode of the numeric and categorical parameters for each of the models, respectively. Shaded blue error bands give the standard deviation uncertainty predicted by the model. . }
     \label{fig:1d-heads}
\end{figure*}

\begin{figure*}
    \centering
    \verb|k|

     \begin{subfigure}[b]{0.1925\textwidth}
         \includegraphics[width=\textwidth,trim={1.09cm 1.04cm 0.86cm 0.35cm},clip]{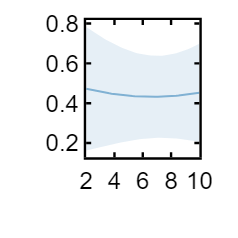}
         \label{fig:ax-0-k}
         \caption{\texttt{GPEI} fold 0}
     \end{subfigure}
     \hfill
     \begin{subfigure}[b]{0.1925\textwidth}
         \includegraphics[width=\textwidth,trim={1.09cm 1.04cm 0.86cm 0.35cm},clip]{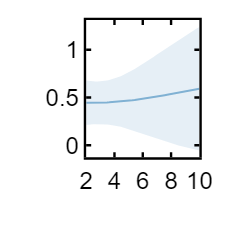}
         \label{fig:ax-1-k}
         \caption{\texttt{GPEI} fold 1}
     \end{subfigure}
     \hfill
     \begin{subfigure}[b]{0.1925\textwidth}
         \includegraphics[width=\textwidth,trim={1.09cm 1.04cm 0.86cm 0.35cm},clip]{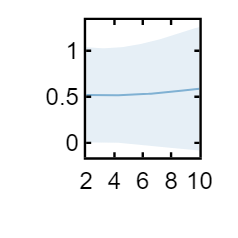}
         \label{fig:ax-2-k}
         \caption{\texttt{GPEI} fold 2}
     \end{subfigure}
     \hfill
     \begin{subfigure}[b]{0.1925\textwidth}
         \includegraphics[width=\textwidth,trim={1.09cm 1.04cm 0.86cm 0.35cm},clip]{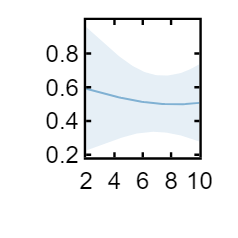}
         \label{fig:ax-3-k}
         \caption{\texttt{GPEI} fold 3}
     \end{subfigure}
     \hfill
     \begin{subfigure}[b]{0.1925\textwidth}
         \includegraphics[width=\textwidth,trim={1.09cm 1.04cm 0.86cm 0.35cm},clip]{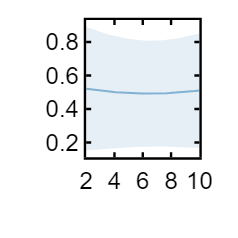}
         \label{fig:ax-4-k}
         \caption{\texttt{GPEI} fold 4}
     \end{subfigure}
     
     \begin{subfigure}[b]{0.1925\textwidth}
         \includegraphics[width=\textwidth,trim={1.09cm 1.04cm 0.86cm 0.35cm},clip]{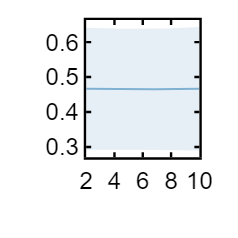}
         \label{fig:saas-0-k}
         \caption{\texttt{SAASBO} fold 0}
     \end{subfigure}
     \hfill
     \begin{subfigure}[b]{0.1925\textwidth}
         \includegraphics[width=\textwidth,trim={1.09cm 1.04cm 0.86cm 0.35cm},clip]{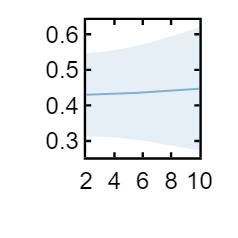}
         \label{fig:saas-1-k}
         \caption{\texttt{SAASBO} fold 1}
     \end{subfigure}
     \hfill
     \begin{subfigure}[b]{0.1925\textwidth}
         \includegraphics[width=\textwidth,trim={1.09cm 1.04cm 0.86cm 0.35cm},clip]{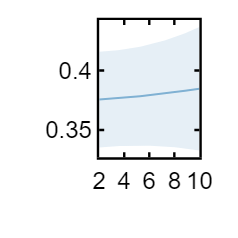}
         \label{fig:saas-2-k}
         \caption{\texttt{SAASBO} fold 2}
     \end{subfigure}
     \hfill
     \begin{subfigure}[b]{0.1925\textwidth}
         \includegraphics[width=\textwidth,trim={1.09cm 1.04cm 0.86cm 0.35cm},clip]{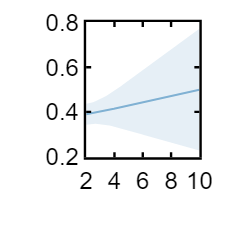}
         \label{fig:saas-3-k}
         \caption{\texttt{SAASBO} fold 3}
     \end{subfigure}
     \hfill
     \begin{subfigure}[b]{0.1925\textwidth}
         \includegraphics[width=\textwidth,trim={1.09cm 1.04cm 0.86cm 0.35cm},clip]{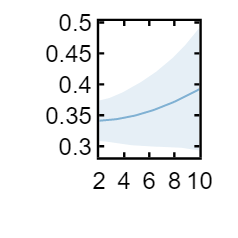}
         \label{fig:saas-4-k}
         \caption{\texttt{SAASBO} fold 4}
     \end{subfigure}

     \caption{One-dimensional (1D) slices of \gls{mae} (eV) vs. \texttt{k} through the \texttt{GPEI} and \texttt{SAASBO} parameter spaces with the rest of the parameters fixed to the mean and mode of the numeric and categorical parameters for each of the models, respectively. Shaded blue error bands give the standard deviation uncertainty predicted by the model. . }
     \label{fig:1d-k}
\end{figure*}

\begin{figure*}
    \centering
    \verb|lr|

     \begin{subfigure}[b]{0.1925\textwidth}
         \includegraphics[width=\textwidth,trim={1.09cm 1.04cm 0.86cm 0.35cm},clip]{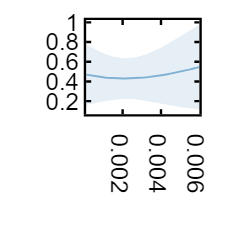}
         \label{fig:ax-0-lr}
         \caption{\texttt{GPEI} fold 0}
     \end{subfigure}
     \hfill
     \begin{subfigure}[b]{0.1925\textwidth}
         \includegraphics[width=\textwidth,trim={1.09cm 1.04cm 0.86cm 0.35cm},clip]{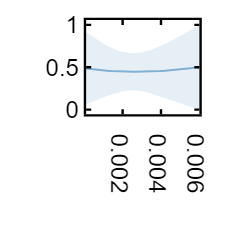}
         \label{fig:ax-1-lr}
         \caption{\texttt{GPEI} fold 1}
     \end{subfigure}
     \hfill
     \begin{subfigure}[b]{0.1925\textwidth}
         \includegraphics[width=\textwidth,trim={1.09cm 1.04cm 0.86cm 0.35cm},clip]{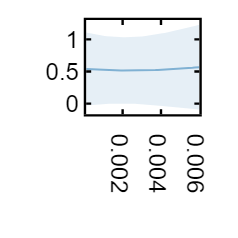}
         \label{fig:ax-2-lr}
         \caption{\texttt{GPEI} fold 2}
     \end{subfigure}
     \hfill
     \begin{subfigure}[b]{0.1925\textwidth}
         \includegraphics[width=\textwidth,trim={1.09cm 1.04cm 0.86cm 0.35cm},clip]{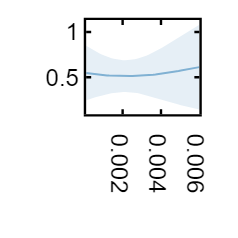}
         \label{fig:ax-3-lr}
         \caption{\texttt{GPEI} fold 3}
     \end{subfigure}
     \hfill
     \begin{subfigure}[b]{0.1925\textwidth}
         \includegraphics[width=\textwidth,trim={1.09cm 1.04cm 0.86cm 0.35cm},clip]{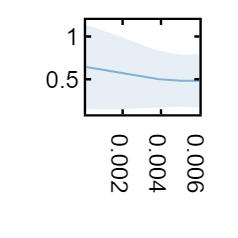}
         \label{fig:ax-4-lr}
         \caption{\texttt{GPEI} fold 4}
     \end{subfigure}
     
     \begin{subfigure}[b]{0.1925\textwidth}
         \includegraphics[width=\textwidth,trim={1.09cm 1.04cm 0.86cm 0.35cm},clip]{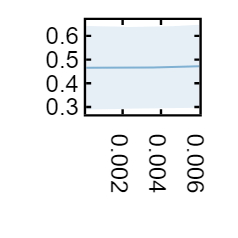}
         \label{fig:saas-0-lr}
         \caption{\texttt{SAASBO} fold 0}
     \end{subfigure}
     \hfill
     \begin{subfigure}[b]{0.1925\textwidth}
         \includegraphics[width=\textwidth,trim={1.09cm 1.04cm 0.86cm 0.35cm},clip]{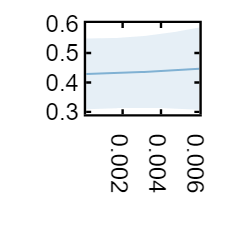}
         \label{fig:saas-1-lr}
         \caption{\texttt{SAASBO} fold 1}
     \end{subfigure}
     \hfill
     \begin{subfigure}[b]{0.1925\textwidth}
         \includegraphics[width=\textwidth,trim={1.09cm 1.04cm 0.86cm 0.35cm},clip]{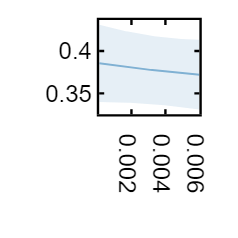}
         \label{fig:saas-2-lr}
         \caption{\texttt{SAASBO} fold 2}
     \end{subfigure}
     \hfill
     \begin{subfigure}[b]{0.1925\textwidth}
         \includegraphics[width=\textwidth,trim={1.09cm 1.04cm 0.86cm 0.35cm},clip]{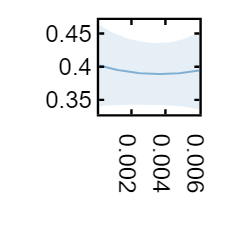}
         \label{fig:saas-3-lr}
         \caption{\texttt{SAASBO} fold 3}
     \end{subfigure}
     \hfill
     \begin{subfigure}[b]{0.1925\textwidth}
         \includegraphics[width=\textwidth,trim={1.09cm 1.04cm 0.86cm 0.35cm},clip]{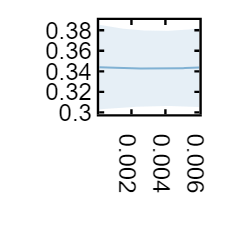}
         \label{fig:saas-4-lr}
         \caption{\texttt{SAASBO} fold 4}
     \end{subfigure}

     \caption{One-dimensional (1D) slices of \gls{mae} (eV) vs. \texttt{lr} through the \texttt{GPEI} and \texttt{SAASBO} parameter spaces with the rest of the parameters fixed to the mean and mode of the numeric and categorical parameters for each of the models, respectively. Shaded blue error bands give the standard deviation uncertainty predicted by the model. . }
     \label{fig:1d-lr}
\end{figure*}

\begin{figure*}
    \centering
    \verb|pe_resolution|

     \begin{subfigure}[b]{0.1925\textwidth}
         \includegraphics[width=\textwidth,trim={1.09cm 1.04cm 0.86cm 0.35cm},clip]{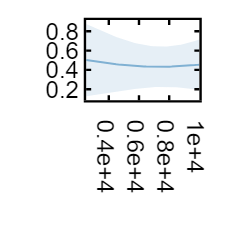}
         \label{fig:ax-0-pe_resolution}
         \caption{\texttt{GPEI} fold 0}
     \end{subfigure}
     \hfill
     \begin{subfigure}[b]{0.1925\textwidth}
         \includegraphics[width=\textwidth,trim={1.09cm 1.04cm 0.86cm 0.35cm},clip]{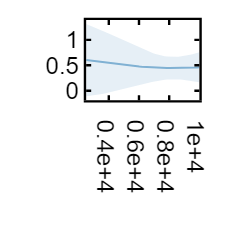}
         \label{fig:ax-1-pe_resolution}
         \caption{\texttt{GPEI} fold 1}
     \end{subfigure}
     \hfill
     \begin{subfigure}[b]{0.1925\textwidth}
         \includegraphics[width=\textwidth,trim={1.09cm 1.04cm 0.86cm 0.35cm},clip]{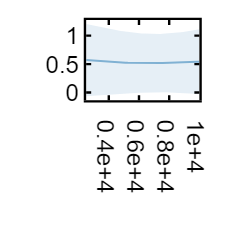}
         \label{fig:ax-2-pe_resolution}
         \caption{\texttt{GPEI} fold 2}
     \end{subfigure}
     \hfill
     \begin{subfigure}[b]{0.1925\textwidth}
         \includegraphics[width=\textwidth,trim={1.09cm 1.04cm 0.86cm 0.35cm},clip]{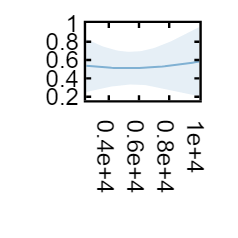}
         \label{fig:ax-3-pe_resolution}
         \caption{\texttt{GPEI} fold 3}
     \end{subfigure}
     \hfill
     \begin{subfigure}[b]{0.1925\textwidth}
         \includegraphics[width=\textwidth,trim={1.09cm 1.04cm 0.86cm 0.35cm},clip]{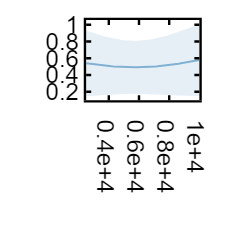}
         \label{fig:ax-4-pe_resolution}
         \caption{\texttt{GPEI} fold 4}
     \end{subfigure}
     
     \begin{subfigure}[b]{0.1925\textwidth}
         \includegraphics[width=\textwidth,trim={1.09cm 1.04cm 0.86cm 0.35cm},clip]{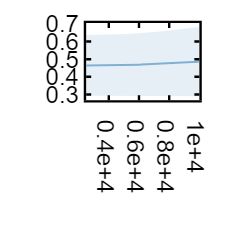}
         \label{fig:saas-0-pe_resolution}
         \caption{\texttt{SAASBO} fold 0}
     \end{subfigure}
     \hfill
     \begin{subfigure}[b]{0.1925\textwidth}
         \includegraphics[width=\textwidth,trim={1.09cm 1.04cm 0.86cm 0.35cm},clip]{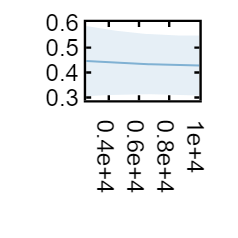}
         \label{fig:saas-1-pe_resolution}
         \caption{\texttt{SAASBO} fold 1}
     \end{subfigure}
     \hfill
     \begin{subfigure}[b]{0.1925\textwidth}
         \includegraphics[width=\textwidth,trim={1.09cm 1.04cm 0.86cm 0.35cm},clip]{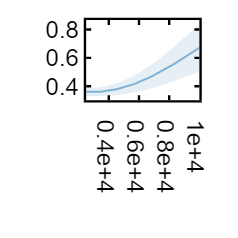}
         \label{fig:saas-2-pe_resolution}
         \caption{\texttt{SAASBO} fold 2}
     \end{subfigure}
     \hfill
     \begin{subfigure}[b]{0.1925\textwidth}
         \includegraphics[width=\textwidth,trim={1.09cm 1.04cm 0.86cm 0.35cm},clip]{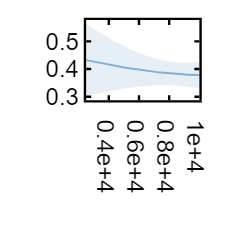}
         \label{fig:saas-3-pe_resolution}
         \caption{\texttt{SAASBO} fold 3}
     \end{subfigure}
     \hfill
     \begin{subfigure}[b]{0.1925\textwidth}
         \includegraphics[width=\textwidth,trim={1.09cm 1.04cm 0.86cm 0.35cm},clip]{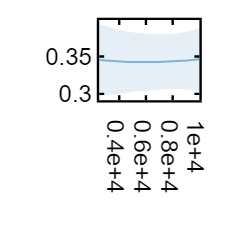}
         \label{fig:saas-4-pe_resolution}
         \caption{\texttt{SAASBO} fold 4}
     \end{subfigure}

     \caption{One-dimensional (1D) slices of \gls{mae} (eV) vs. \texttt{pe\_resolution} through the \texttt{GPEI} and \texttt{SAASBO} parameter spaces with the rest of the parameters fixed to the mean and mode of the numeric and categorical parameters for each of the models, respectively. Shaded blue error bands give the standard deviation uncertainty predicted by the model. . }
     \label{fig:1d-pe_resolution}
\end{figure*}

\begin{figure*}
    \centering
    \verb|ple_resolution|

     \begin{subfigure}[b]{0.1925\textwidth}
         \includegraphics[width=\textwidth,trim={1.09cm 1.04cm 0.86cm 0.35cm},clip]{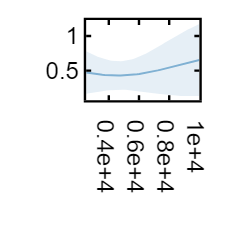}
         \label{fig:ax-0-ple_resolution}
         \caption{\texttt{GPEI} fold 0}
     \end{subfigure}
     \hfill
     \begin{subfigure}[b]{0.1925\textwidth}
         \includegraphics[width=\textwidth,trim={1.09cm 1.04cm 0.86cm 0.35cm},clip]{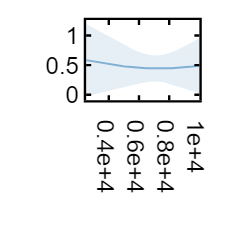}
         \label{fig:ax-1-ple_resolution}
         \caption{\texttt{GPEI} fold 1}
     \end{subfigure}
     \hfill
     \begin{subfigure}[b]{0.1925\textwidth}
         \includegraphics[width=\textwidth,trim={1.09cm 1.04cm 0.86cm 0.35cm},clip]{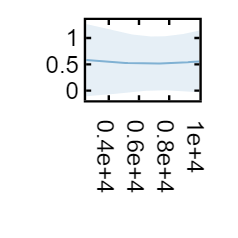}
         \label{fig:ax-2-ple_resolution}
         \caption{\texttt{GPEI} fold 2}
     \end{subfigure}
     \hfill
     \begin{subfigure}[b]{0.1925\textwidth}
         \includegraphics[width=\textwidth,trim={1.09cm 1.04cm 0.86cm 0.35cm},clip]{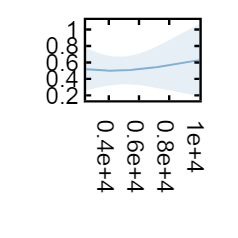}
         \label{fig:ax-3-ple_resolution}
         \caption{\texttt{GPEI} fold 3}
     \end{subfigure}
     \hfill
     \begin{subfigure}[b]{0.1925\textwidth}
         \includegraphics[width=\textwidth,trim={1.09cm 1.04cm 0.86cm 0.35cm},clip]{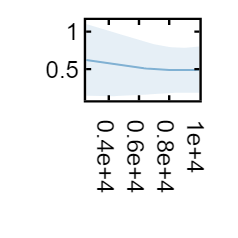}
         \label{fig:ax-4-ple_resolution}
         \caption{\texttt{GPEI} fold 4}
     \end{subfigure}
     
     \begin{subfigure}[b]{0.1925\textwidth}
         \includegraphics[width=\textwidth,trim={1.09cm 1.04cm 0.86cm 0.35cm},clip]{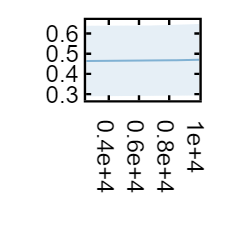}
         \label{fig:saas-0-ple_resolution}
         \caption{\texttt{SAASBO} fold 0}
     \end{subfigure}
     \hfill
     \begin{subfigure}[b]{0.1925\textwidth}
         \includegraphics[width=\textwidth,trim={1.09cm 1.04cm 0.86cm 0.35cm},clip]{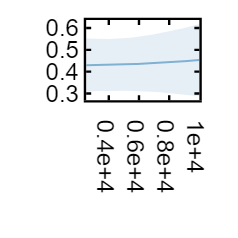}
         \label{fig:saas-1-ple_resolution}
         \caption{\texttt{SAASBO} fold 1}
     \end{subfigure}
     \hfill
     \begin{subfigure}[b]{0.1925\textwidth}
         \includegraphics[width=\textwidth,trim={1.09cm 1.04cm 0.86cm 0.35cm},clip]{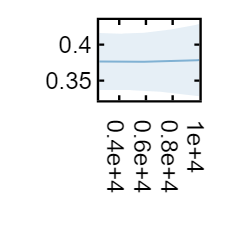}
         \label{fig:saas-2-ple_resolution}
         \caption{\texttt{SAASBO} fold 2}
     \end{subfigure}
     \hfill
     \begin{subfigure}[b]{0.1925\textwidth}
         \includegraphics[width=\textwidth,trim={1.09cm 1.04cm 0.86cm 0.35cm},clip]{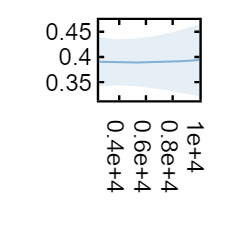}
         \label{fig:saas-3-ple_resolution}
         \caption{\texttt{SAASBO} fold 3}
     \end{subfigure}
     \hfill
     \begin{subfigure}[b]{0.1925\textwidth}
         \includegraphics[width=\textwidth,trim={1.09cm 1.04cm 0.86cm 0.35cm},clip]{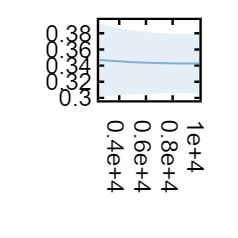}
         \label{fig:saas-4-ple_resolution}
         \caption{\texttt{SAASBO} fold 4}
     \end{subfigure}

     \caption{One-dimensional (1D) slices of \gls{mae} (eV) vs. \texttt{ple\_resolution} through the \texttt{GPEI} and \texttt{SAASBO} parameter spaces with the rest of the parameters fixed to the mean and mode of the numeric and categorical parameters for each of the models, respectively. Shaded blue error bands give the standard deviation uncertainty predicted by the model. . }
     \label{fig:1d-ple_resolution}
\end{figure*}

\begin{figure*}
    \centering
    \verb|pos_scaler|

     \begin{subfigure}[b]{0.1925\textwidth}
         \includegraphics[width=\textwidth,trim={1.09cm 1.04cm 0.86cm 0.35cm},clip]{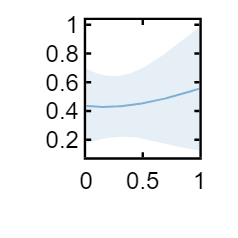}
         \label{fig:ax-0-pos_scaler}
         \caption{\texttt{GPEI} fold 0}
     \end{subfigure}
     \hfill
     \begin{subfigure}[b]{0.1925\textwidth}
         \includegraphics[width=\textwidth,trim={1.09cm 1.04cm 0.86cm 0.35cm},clip]{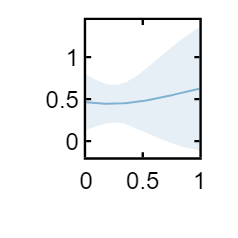}
         \label{fig:ax-1-pos_scaler}
         \caption{\texttt{GPEI} fold 1}
     \end{subfigure}
     \hfill
     \begin{subfigure}[b]{0.1925\textwidth}
         \includegraphics[width=\textwidth,trim={1.09cm 1.04cm 0.86cm 0.35cm},clip]{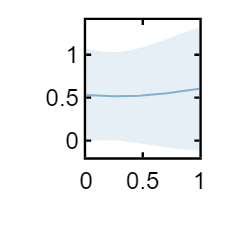}
         \label{fig:ax-2-pos_scaler}
         \caption{\texttt{GPEI} fold 2}
     \end{subfigure}
     \hfill
     \begin{subfigure}[b]{0.1925\textwidth}
         \includegraphics[width=\textwidth,trim={1.09cm 1.04cm 0.86cm 0.35cm},clip]{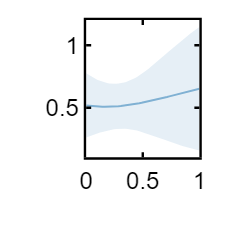}
         \label{fig:ax-3-pos_scaler}
         \caption{\texttt{GPEI} fold 3}
     \end{subfigure}
     \hfill
     \begin{subfigure}[b]{0.1925\textwidth}
         \includegraphics[width=\textwidth,trim={1.09cm 1.04cm 0.86cm 0.35cm},clip]{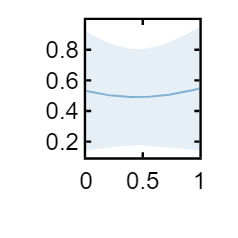}
         \label{fig:ax-4-pos_scaler}
         \caption{\texttt{GPEI} fold 4}
     \end{subfigure}
     
     \begin{subfigure}[b]{0.1925\textwidth}
         \includegraphics[width=\textwidth,trim={1.09cm 1.04cm 0.86cm 0.35cm},clip]{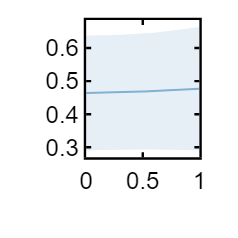}
         \label{fig:saas-0-pos_scaler}
         \caption{\texttt{SAASBO} fold 0}
     \end{subfigure}
     \hfill
     \begin{subfigure}[b]{0.1925\textwidth}
         \includegraphics[width=\textwidth,trim={1.09cm 1.04cm 0.86cm 0.35cm},clip]{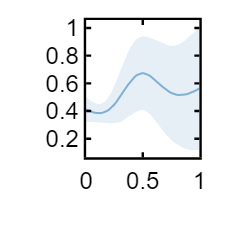}
         \label{fig:saas-1-pos_scaler}
         \caption{\texttt{SAASBO} fold 1}
     \end{subfigure}
     \hfill
     \begin{subfigure}[b]{0.1925\textwidth}
         \includegraphics[width=\textwidth,trim={1.09cm 1.04cm 0.86cm 0.35cm},clip]{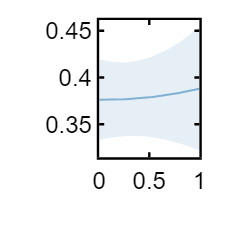}
         \label{fig:saas-2-pos_scaler}
         \caption{\texttt{SAASBO} fold 2}
     \end{subfigure}
     \hfill
     \begin{subfigure}[b]{0.1925\textwidth}
         \includegraphics[width=\textwidth,trim={1.09cm 1.04cm 0.86cm 0.35cm},clip]{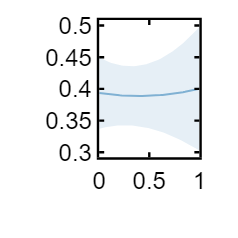}
         \label{fig:saas-3-pos_scaler}
         \caption{\texttt{SAASBO} fold 3}
     \end{subfigure}
     \hfill
     \begin{subfigure}[b]{0.1925\textwidth}
         \includegraphics[width=\textwidth,trim={1.09cm 1.04cm 0.86cm 0.35cm},clip]{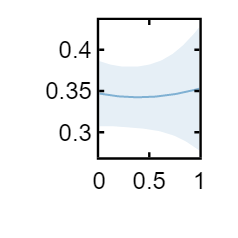}
         \label{fig:saas-4-pos_scaler}
         \caption{\texttt{SAASBO} fold 4}
     \end{subfigure}

     \caption{One-dimensional (1D) slices of \gls{mae} (eV) vs. \texttt{pos\_scaler} through the \texttt{GPEI} and \texttt{SAASBO} parameter spaces with the rest of the parameters fixed to the mean and mode of the numeric and categorical parameters for each of the models, respectively. Shaded blue error bands give the standard deviation uncertainty predicted by the model. . }
     \label{fig:1d-pos_scaler}
\end{figure*}

\begin{figure*}
    \centering
    \verb|weight_decay|

     \begin{subfigure}[b]{0.1925\textwidth}
         \includegraphics[width=\textwidth,trim={1.09cm 1.04cm 0.86cm 0.35cm},clip]{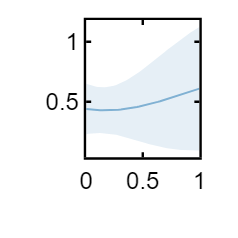}
         \label{fig:ax-0-weight_decay}
         \caption{\texttt{GPEI} fold 0}
     \end{subfigure}
     \hfill
     \begin{subfigure}[b]{0.1925\textwidth}
         \includegraphics[width=\textwidth,trim={1.09cm 1.04cm 0.86cm 0.35cm},clip]{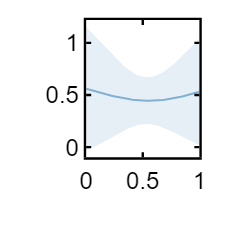}
         \label{fig:ax-1-weight_decay}
         \caption{\texttt{GPEI} fold 1}
     \end{subfigure}
     \hfill
     \begin{subfigure}[b]{0.1925\textwidth}
         \includegraphics[width=\textwidth,trim={1.09cm 1.04cm 0.86cm 0.35cm},clip]{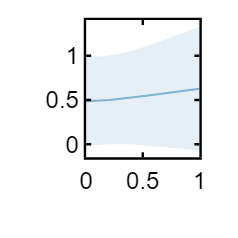}
         \label{fig:ax-2-weight_decay}
         \caption{\texttt{GPEI} fold 2}
     \end{subfigure}
     \hfill
     \begin{subfigure}[b]{0.1925\textwidth}
         \includegraphics[width=\textwidth,trim={1.09cm 1.04cm 0.86cm 0.35cm},clip]{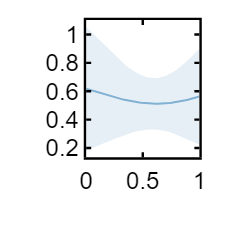}
         \label{fig:ax-3-weight_decay}
         \caption{\texttt{GPEI} fold 3}
     \end{subfigure}
     \hfill
     \begin{subfigure}[b]{0.1925\textwidth}
         \includegraphics[width=\textwidth,trim={1.09cm 1.04cm 0.86cm 0.35cm},clip]{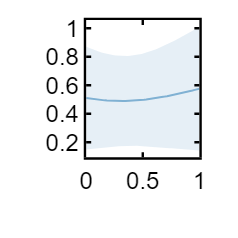}
         \label{fig:ax-4-weight_decay}
         \caption{\texttt{GPEI} fold 4}
     \end{subfigure}
     
     \begin{subfigure}[b]{0.1925\textwidth}
         \includegraphics[width=\textwidth,trim={1.09cm 1.04cm 0.86cm 0.35cm},clip]{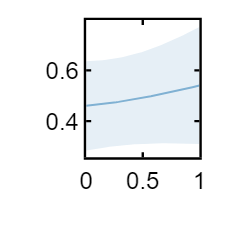}
         \label{fig:saas-0-weight_decay}
         \caption{\texttt{SAASBO} fold 0}
     \end{subfigure}
     \hfill
     \begin{subfigure}[b]{0.1925\textwidth}
         \includegraphics[width=\textwidth,trim={1.09cm 1.04cm 0.86cm 0.35cm},clip]{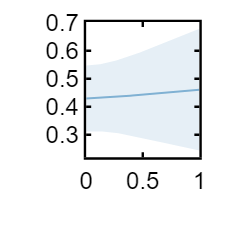}
         \label{fig:saas-1-weight_decay}
         \caption{\texttt{SAASBO} fold 1}
     \end{subfigure}
     \hfill
     \begin{subfigure}[b]{0.1925\textwidth}
         \includegraphics[width=\textwidth,trim={1.09cm 1.04cm 0.86cm 0.35cm},clip]{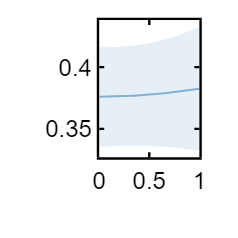}
         \label{fig:saas-2-weight_decay}
         \caption{\texttt{SAASBO} fold 2}
     \end{subfigure}
     \hfill
     \begin{subfigure}[b]{0.1925\textwidth}
         \includegraphics[width=\textwidth,trim={1.09cm 1.04cm 0.86cm 0.35cm},clip]{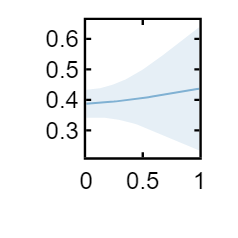}
         \label{fig:saas-3-weight_decay}
         \caption{\texttt{SAASBO} fold 3}
     \end{subfigure}
     \hfill
     \begin{subfigure}[b]{0.1925\textwidth}
         \includegraphics[width=\textwidth,trim={1.09cm 1.04cm 0.86cm 0.35cm},clip]{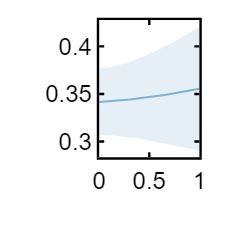}
         \label{fig:saas-4-weight_decay}
         \caption{\texttt{SAASBO} fold 4}
     \end{subfigure}

     \caption{One-dimensional (1D) slices of \gls{mae} (eV) vs. \texttt{weight\_decay} through the \texttt{GPEI} and \texttt{SAASBO} parameter spaces with the rest of the parameters fixed to the mean and mode of the numeric and categorical parameters for each of the models, respectively. Shaded blue error bands give the standard deviation uncertainty predicted by the model. . }
     \label{fig:1d-weight_decay}
\end{figure*}

\begin{figure*}
    \centering
    \verb|batch_size|

     \begin{subfigure}[b]{0.1925\textwidth}
         \includegraphics[width=\textwidth,trim={1.09cm 1.04cm 0.86cm 0.35cm},clip]{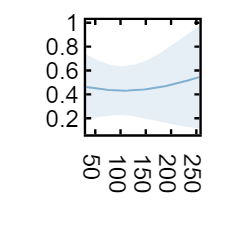}
         \label{fig:ax-0-batch_size}
         \caption{\texttt{GPEI} fold 0}
     \end{subfigure}
     \hfill
     \begin{subfigure}[b]{0.1925\textwidth}
         \includegraphics[width=\textwidth,trim={1.09cm 1.04cm 0.86cm 0.35cm},clip]{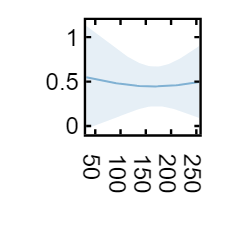}
         \label{fig:ax-1-batch_size}
         \caption{\texttt{GPEI} fold 1}
     \end{subfigure}
     \hfill
     \begin{subfigure}[b]{0.1925\textwidth}
         \includegraphics[width=\textwidth,trim={1.09cm 1.04cm 0.86cm 0.35cm},clip]{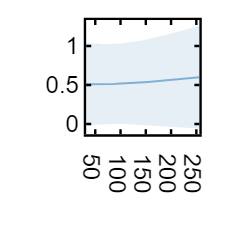}
         \label{fig:ax-2-batch_size}
         \caption{\texttt{GPEI} fold 2}
     \end{subfigure}
     \hfill
     \begin{subfigure}[b]{0.1925\textwidth}
         \includegraphics[width=\textwidth,trim={1.09cm 1.04cm 0.86cm 0.35cm},clip]{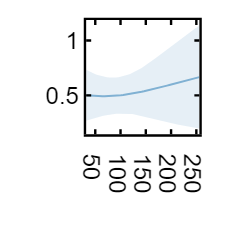}
         \label{fig:ax-3-batch_size}
         \caption{\texttt{GPEI} fold 3}
     \end{subfigure}
     \hfill
     \begin{subfigure}[b]{0.1925\textwidth}
         \includegraphics[width=\textwidth,trim={1.09cm 1.04cm 0.86cm 0.35cm},clip]{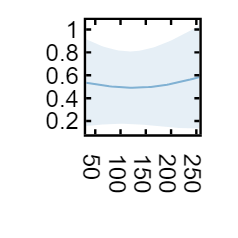}
         \label{fig:ax-4-batch_size}
         \caption{\texttt{GPEI} fold 4}
     \end{subfigure}
     
     \begin{subfigure}[b]{0.1925\textwidth}
         \includegraphics[width=\textwidth,trim={1.09cm 1.04cm 0.86cm 0.35cm},clip]{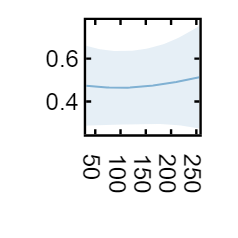}
         \label{fig:saas-0-batch_size}
         \caption{\texttt{SAASBO} fold 0}
     \end{subfigure}
     \hfill
     \begin{subfigure}[b]{0.1925\textwidth}
         \includegraphics[width=\textwidth,trim={1.09cm 1.04cm 0.86cm 0.35cm},clip]{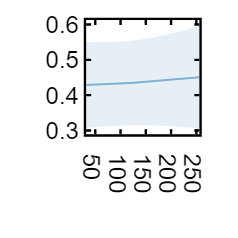}
         \label{fig:saas-1-batch_size}
         \caption{\texttt{SAASBO} fold 1}
     \end{subfigure}
     \hfill
     \begin{subfigure}[b]{0.1925\textwidth}
         \includegraphics[width=\textwidth,trim={1.09cm 1.04cm 0.86cm 0.35cm},clip]{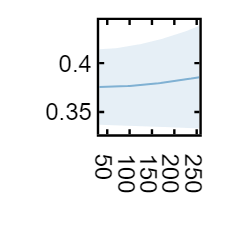}
         \label{fig:saas-2-batch_size}
         \caption{\texttt{SAASBO} fold 2}
     \end{subfigure}
     \hfill
     \begin{subfigure}[b]{0.1925\textwidth}
         \includegraphics[width=\textwidth,trim={1.09cm 1.04cm 0.86cm 0.35cm},clip]{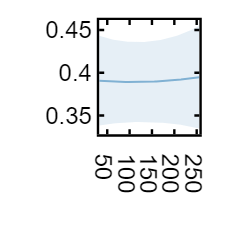}
         \label{fig:saas-3-batch_size}
         \caption{\texttt{SAASBO} fold 3}
     \end{subfigure}
     \hfill
     \begin{subfigure}[b]{0.1925\textwidth}
         \includegraphics[width=\textwidth,trim={1.09cm 1.04cm 0.86cm 0.35cm},clip]{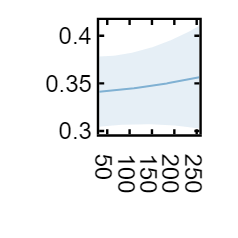}
         \label{fig:saas-4-batch_size}
         \caption{\texttt{SAASBO} fold 4}
     \end{subfigure}

     \caption{One-dimensional (1D) slices of \gls{mae} (eV) vs. \texttt{batch\_size} through the \texttt{GPEI} and \texttt{SAASBO} parameter spaces with the rest of the parameters fixed to the mean and mode of the numeric and categorical parameters for each of the models, respectively. Shaded blue error bands give the standard deviation uncertainty predicted by the model. . }
     \label{fig:1d-batch_size}
\end{figure*}

\begin{figure*}
    \centering
    \verb|out_hidden4|

     \begin{subfigure}[b]{0.1925\textwidth}
         \includegraphics[width=\textwidth,trim={1.09cm 1.04cm 0.86cm 0.35cm},clip]{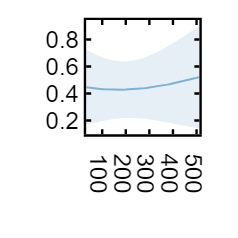}
         \label{fig:ax-0-out_hidden4}
         \caption{\texttt{GPEI} fold 0}
     \end{subfigure}
     \hfill
     \begin{subfigure}[b]{0.1925\textwidth}
         \includegraphics[width=\textwidth,trim={1.09cm 1.04cm 0.86cm 0.35cm},clip]{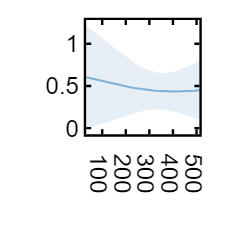}
         \label{fig:ax-1-out_hidden4}
         \caption{\texttt{GPEI} fold 1}
     \end{subfigure}
     \hfill
     \begin{subfigure}[b]{0.1925\textwidth}
         \includegraphics[width=\textwidth,trim={1.09cm 1.04cm 0.86cm 0.35cm},clip]{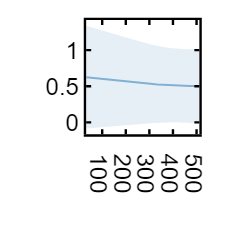}
         \label{fig:ax-2-out_hidden4}
         \caption{\texttt{GPEI} fold 2}
     \end{subfigure}
     \hfill
     \begin{subfigure}[b]{0.1925\textwidth}
         \includegraphics[width=\textwidth,trim={1.09cm 1.04cm 0.86cm 0.35cm},clip]{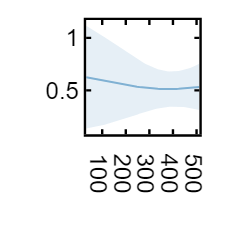}
         \label{fig:ax-3-out_hidden4}
         \caption{\texttt{GPEI} fold 3}
     \end{subfigure}
     \hfill
     \begin{subfigure}[b]{0.1925\textwidth}
         \includegraphics[width=\textwidth,trim={1.09cm 1.04cm 0.86cm 0.35cm},clip]{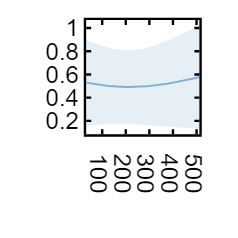}
         \label{fig:ax-4-out_hidden4}
         \caption{\texttt{GPEI} fold 4}
     \end{subfigure}
     
     \begin{subfigure}[b]{0.1925\textwidth}
         \includegraphics[width=\textwidth,trim={1.09cm 1.04cm 0.86cm 0.35cm},clip]{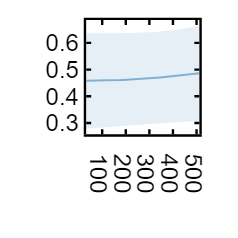}
         \label{fig:saas-0-out_hidden4}
         \caption{\texttt{SAASBO} fold 0}
     \end{subfigure}
     \hfill
     \begin{subfigure}[b]{0.1925\textwidth}
         \includegraphics[width=\textwidth,trim={1.09cm 1.04cm 0.86cm 0.35cm},clip]{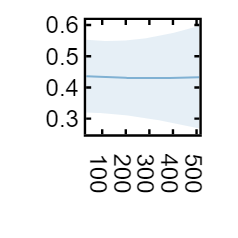}
         \label{fig:saas-1-out_hidden4}
         \caption{\texttt{SAASBO} fold 1}
     \end{subfigure}
     \hfill
     \begin{subfigure}[b]{0.1925\textwidth}
         \includegraphics[width=\textwidth,trim={1.09cm 1.04cm 0.86cm 0.35cm},clip]{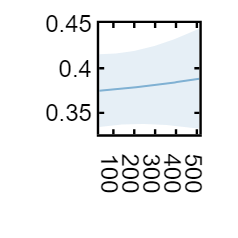}
         \label{fig:saas-2-out_hidden4}
         \caption{\texttt{SAASBO} fold 2}
     \end{subfigure}
     \hfill
     \begin{subfigure}[b]{0.1925\textwidth}
         \includegraphics[width=\textwidth,trim={1.09cm 1.04cm 0.86cm 0.35cm},clip]{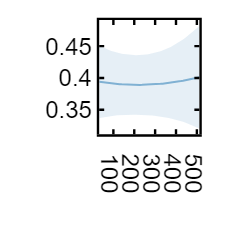}
         \label{fig:saas-3-out_hidden4}
         \caption{\texttt{SAASBO} fold 3}
     \end{subfigure}
     \hfill
     \begin{subfigure}[b]{0.1925\textwidth}
         \includegraphics[width=\textwidth,trim={1.09cm 1.04cm 0.86cm 0.35cm},clip]{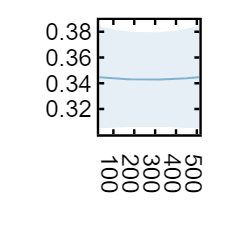}
         \label{fig:saas-4-out_hidden4}
         \caption{\texttt{SAASBO} fold 4}
     \end{subfigure}

     \caption{One-dimensional (1D) slices of \gls{mae} (eV) vs. \texttt{out\_hidden4} through the \texttt{GPEI} and \texttt{SAASBO} parameter spaces with the rest of the parameters fixed to the mean and mode of the numeric and categorical parameters for each of the models, respectively. Shaded blue error bands give the standard deviation uncertainty predicted by the model. . }
     \label{fig:1d-out_hidden4}
\end{figure*}

\begin{figure*}
    \centering
    \verb|betas1|

     \begin{subfigure}[b]{0.1925\textwidth}
         \includegraphics[width=\textwidth,trim={1.09cm 1.04cm 0.86cm 0.35cm},clip]{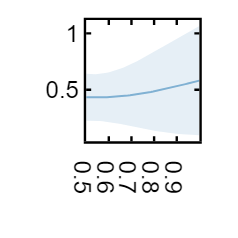}
         \label{fig:ax-0-betas1}
         \caption{\texttt{GPEI} fold 0}
     \end{subfigure}
     \hfill
     \begin{subfigure}[b]{0.1925\textwidth}
         \includegraphics[width=\textwidth,trim={1.09cm 1.04cm 0.86cm 0.35cm},clip]{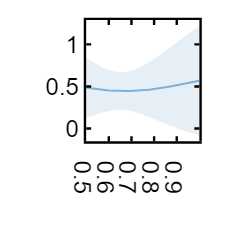}
         \label{fig:ax-1-betas1}
         \caption{\texttt{GPEI} fold 1}
     \end{subfigure}
     \hfill
     \begin{subfigure}[b]{0.1925\textwidth}
         \includegraphics[width=\textwidth,trim={1.09cm 1.04cm 0.86cm 0.35cm},clip]{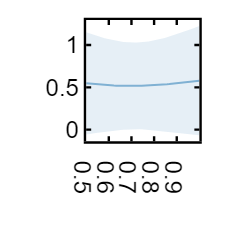}
         \label{fig:ax-2-betas1}
         \caption{\texttt{GPEI} fold 2}
     \end{subfigure}
     \hfill
     \begin{subfigure}[b]{0.1925\textwidth}
         \includegraphics[width=\textwidth,trim={1.09cm 1.04cm 0.86cm 0.35cm},clip]{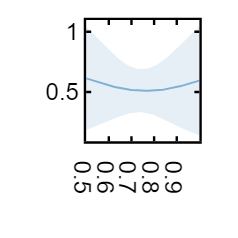}
         \label{fig:ax-3-betas1}
         \caption{\texttt{GPEI} fold 3}
     \end{subfigure}
     \hfill
     \begin{subfigure}[b]{0.1925\textwidth}
         \includegraphics[width=\textwidth,trim={1.09cm 1.04cm 0.86cm 0.35cm},clip]{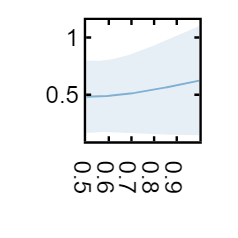}
         \label{fig:ax-4-betas1}
         \caption{\texttt{GPEI} fold 4}
     \end{subfigure}
     
     \begin{subfigure}[b]{0.1925\textwidth}
         \includegraphics[width=\textwidth,trim={1.09cm 1.04cm 0.86cm 0.35cm},clip]{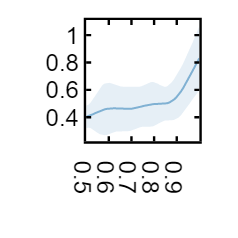}
         \label{fig:saas-0-betas1}
         \caption{\texttt{SAASBO} fold 0}
     \end{subfigure}
     \hfill
     \begin{subfigure}[b]{0.1925\textwidth}
         \includegraphics[width=\textwidth,trim={1.09cm 1.04cm 0.86cm 0.35cm},clip]{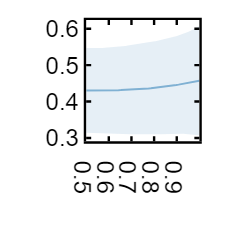}
         \label{fig:saas-1-betas1}
         \caption{\texttt{SAASBO} fold 1}
     \end{subfigure}
     \hfill
     \begin{subfigure}[b]{0.1925\textwidth}
         \includegraphics[width=\textwidth,trim={1.09cm 1.04cm 0.86cm 0.35cm},clip]{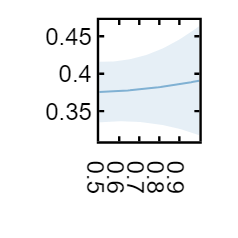}
         \label{fig:saas-2-betas1}
         \caption{\texttt{SAASBO} fold 2}
     \end{subfigure}
     \hfill
     \begin{subfigure}[b]{0.1925\textwidth}
         \includegraphics[width=\textwidth,trim={1.09cm 1.04cm 0.86cm 0.35cm},clip]{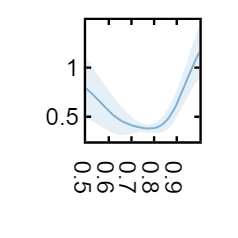}
         \label{fig:saas-3-betas1}
         \caption{\texttt{SAASBO} fold 3}
     \end{subfigure}
     \hfill
     \begin{subfigure}[b]{0.1925\textwidth}
         \includegraphics[width=\textwidth,trim={1.09cm 1.04cm 0.86cm 0.35cm},clip]{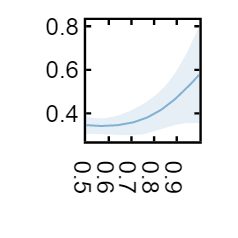}
         \label{fig:saas-4-betas1}
         \caption{\texttt{SAASBO} fold 4}
     \end{subfigure}

     \caption{One-dimensional (1D) slices of \gls{mae} (eV) vs. \texttt{betas1} through the \texttt{GPEI} and \texttt{SAASBO} parameter spaces with the rest of the parameters fixed to the mean and mode of the numeric and categorical parameters for each of the models, respectively. Shaded blue error bands give the standard deviation uncertainty predicted by the model. . }
     \label{fig:1d-betas1}
\end{figure*}

\begin{figure*}
    \centering
    \verb|betas2|

     \begin{subfigure}[b]{0.1925\textwidth}
         \includegraphics[width=\textwidth,trim={1.09cm 1.04cm 0.86cm 0.35cm},clip]{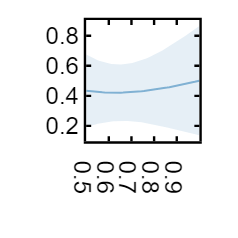}
         \label{fig:ax-0-betas2}
         \caption{\texttt{GPEI} fold 0}
     \end{subfigure}
     \hfill
     \begin{subfigure}[b]{0.1925\textwidth}
         \includegraphics[width=\textwidth,trim={1.09cm 1.04cm 0.86cm 0.35cm},clip]{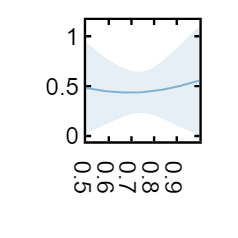}
         \label{fig:ax-1-betas2}
         \caption{\texttt{GPEI} fold 1}
     \end{subfigure}
     \hfill
     \begin{subfigure}[b]{0.1925\textwidth}
         \includegraphics[width=\textwidth,trim={1.09cm 1.04cm 0.86cm 0.35cm},clip]{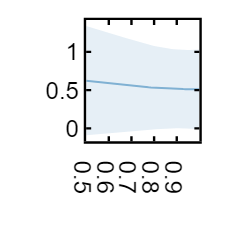}
         \label{fig:ax-2-betas2}
         \caption{\texttt{GPEI} fold 2}
     \end{subfigure}
     \hfill
     \begin{subfigure}[b]{0.1925\textwidth}
         \includegraphics[width=\textwidth,trim={1.09cm 1.04cm 0.86cm 0.35cm},clip]{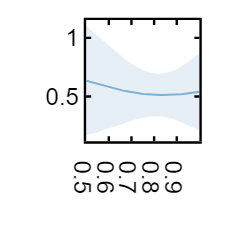}
         \label{fig:ax-3-betas2}
         \caption{\texttt{GPEI} fold 3}
     \end{subfigure}
     \hfill
     \begin{subfigure}[b]{0.1925\textwidth}
         \includegraphics[width=\textwidth,trim={1.09cm 1.04cm 0.86cm 0.35cm},clip]{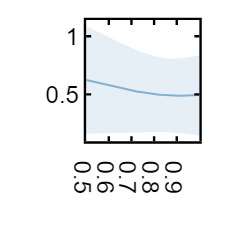}
         \label{fig:ax-4-betas2}
         \caption{\texttt{GPEI} fold 4}
     \end{subfigure}
     
     \begin{subfigure}[b]{0.1925\textwidth}
         \includegraphics[width=\textwidth,trim={1.09cm 1.04cm 0.86cm 0.35cm},clip]{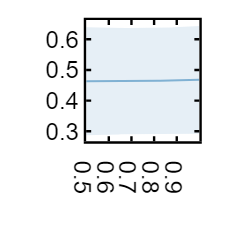}
         \label{fig:saas-0-betas2}
         \caption{\texttt{SAASBO} fold 0}
     \end{subfigure}
     \hfill
     \begin{subfigure}[b]{0.1925\textwidth}
         \includegraphics[width=\textwidth,trim={1.09cm 1.04cm 0.86cm 0.35cm},clip]{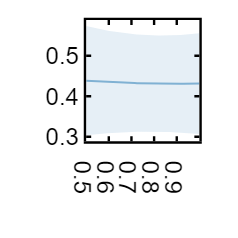}
         \label{fig:saas-1-betas2}
         \caption{\texttt{SAASBO} fold 1}
     \end{subfigure}
     \hfill
     \begin{subfigure}[b]{0.1925\textwidth}
         \includegraphics[width=\textwidth,trim={1.09cm 1.04cm 0.86cm 0.35cm},clip]{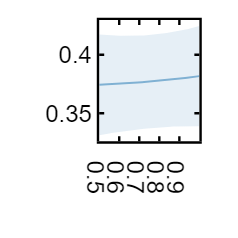}
         \label{fig:saas-2-betas2}
         \caption{\texttt{SAASBO} fold 2}
     \end{subfigure}
     \hfill
     \begin{subfigure}[b]{0.1925\textwidth}
         \includegraphics[width=\textwidth,trim={1.09cm 1.04cm 0.86cm 0.35cm},clip]{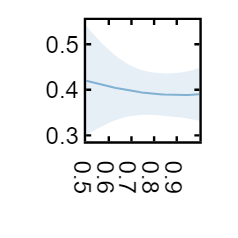}
         \label{fig:saas-3-betas2}
         \caption{\texttt{SAASBO} fold 3}
     \end{subfigure}
     \hfill
     \begin{subfigure}[b]{0.1925\textwidth}
         \includegraphics[width=\textwidth,trim={1.09cm 1.04cm 0.86cm 0.35cm},clip]{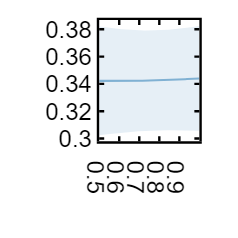}
         \label{fig:saas-4-betas2}
         \caption{\texttt{SAASBO} fold 4}
     \end{subfigure}

     \caption{One-dimensional (1D) slices of \gls{mae} (eV) vs. \texttt{betas2} through the \texttt{GPEI} and \texttt{SAASBO} parameter spaces with the rest of the parameters fixed to the mean and mode of the numeric and categorical parameters for each of the models, respectively. Shaded blue error bands give the standard deviation uncertainty predicted by the model. . }
     \label{fig:1d-betas2}
\end{figure*}

